\documentclass{aa}  

\usepackage{graphicx}
\usepackage{txfonts}
\usepackage{mathtools}	% additional packages
\usepackage{hyperref}
\begin{document} 

   \title{Eccentricity excitation and merging of planetary \\
   embryos heated by pebble accretion}
   \titlerunning{Eccentricity excitation and merging of planetary
   embryos heated by pebble accretion}
   \authorrunning{O.~Chrenko~et~al.}

   \subtitle{}

   \author{O. Chrenko\inst{1}
	  \and
          M. Bro\v{z}\inst{1}
          \and
	  M. Lambrechts\inst{2}
          }

   \institute{Institute of Astronomy, Charles University in Prague,
	      V Hole\v sovi\v ck\'ach 2, CZ--18000 Prague~8, Czech Republic\\
              \email{chrenko@sirrah.troja.mff.cuni.cz}
	      \and
              Laboratoire Lagrange, UMR7293, Universit\'e C\^ote d'Azur, CNRS, Observatoire
	      de la C\^ote d'Azur, Boulevard de l'Observatoire, 06304 Nice Cedex 4, France\\
             }

   \date{Received April 25, 2017 / Accepted June 17, 2017}

% \abstract{}{}{}{}{} 
% 5 {} token are mandatory
 
  \abstract
  {
  Planetary embryos can continue to grow by pebble accretion
  until they become giant planet cores. Simultaneously, these
  embryos mutually interact and also migrate due to torques arising from
  the protoplanetary disk.
  }
  {
  Our aim is to investigate how pebble accretion alters the orbital
  evolution of embryos undergoing the Type-I migration.
  In particular, we study whether they establish resonant chains,
  whether these chains are prone to instabilities and if giant planet
  cores form through embryo merging, thus occurring more rapidly
  than by pebble accretion alone.
  }
  {For the first time, we perform self-consistent global-scale 
   radiative hydrodynamic simulations
   of a two-fluid protoplanetary disk consisting of gas and pebbles,
   the latter being accreted by embedded embryos. Accretion heating,
   along with other radiative processes, is accounted for to
   correctly model the Type-I migration.
  }
  {
   We track the evolution of four super-Earth-like embryos,
   initially located in a region where the disk structure allows for a convergent migration.
   Generally, embryo merging is facilitated by rapidly increasing embryo masses
   and breaks the otherwise oligarchic growth.
   Moreover, we find that the orbital eccentricity of each embryo
   is considerably excited ($\simeq0.03$) due to the presence of an asymmetric
   underdense lobe of gas, a so-called `hot trail', produced by accretion heating
   of the embryo's vicinity.
   Eccentric orbits lead the embryos to frequent close encounters
   and make resonant locking more difficult.
  }
  {
    Embryo merging typically produces
    one massive core ($\gtrsim 10\,\mathrm{M_{E}}$) in our simulations,
    orbiting near $10\,\mathrm{AU}$.
    Pebble accretion is naturally accompanied
    by occurrence of eccentric orbits
    which should be considered
    in future efforts to explain the structure
    of exoplanetary systems.
  }

   \keywords{Hydrodynamics -- Planets and satellites: formation --
     Planet-disk interactions --
     Protoplanetary disks --
     Planets and satellites: gaseous planets}

   \maketitle
%
%________________________________________________________________

\section{Introduction}
\label{sec:intro}

Interactions of gas and solids in protoplanetary disks
are the basis for subsequent growth of all kinds of planets,
whether they will finally become
terrestrial, super-Earths, ice giants or gas giants.
These interactions have to be computed with an appropriate feedback,
as there are a number of relatively complicated but inevitable phenomena.
Setting the classical in-spiralling of solids due to gas drag aside,
there are processes like
streaming instability and local collapse \citep{Johansen_etal_2007Natur.448.1022J},
pebble accretion assisted by aerodynamic drag
\citep{Lambrechts_Johansen_2012A&A...544A..32L,Morbidelli_Nesvorny_2012A&A...546A..18M},
accretion heating of planetary embryos and surrounding gas
\citep{Benitez-Llambay_etal_2015Natur.520...63B}, or
embryo-disk interactions in general \citep[e.g][]{Kley_etal_2009A&A...506..971K}.
Sufficiently complex hydrodynamic models with radiative transfer (RHD)
are usually needed for realistic treatment of these processes.

The radiative properties of the protoplanetary
disk are mostly determined by the opacity~$\kappa$.
As a flux-mean (Rosseland) value,
$\kappa$ is mostly caused by icy, silicate or carbonaceous dust grains 
\citep{Mathis_etal_1977ApJ...217..425M,Bell_Lin_1994ApJ...427..987B}
that have different wavelength-dependent optical constants \citep{Jager_etal_2003A&A...408..193J}.
The size-frequency distribution of dust grains is often assumed shallow,
with a cumulative slope $q = -2.5$ \citep{Mathis_etal_1977ApJ...217..425M,Birnstiel_etal_2012A&A...539A.148B}.
Any sudden transition in the composition of the dust component
(e.g. grain evaporation or `rain out') affects local heating and cooling
properties of the gas disk. 
Consequently, variations of the scale height $H(r)$ might occur
and moreover, the pressure gradient might exhibit a reversal, $\nabla P > 0$, which
leads to accumulation of solids
(and even planetary embryos).
Typical transitions are located e.g. at the inner rim of the disk
due to UV photoionisation and corotation with stellar magnetic field,
at the evaporation line of silicates \citep{Flock_etal_2016ApJ...827..144F}, and
at the snowline corresponding to water evaporation
\citep{Morbidelli_etal_2015Icar..258..418M}.
Important heating sources are provided by 
viscous dissipation, especially in the inner disk, and
stellar irradiation of the inclined/flared disk atmosphere
\citep{Bitsch_etal_2014A&A...570A..75B}.

While small ($\mu{\mathrm m}$-sized) grains usually influence overall optical properties,
large ($\mathrm{mm}$-sized) dust particles or ($\mathrm{cm}$-sized) pebbles -- if already
present -- dominate the mass distribution.
According to recent developments in the theory of planet formation,
pebbles can be efficiently accreted by
larger seed masses, e.g. planetesimals or embryos,
with high enough accretion rate to finally produce giant planet cores
\citep{Lambrechts_Johansen_2012A&A...544A..32L,Lambrechts_Johansen_2014A&A...572A.107L}
with masses $\gtrsim10\,\mathrm{M_{E}}$,
well within the protoplanetary disk lifetime
  which is typically $\simeq10\,\mathrm{Myr}$ \citep{Fedele_etal_2010A&A...510A..72F}.
Global-scale $N$-body simulations
demonstrated that the giant planets
of the Solar System can be reproduced
by pebble accretion \citep{Levison_etal_2015Natur.524..322L},
provided that dynamical stirring of orbital inclinations
breaks the oligarchic growth of the seed masses
\citep{Kretke_Levison_2014AJ....148..109K}.

A downside of the aforementioned global-scale
simulations with pebble accretion is that they
do not model the interactions between the protoplanets
and the surrounding gaseous disk in a self-consistent way
because no hydrodynamics is employed.
However, during the evolutionary phase when multiple
low-mass embryos are present, it is inevitable
that these embryos interact gravitationally with
the disk and undergo the Type-I migration, when no gap
is opened. There are many purely hydrodynamical effects
contributing to the resulting torque acting on the planets:
spiral arms (launched at the Lindblad resonances and independent of
viscosity~$\nu$),
the corotation torque from the asymmetric gas structures
formed in the corotation regions of embryos \citep{Masset_2002A&A...387..605M}
and additional forcing produced by asymmetries
related to radiative effects operating in the vicinity
of the embryos, e.g. the cold finger \citep{Lega_etal_2014MNRAS.440..683L}
or the heating torque \citep{Benitez-Llambay_etal_2015Natur.520...63B}.

The embryos -- albeit having generally different migration rates --
can accumulate near some of the pressure gradient reversals, mutually interact,
get locked in a resonant configuration and
create a `convoy' \citep{Pierens_etal_2013A&A...558A.105P}.
Such a configuration naturally prevents any merging.
Stability of the resonant chain can be possibly 
reduced by larger number of embryos present in the system \citep{Pierens_etal_2013A&A...558A.105P},
also when the disk is massive and exhibits large accretion rates,
\citep[$10^{-7}\,\mathrm{M_{\sun}}\,{\mathrm{yr}}^{-1}$ according to][]{Zhang_etal_2014ApJ...797...20Z},
or if some of the embryos enter a fast migration regime due to
strong corotation torque when the initially
librating gaseous material is contracted
into the tadpole region \citep{Pierens_2015MNRAS.454.2003P}.
To the present knowledge, it is unclear how pebble accretion
and accretion heating affect the convergent migration and resonant chain stability
and we address these particular issues in this paper.
We aim to find whether the migrating embryos merge
or remain in the chain while they continue to grow.
The resonant chain (in)stability is important
also with respect to the observed exoplanetary systems
because these are often non-resonant \citep[e.g.][]{Winn_Fabrycky_2015ARA&A..53..409W}.

The embryo growth and/or merging
closely precede an evolutionary epoch which
provides important observational evidence
of the planet-forming processes.
Once a giant planet core is formed 
it can clear a gap in the disk along its orbit
and its further migration is driven by the viscous
evolution of the disk
\citep[the Type-II migration, e.g.][]{Lin_Papaloizou_1986ApJ...309..846L,Crida_Bitsch_2017Icar..285..145C}.
Such a gap may become observable and the disk is then classified
as pre-transitional \citep[according to][]{Espaillat_etal_2010ApJ...717..441E,Espaillat_etal_2014prpl.conf..497E}.

To summarize, the protoplanetary system within the scope of this paper is 
assumed to consist of the gas disk with opacities dominated by fully
coupled dust, pebble disk (strongly but not fully coupled)
and already formed low-mass embryos ($\sim1\,\mathrm{M_E}$)
which continue to grow by pebble accretion.
Our hydrodynamic simulations aim to check whether different
migration rates, evolving embryo masses, accretion heating and
mutual perturbations between embryos
can break the resonant chains
and create a giant-planet core, capable to open a gap.

Our paper is organised as follows.
In Sec.~\ref{sec:model} we summarise all the equations and approximations
of our 2-dimensional RHD model.
We also describe relevant initial and boundary conditions.
Technical details of the model and useful explanatory
derivations are given in Appendices~\ref{sec:implicit_scheme},
\ref{sec:motion_pebble} and \ref{sec:semiimp_source_step}.
A validation of our model is given later in Appendix~\ref{sec:verification}.
In Sec.~\ref{sec:results} we present results of our 
global-scale simulations
focused on the migration of several pebble-accreting and heated embryos.
Sec.~\ref{sec:hot_trail} desribes how the accretion
heating affects the orbital eccentricities and 
disk torques acting on the embryos.
We discuss possible future model improvements
and also possibilities to relate our results
with observations in Sec.~\ref{sec:discussion}.
Sec.~\ref{sec:conclusions} is devoted to conclusions.

%__________________________________________________________________

\section{Protoplanetary system modelling}
\label{sec:model}

The model we present is based on the publicly available
2D hydrodynamic code \textsc{fargo} \citep{Masset_2000A&AS..141..165M,Baruteau_Masset_2008ApJ...672.1054B}
which we extensively modified in order to follow
the evolution and mutual interactions between three components
of protoplanetary systems: a differentially rotating disk of the nebular gas,
a partially coupled disk of pebbles
and several embedded planetary embryos.
The \textsc{fargo} code is designed as an Eulerian solver on a polar
staggered mesh. The numerical scheme relies on the operator-splitting
technique according to \cite{Stone_Norman_1992ApJS...80..753S}, with
a modified transport sub-step which utilizes van Leer's second-order upwind
interpolation \citep{VanLeer_1977JCoPh..23..276V} for radial advection
and the \textsc{fargo} algorithm \citep{Masset_2000A&AS..141..165M}
in the azimuthal direction.
Let us briefly summarize new physical modules that were implemented
in our modified version of the code\footnote{The
code will be publicly available in the near future
at \url{http://sirrah.troja.mff.cuni.cz/~chrenko/}.
}.

Considering the gaseous disk, we relax the isothermal
approximation and account for the evolution of temperature
within the disk. The extended set of hydrodynamic equations thus
contains the energy equation with multiple relevant source
terms, in particular: compressional heating, viscous heating, stellar irradiation,
vertical escape of radiation, radiative diffusion in the midplane
and radiative feedback to accretion heating of embryos.

Regarding the pebble disk, we assume it consists
of mm-cm sized pebbles \citep{Lambrechts_Johansen_2012A&A...544A..32L}.
Pebbles orbiting within the nebular gas
are subject to the aerodynamic drag which
changes their angular momentum.
The characteristic time scale of the angular
momentum change is usually described by the stopping time $t_{\mathrm{s}}$
\citep{Adachi_etal_1976PThPh..56.1756A,Weidenschilling_1977MNRAS.180...57W}.
Its dimensionless form, the Stokes number, is 
defined as $\tau \equiv \Omega_{\mathrm{K}}t_{\mathrm{s}}$,
where $\Omega_{\mathrm{K}}$ denotes the Keplerian angular frequency.
It is an important quantity encapsulating the particle
size and coupling to the nebular gas.
In this study, we follow \cite{Lambrechts_Johansen_2014A&A...572A.107L}
and consider particles smaller than the mean free path
in the nebular gas, typically with $\tau\lesssim0.1$.
The friction then arises due to anisotropic collisions
between individual gas molecules and pebbles
and the drag operates in the Epstein regime.
Due to parametrization by $\tau$, we practically neglect
drag regimes relative to the local Reynolds number.
Because of their aerodynamic properties, pebbles are strongly coupled with
the gas flow and thus we study their evolution using a two-fluid model in which 
the pebble disk is modeled as another Eulerian, but pressureless 
and inviscid, fluid \citep[e.g.][]{Youding_Goodman_2005ApJ...620..459Y}.

The embedded embryos are evolved in 3D 
using a high-accuracy integration technique, accounting
for close encounters, possible collisions and merging.
An artificial vertical force acting on the embryos
is applied to damp their inclinations as predicted
for 3D disks \citep{Tanaka_Ward_2004ApJ...602..388T}.
The embryos are allowed to
grow by the drag-assisted pebble accretion,
capturing pebbles from the circumplanetary flow.
We also consider that the embryos can be heated by
this vigorous material deposition and
consequently radiate the excessive energy into 
the surrounding gas.

The mutual interactions accounted for in the model are as follows.
Both the gas and pebbles evolve in the gravitational potential
of the protostar and embryos. The potential is computed by
an averaging procedure in a direction perpendicular
to the midplane to avoid unrealistic potential smoothing and spreading
\citep{Muller_etal_2012A&A...541A.123M}. All the embryos participate
in mutual $N$-body interactions and they also feel the
gravitational pull of the gas disk, but the gravity of the
pebble disk is ignored due to its relatively low mass.
The gas disk and pebbles are only coupled through
the linear drag term and no self-gravity is taken into account.
The detailed aspects of the model implementation into \textsc{fargo}
are elaborated in the following individual subsections.

\subsection{Two-fluid model of the gas-pebble disk}
\label{sec:hydromodel}

In our hydrodynamic model, we study the evolution of
the gas surface density $\Sigma$,
the vertically averaged gas flow velocity $\vec{v}=\left(v_{r},v_{\theta}\right)$,
the specific internal energy of the gas $E$, the surface
density of the pebble disk $\Sigma_{\mathrm{p}}$ and
its velocity field $\vec{V}=\left(V_{r},V_{\theta}\right)$.
The fundamental fluid equations to be solved
can be written by means of
the vertically integrated quantities as follows:

\begin{equation}
  \frac{ \partial \Sigma}{\partial t} + \nabla\cdot\left( \Sigma\vec{v} \right) = 0 \, ,
  \label{eq:continuity}
\end{equation}
\begin{equation}
  \frac{ \partial \vec{v}}{\partial t} + \vec{v}\cdot\nabla\vec{v} = - \frac{1}{\Sigma}\nabla P  + \frac{1}{\Sigma}\nabla\cdot\tens{T} - \frac{\int{\rho\nabla\phi\mathrm{d}z}}{\Sigma}  +
  \frac{\Sigma_{\mathrm{p}}}{\Sigma}\frac{\Omega_{\mathrm{K}}}{\tau}\left(\vec{V}-\vec{v}\right) \, ,
  \label{eq:navierestokes}
\end{equation}
\begin{equation}
  \frac{ \partial E}{ \partial t} + \nabla\cdot\left( E\vec{v} \right) = - P\nabla\cdot\vec{v}
  + Q_{\mathrm{visc}} + Q_{\mathrm{irr}} + Q_{\mathrm{acc}} - Q_{\mathrm{rad}} \, ,
\label{eq:energy}
\end{equation}
\begin{equation}
  \frac{ \partial \Sigma_{\mathrm{p}}}{\partial t} + \nabla\cdot\left( \Sigma_{\mathrm{p}}\vec{V} \right) = -\left(\frac{\partial \Sigma_{\mathrm{p}}}{\partial t}\right)_{\mathrm{acc}} \, ,
  \label{eq:continuity_peb}
\end{equation}
\begin{equation}
  \frac{ \partial \vec{V}}{\partial t} + \vec{V}\cdot\nabla\vec{V} = - \frac{\int{\rho_{\mathrm{p}}\nabla\phi\mathrm{d}z}}{\Sigma_{\mathrm{p}}} - \frac{\Omega_{\mathrm{K}}}{\tau}\left(\vec{V}-\vec{v}\right) \, .
  \label{eq:navierestokes_peb}
\end{equation}
Here $P$ denotes the vertically integrated pressure, $\tens{T}$
is the viscous stress tensor
\citep[see e.g.][]{Masset_2002A&A...387..605M},
$\phi$ is the gravitational potential
arising from the protostar and planetary embryos,
$\rho$ and $\rho_{\mathrm{p}}$ are the volume
densities of the gas and pebbles, respectively.
The individual source terms on the right-hand side 
of the energy equation represent the compressional heating,
the viscous heating $Q_{\mathrm{visc}}$, the stellar
irradiation $Q_{\mathrm{irr}}$, the radiative diffusion $Q_{\mathrm{rad}}$
and the heating $Q_{\mathrm{acc}}$
arising from pebble accretion which is symbolically
considered in the pebble mass continuity equation as
the $-\left({\partial \Sigma_{\mathrm{p}}}/{\partial t}\right)_{\mathrm{acc}}$ term.
We emphasize that the gradient and divergence operators
are always 2D in our model.

The following ideal gas equation of state is introduced
as the thermodynamic closing relation
\begin{equation}
  P = \Sigma \frac{RT}{\mu} = \left(\gamma-1\right)E \, ,
  \label{eq:state}
\end{equation}
with $R$ being the universal gas constant, $\mu=2.4\,\mathrm{g\,mol^{-1}}$
being the mean molecular weight and $\gamma=1.4$ denoting
the adiabatic index (specific heat ratio).

Before proceeding to the description of all
the individual source terms,
let us remark that we assume a simple vertical stratification
of the disk in order to approximate certain
effects that are expected to operate in realistic 3D disks.
The gas volume density $\rho\left(r,\theta,z\right)$ follows
a Gaussian form
\begin{equation}
  \rho\left(r,\theta,z\right)=\frac{\Sigma(r,\theta)}{\sqrt{2\pi}H(r,\theta)}\exp{\left(-\frac{z^{2}}{2H(r,\theta)^{2}}\right)} \, ,
  \label{eq:density_distribution}
\end{equation}
where $H=c_{\mathrm{s,iso}}/\Omega_{\mathrm{K}}=c_{\mathrm{s}}/ ( \sqrt{\gamma} \Omega_{\mathrm{K}} )$ 
is the local pressure scale height and
$c_{\mathrm{s}}=\sqrt{\gamma P/\Sigma}$
is the adiabatic sound speed which differs
  from the isothermal sound speed $c_{\mathrm{s,iso}}$
  by a factor $\sqrt{\gamma}$.
The normalization constant $\Sigma/(\sqrt{2\pi}H)$
actually represents the gas volume density $\rho_{0}$ in the midplane.
In principle, Eq.~(\ref{eq:density_distribution}) holds
only for vertically isothermal disks
which is an assumption we do not impose
when discussing the energy source terms
in Sec.~\ref{sec:energy_source_terms}.
But because recent 3D simulations demonstrated
that the optically thick parts of protoplanetary
disks have a flat vertical temperature distribution
\citep{Flock_etal_2013A&A...560A..43F}, we decided to use
Eq.~(\ref{eq:density_distribution})
as a viable first approximation of the vertical stratification.

\subsection{Energy source terms}
\label{sec:energy_source_terms}

Let us first describe how the radiation transport
is treated in our model. The corresponding term $Q_{\mathrm{rad}}$
is given by the vertically integrated divergence of the 3D radiative flux $\vec{F}_{\mathrm{3D}}$:
\begin{equation}
  \begin{split}
  Q_{\mathrm{rad}}=\int\limits_{-\infty}^{\infty}\nabla_{\mathrm{3D}}\cdot\vec{F}_{\mathrm{3D}}\mathrm{d}z
  & \simeq \int\limits_{-H}^{H}\frac{\partial F_{z}}{\partial z}\mathrm{d}z + 2H\nabla\cdot\vec{F} \\
     & \equiv Q_{\mathrm{vert}} + 2H\nabla\cdot\vec{F} \, ,
  \end{split}
  \label{eq:qrad_split}
\end{equation}
where we assumed that the vertical outward radiation is liberated
at $H$ which is expected to be much smaller than the
radial extent of the disk. The amount of energy which is
transported by radiation
is therefore dominant in the vertical direction \citep{DAngelo_etal_2003ApJ...599..548D}.
We estimate these radiative losses
caused by the vertical escape of radiation
from both sides of the disk as
\begin{equation}
  Q_{\mathrm{vert}} \simeq 2\sigma_{\mathrm{R}}T_{\mathrm{eff}}^{4}=\frac{2\sigma_{\mathrm{R}}T^{4}}{\tau_{\mathrm{eff}}} \, ,
  \label{eq:qminus}
\end{equation}
where $\sigma_{\mathrm{R}}$ is the Stefan-Boltzmann constant, $T$
stands for the midplane temperature and $\tau_{\mathrm{eff}}$ is
the effective optical depth. \cite{Hubeny_1990ApJ...351..632H}
generalized the gray model of stellar
atmospheres in LTE for the case of accretion disks and found
\begin{equation}
  \tau_{\mathrm{eff}} = \frac{3}{8}\tau_{\mathrm{opt}} + \frac{1}{2} + \frac{1}{4\tau_{\mathrm{opt}}} \, ,
  \label{eq:taueff}
\end{equation}
where we implicitly assumed that the disk is stellar
irradiated
\citep[otherwise $1/2$ term
should be replaced with $\sqrt{3}/4$;][]{DAngelo_Marzari_2012ApJ...757...50D} and that
the mean Rosseland opacity and the Planck opacity
are identical which is a viable approximation
as discussed e.g. by \cite{Bitsch_etal_2013A&A...549A.124B}.
The relation (\ref{eq:taueff}) is highly convenient in
case of a protoplanetary disk
because it can characterize both optically thin and thick environment.

The optical depth $\tau_{\mathrm{opt}}$ is measured from the midplane to the disk surface
and we estimate it as
\begin{equation}
  \begin{split}
  \tau_{\mathrm{opt}} = \int\limits_{0}^{\infty}\kappa(r,\theta,z)\rho(r,\theta,z)\mathrm{d}z & \simeq c_{\kappa}\kappa(r,\theta)\int\limits_{0}^{\infty}\rho(r,\theta,z)\mathrm{d}z \\
    & = c_{\kappa}\frac{\kappa(r,\theta)\Sigma(r,\theta)}{2} \, ,
  \end{split}
  \label{eq:optical_depth}
\end{equation}
where $c_{\kappa}=0.6$ is a correction factor which accounts for
the opacity drop in the layers above the midplane
(see \cite{Muller_Kley_2012A&A...539A..18M} for a similar approach).
This parametric factor in fact sets the local efficiency of vertical cooling
and can be tuned so that the resulting disk structure resembles the one
obtained in 3D models.

We adopt the power-law mean Rosseland opacity
$\kappa=\kappa_{0}\rho^{a}T^{b}$ with the coefficients $a$ and $b$
derived by \cite{Lin_Papaloizou_1985} and further refined by
\cite{Bell_Lin_1994ApJ...427..987B}
for various temperature intervals and corresponding opacity regimes.
The transitions between individual opacity regimes are smoothed
out as in \cite{Lin_Papaloizou_1985}
\citep[see also][]{Keith_Wardle_2014MNRAS.440...89K}.

Coming back to the midplane radiative flux (see Eq.~\ref{eq:qrad_split}),
we utilize
the flux-limited diffusion approximation
\citep{Levermoe_Pomraning_1981ApJ...248..321L,Klahr_Kley_2006A&A...445..747K} to express
\begin{equation}
  \vec{F} = - \lambda_{\mathrm{lim}} \frac{16\sigma_{\mathrm{R}}}{\rho_{0}\kappa}T^{3}\nabla T \, .
  \label{eq:flux}
\end{equation}
In this approximation, scattering effects are neglected and $\lambda_{\mathrm{lim}}$
denotes the flux limiter, which is calculated according to
\cite{Kley_1989A&A...208...98K}.
The radiative transport is treated by means of
the one-temperature approach 
\citep{Kley_etal_2009A&A...506..971K}.
This means that
the internal energy of the gas is presumed to be dominated
by the thermal energy whereas the radiative energy is relatively small.
The radiation field is thermalized to the same temperature
$T$ as the gas.

The stellar irradiation is governed by $Q_{\mathrm{irr}}$ term
which is complementary to $Q_{\mathrm{vert}}$ and reads
\begin{equation}
  Q_{\mathrm{irr}} = \frac{2\sigma_{\mathrm{R}}T_{\mathrm{irr}}^{4}}{\tau_{\mathrm{eff}}} \, .
  \label{eq:qirr}
\end{equation}
The irradiation temperature $T_{\mathrm{irr}}$ can be obtained
from the projection of the stellar radiation flux onto the disk
surface 
\citep{Chiang_Goldreich_1997ApJ...490..368C,Menou_Goodman_2004ApJ...606..520M,
Pierens_2015MNRAS.454.2003P}
\begin{equation}
  T_{\mathrm{irr}}^{4} = (1-A)\left(\frac{R_{\star}}{r}\right)^{2}T_{\star}^{4}\sin\alpha \, .
  \label{eq:Tirr}
\end{equation}
Here $A=0.5$ is the disk albedo,
assumed to be a mean value
implicitly averaged over the stellar flux,
and $T_{\star}=4370\,\mathrm{K}$
is the effective temperature of the protostar
with the stellar radius $R_{\star}=1.5\,\mathrm{R_{\sun}}$. 
Together with the stellar mass $M_{\star}=1.0\,\mathrm{M_{\sun}}$,
the given parameters represent a protostar similar to T~Tauri type
\citep{Paxton_etal_2015ApJS..220...15P}.
Finally, $\alpha$
is the grazing angle at which the starlight strikes
the disk. The grazing angle can be approximated by
reconstructing the disk surface from the local pressure
scale height $H$.
Adopting the geometric formulation of
\cite{Baillie_Charnoz_2014ApJ...786...35B}, we use
\begin{equation}
  \alpha = \arctan\left(\frac{\mathrm{d}H}{\mathrm{d}r}\right) -
           \arctan\left(\frac{H-0.4R_{\star}}{r}\right) \, .
  \label{eq:grazing_angle}
\end{equation}
If $\alpha < 0$, the corresponding surface facet
is not oriented towards the incident irradiating flux
thus we set $Q_{\mathrm{irr}}=0$ in this case.
Unlike in an isothermal model,
the aspect ratio $h(r)=H(r)/r$ is not time independent
but it evolves instead. Therefore the disk can
flare in its outer parts where the stellar
irradiation dominates the energy budget
\citep{DAlessio_etal_1998ApJ...500..411D,
Dullemond_2002A&A...395..853D,Bitsch_etal_2013A&A...549A.124B}.

The viscous dissipation heating $Q_{\mathrm{visc}}$ is calculated 
according to \cite{Mihalas_WeibelMihalas_1984frh..book.....M}
\begin{equation}
  Q_{\mathrm{visc}} = \frac{1}{2\nu\Sigma}\left(\tau_{rr}^2 + 2\tau_{r\theta}^2 + \tau_{\theta\theta}^2\right) + \frac{2\nu\Sigma}{9}\left(\nabla\cdot\vec{v}\right)^{2}.
\label{eq:qplus}
\end{equation}
Here $\nu=5\times10^{14}\,\mathrm{cm^{2}\,s^{-1}}$
is the kinematic viscosity and $\tau_{ij}$ corresponds
to the individual components of the viscous stress tensor $\tens{T}$.
We emphasize that the viscosity is fixed and not solved explicitly 
in the model.

The accretion heating term $Q_{\mathrm{acc}}$ is nonzero
only in the nearest vicinity of embedded planetary embryos and
it depends on their accretion rate.
The luminosity of an accreting embryo with the mass $M_{\mathrm{em}}$
and the radius $R_{\mathrm{em}}$
is given by
\begin{equation}
  L = \frac{GM_{\mathrm{em}}}{R_{\mathrm{em}}}\frac{\mathrm{d}M_{\mathrm{em}}}{\mathrm{d}t} \, .
  \label{eq:accretion_luminosity}
\end{equation}
The resulting heating of the surrounding gas is 
provided by placing an inner heat source
into the grid cell which contains the respective embryo.
The specific power of this source reads
\begin{equation}
  Q_{\mathrm{acc}} = \frac{L}{S} \, ,
  \label{eq:accretion_heating}
\end{equation}
where $S$ is the cell area.
In this work, we assume that the mass growth of embryos is
driven solely by pebble accretion. The accretion
rate $\mathrm{d}M_{\mathrm{em}}/\mathrm{d}t$ is computed
self-consistently as described in Sec.~\ref{sec:pebble_accretion}.
We emphasize that the accretion heating term $Q_{\mathrm{acc}}$
is not always switched on in our simulations and we will remind the reader
in such cases.

The numerical solution of the energy equation (Eq.~\ref{eq:energy})
is described in Appendix~\ref{sec:implicit_scheme}.

\subsection{Initial state of the gas disk}
\label{sec:init_gas}

The thermal equilibrium of any gaseous disk studied in our
model is achieved by a rather complicated interplay between
the heating and cooling sources introduced above.
Therefore it would be difficult to search for an analytic formula 
describing the initial state of an isolated disk in equilibrium.
In order to initialize the hydrodynamic fields over the computational
domain, we use either simple power-law functions or equilibrium
solutions known from less sophisticated models.
The resulting gas disk, which lacks the pebble component and
embedded objects at this point, is then numerically relaxed
towards its stationary state. 
This serves as a preparation stage for the following
complete simulations.

The non-relaxed hydrodynamic profiles
are assumed to be symmetric in $\theta$.
The surface density is described by the power-law profile
$\Sigma = 750\left(r/(1\,\mathrm{AU})\right)^{-0.5}\,\mathrm{g\,cm^{-2}}$.
We start with an initially non-flaring disk, having the aspect ratio
$h=H/r=0.05$. In accordance with this setup, we can
subsequently initialize $c_{\mathrm{s}}$, $P$ and $T$.
We verified that the choice of
initially non-flaring disk
does not prevent flaring during the relaxation.
The radial velocity $v_{r}$ is initially set to zero,
while $v_{\theta}$ is set by imposing the equilibrium
between the central gravity, pressure gradient and centrifugal
acceleration.
The disk is fully extended in azimuth and radially bordered
by the inner boundary $r_{\mathrm{min}}=2.8\,\mathrm{AU}$
and the outer boundary $r_{\mathrm{max}}=14\,\mathrm{AU}$.
The polar computational domain is
divided into 1536 azimuthal sectors and 1024 evenly spaced radial rings.
The grid sampling should be sufficient to reasonably resolve
the corotation region of low-mass embryos and properly
reproduce the related torques \citep[e.g.][]{Lega_etal_2014MNRAS.440..683L}.

\subsection{Initial state of the pebble disk}
\label{sec:init_pebbles}

We use the hydrodynamic polar grid to insert a sea
of pebbles within the gaseous disk which has been
already relaxed towards its equilibrium state
in the absence of planetary embryos.
Using solely the hydrodynamic quantities
together with several parameters introduced in this section,
we initialize $\Sigma_{\mathrm{p}}$, $V_{r}$ and $V_{\theta}$
over the computational domain and evolve the fluid of pebbles
in the course of simulation.

The aerodynamic properties of pebbles which 
interact with the gas in the Epstein regime are
characterized by the Stokes number
\begin{equation}
  \tau = \frac{\rho_{\mathrm{b}}R_{\mathrm{p}}}{\rho_{0} c_{\mathrm{s}}}\Omega_{\mathrm{K}} \, ,
  \label{eq:stokes_epstein}
\end{equation}
where $\rho_{\mathrm{b}}=1\,\mathrm{g\,cm^{-3}}$ is the pebble bulk density,
$R_{\mathrm{p}}$ is the pebble size and $\rho_{0}$
is the midplane volume density of the nebular gas.
Then the initial velocity field can be described
by an analytic estimate for a pebble drifting in a
steady-state gaseous disk while neglecting
the presence of any massive perturbers besides
the protostar
\citep[e.g.][see also Appendix \ref{sec:motion_pebble}]{Nakagawa_etal_1986Icar...67..375N,Guillot_etal_2014A&A...572A..72G}
\begin{equation}
  V_{r} = - \frac{2\tau}{1+\tau^{2}}\left(\eta v_{\mathrm{K}} - \frac{1}{2\tau}v_{r}\right) \, , 
  \label{eq:vr_pebbles}
\end{equation}
\begin{equation}
  V_{\theta} = v_{\mathrm{K}} - \frac{1}{1+\tau^{2}}\left(\eta v_{\mathrm{K}} - \frac{\tau}{2}v_{r}\right) \, ,
  \label{eq:vt_pebbles}
\end{equation}
where $v_{\mathrm{K}}$ is the local Keplerian
velocity and $\eta$ measures
how much the gas departs from local Keplerian rotation
\begin{equation}
  v_{\theta} = (1-\eta)v_{\mathrm{\mathrm{K}}} \, .
  \label{eq:eta}
\end{equation}
In simple stationary disks, $\eta$
is a monotonic function reflecting the sub-Keplerian rotation
of the pressure-supported nebular gas.
In realistic disks, however, the situation is
more complicated -- the $\eta$ profile is affected
e.g. by the pressure dips and bumps which
can occur at the opacity transitions
\citep{Bitsch_etal_2014A&A...570A..75B} and also
by viscous shear.

As mentioned above, we aim to describe the pebble disk
by a single fluid while in reality, protoplanetary systems
are certainly populated by pebbles of various sizes.
Despite of our simplification, we would like
the material delivery towards the accreting embryos to be realistic.
It is thus important to discuss the choice of the particle size and
Stokes number.
As argued by \cite{Birnstiel_etal_2012A&A...539A.148B},
the most of the pebble mass is concentrated towards the upper end of
the size spectrum and, at the same time, the largest
pebbles are the fastest drifters.
At a given radial distance, it is reasonable to assume
that the pebble size distribution has a steep upper cutoff
and all the particles larger than this cutoff
are swiftly removed by the drift, while
particles smaller than this cutoff do not contribute to
the total mass of solids significantly.
In this work we presume that such a dominant size
is also the best choice for characterizing the
pebble disk by a single fluid so that its resulting
hydrodynamic behaviour is the most similar to a real pebble disk
which is a mixture of many particle species.
In other words, the dominant pebble size can be viewed as an
effective workaround to avoid using a numerically demanding
multi-fluid model and obtain a reasonably evolving disk
of solids at the same time.
Please note that $R_{\mathrm{p}}$
is always understood as the dominant drift-limited
size in what follows and that we also neglect other size-limiting
processes such as fragmentation.

The Stokes number $\tau_{\mathrm{d}}$ of the dominant
pebble size can be found by balancing
the characteristic time scale for the particle
growth $t_{\mathrm{grow}}=R_{\mathrm{p}}/\dot{R}_{\mathrm{p}}$
and the time scale of the particle removal by the drift
$t_{\mathrm{drift}}=r/V_{\mathrm{r}}$.
Following \cite{Garaud_2007ApJ...671.2091G}
and staying within
the limits of the Epstein regime, the growth time scale is
\begin{equation}
  t_{\mathrm{grow}} = \frac{4}{\sqrt{3}\epsilon_{\mathrm{p}}\left(\Sigma_{\mathrm{p}}/\Sigma\right)\Omega_{\mathrm{K}}} \, ,
  \label{eq:growth_timescale}
\end{equation}
and depends only on the local solid-to-gas ratio, orbital frequency
and the pebble coagulation efficiency, assumed $\epsilon_{\mathrm{p}}=0.5$.
Because $\tau<1$, we approximate $V_{r}\approx-2\tau\eta r\Omega_{\mathrm{K}}$
and, by equating the characteristic time scales, we write
\begin{equation}
  \tau_{\mathrm{d}} = \frac{\sqrt{3}}{8}\frac{\epsilon_{\mathrm{p}}}{\eta}\frac{\Sigma_{\mathrm{p}}}{\Sigma} \, .
  \label{eq:tau_dominant_1}
\end{equation}

Up to this point, the pebble surface density $\Sigma_{\mathrm{p}}$
was unconstrained. When studying pebble accretion,
it is useful to keep track
of the total radial mass flux $\dot{M}_{\mathrm{F}}$
of solids through the system. In the following, we set
the initial $\dot{M}_{\mathrm{F}}=2\times10^{-4}\,\mathrm{M_{E}\,yr^{-1}}$
\citep{Lambrechts_Johansen_2014A&A...572A.107L}
as an input parameter and assuming an equilibrium situation,
we impose the following continuity requirement
\citep{Lambrechts_Johansen_2014A&A...572A.107L}
\begin{equation}
  \Sigma_{\mathrm{p}} = \frac{\dot{M}_{\mathrm{F}}}{2\pi rV_{r}} \, .
  \label{eq:radial_massflux}
\end{equation}
Plugging Eq.~(\ref{eq:radial_massflux}) in (\ref{eq:tau_dominant_1}) and using
the approximate expression for $V_{r}$ again, one finds
\begin{equation}
  \tau_{\mathrm{d}} = \frac{1}{r\eta}\sqrt{\frac{\sqrt{3}\epsilon_{\mathrm{p}}\dot{M}_{\mathrm{F}}}{32\pi\Omega_{\mathrm{K}}\Sigma}} \, .
  \label{eq:tau_dominant_2}
\end{equation}
The corresponding dominant particle size can be easily obtained when utilizing
the inverse of Eq.~(\ref{eq:stokes_epstein}).
In the last expression, $\tau_{\mathrm{d}}$ depends only on two model
parameters ($\epsilon_{\mathrm{p}}$ and $\dot{M}_{\mathrm{F}}$)
and the hydrodynamic state of the gaseous background.
Therefore it is a convenient starting point for
the pebble disk initialization.

To summarize the initial conditions, we first use the combination
of Eq.~(\ref{eq:stokes_epstein}) and (\ref{eq:tau_dominant_2})
to find $R_{\mathrm{p}}(r)$. Because the relaxed gaseous disk
is very close to axial symmetry (within
discretization errors and numerical artefacts)
when we incorporate the pebble disk, it is reasonable to
consider that the pebble size changes only radially.
We further assume that once the planetary embryos are present,
they do not cause global-scale changes of $\eta$,
thus the initial $R_{\mathrm{p}}(r)$ profile is kept fixed during
our simulations. Subsequently,
we calculate the initial $(V_{r},V_{\theta})$ field
(Eqs. \ref{eq:vr_pebbles} and
\ref{eq:vt_pebbles}) which
sets $\Sigma_{\mathrm{p}}$ from the
mass flux conservation law (\ref{eq:radial_massflux}).
We emphasize that unlike $R_{\mathrm{p}}(r)$, the Stokes
number $\tau(r,\theta)$ is considered a cell-dependent quantity
during the simulations and it is recalculated
each time step to obtain proper aerodynamics for a given particle
size moving in the evolving gaseous background.
This is to account for situations when pebbles
suddenly enter gas clumps or underdense regions.

\subsection{Pebble accretion}
\label{sec:pebble_accretion}

Pebble accretion enters our model
through Eq.~(\ref{eq:continuity_peb}) in which
it acts like a mass sink.
At the same time, the mass removed from the pebble
component is accreted by the growing embryos.
According to \cite{Lambrechts_Johansen_2012A&A...544A..32L},
two fundamental regimes of pebble
accretion have to be considered,
namely the Bondi\footnote{In the original work of
\cite{Lambrechts_Johansen_2012A&A...544A..32L},
the Bondi regime is referred to as the drift regime.}
and the Hill regimes,
while the transition between the two
occurs when the pebble accretion
Bondi radius $R_{\mathrm{B}}$ becomes
comparable to the Hill sphere
radius $R_{\mathrm{H}}$ of the accreting body.
The former radius corresponds to the distance
such that a pebble with impact parameter $b\leq R_{\mathrm{B}}$
will suffer a $\geq1\,\mathrm{rad}$ deflection,
while the latter radius defines the
region in which the gravitational
pull of the accreting body dominates over
the primary field. The defining equations are
\begin{equation}
  R_{\mathrm{B}} = \frac{GM_{\mathrm{em}}}{v_{\mathrm{rel}}^{2}} \, ,
  \label{eq:bondi}
\end{equation}
and
\begin{equation}
  R_{\mathrm{H}} = \left(\frac{GM_{\mathrm{em}}}{3\Omega_{\mathrm{K}}^{2}}\right)^{1/3} \, ,
  \label{eq:hill}
\end{equation}
where $v_{\mathrm{rel}}$ denotes the relative velocity
between the pebble and the accreting body
with mass $M_{\mathrm{em}}$. 

In the Bondi regime, if
$R_{\mathrm{B}}\lesssim R_{\mathrm{H}}$,
the only pebbles that experience a significant
deflection arrive through a small fraction
of the Hill sphere thus they enter the encounter
region with the relative velocity which is
set by the local headwind experienced by the embryo,
therefore $v_{\mathrm{rel}}\simeq v_{\mathrm{head}}$.

On the other hand, if $R_{\mathrm{B}}\gtrsim R_{\mathrm{H}}$,
the relative encounter velocity for most
of the pebbles is dominated by the Keplerian
shear which becomes more important
than headwind on orbital separations
comparable to $R_{\mathrm{H}}$.
In such a case the Hill regime is triggered.
It is obvious that the equality of
$R_{\mathrm{B}}$ and $R_{\mathrm{H}}$
is reached for a specific value of $M_{\mathrm{em}}$
called the transition mass
\begin{equation}
  M_{\mathrm{t}} =
  \sqrt{\frac{1}{3}}\frac{v_{\mathrm{head}}^{3}}{G\Omega_{\mathrm{K}}} \, .
  \label{eq:transition_mass}
\end{equation}
Super-Earth-like embryos which we investigate
in this paper usually grow in the Hill regime.

\cite{Lambrechts_Johansen_2012A&A...544A..32L}
also found that there is a well-defined
maximum distance at which the pebbles must approach
the embryo in order to be accreted. This effective
accretion radius for both regimes
is given by
\begin{equation}
  R_{\mathrm{eff}} =
  \begin{dcases*}
    R_{\mathrm{B}}\sqrt{\frac{\tau}{t_{\mathrm{B}}\Omega_{\mathrm{K}}}} \,, &
    Bondi regime $\left(M_{\mathrm{em}} < M_{\mathrm{t}}\right)$ \\
    \min\left[R_{\mathrm{H}}\left(\frac{\tau}{0.1}\right)^{1/3}, R_{\mathrm{H}}\right] \,, &
    Hill regime $\left(M_{\mathrm{em}}\geq M_{\mathrm{t}}\right)$
  \end{dcases*}
  \label{eq:effective_radii}
\end{equation}
where $t_{\mathrm{B}}=R_{\mathrm{B}}/v_{\mathrm{rel}}$
is the Bondi radius crossing time.

Because our simulations cover a relatively large portion
of the protoplanetary disk, the grid resolution near embryos
is not detailed enough to capture the final
stage of the in-spiraling motion of pebbles.
Thus the fluid model does not allow for fully self-consistent
pebble accretion calculation because we are not able to resolve
the flow of pebbles falling on the embryo's surface.
We instead rely on the knowledge of the effective
accretion radius $R_{\mathrm{eff}}$ and we
employ a recipe which is somewhat similar to the usual gas accretion 
treatment in 2D hydrodynamic models \citep{Kley_1999MNRAS.303..696K}.

First, we identify all the grid cells which have their
midplane distance from the
embryo smaller than $R_{\mathrm{eff}}$. Second,
we compute the following mass-related quantities:
\begin{itemize}
  \item The expected embryo mass increase $\Delta M_{\mathrm{expec}}$.
    Here we utilize the analytic accretion rates
    derived from detailed pebble accretion models
    \citep{Lambrechts_Johansen_2012A&A...544A..32L}.
    Following \cite{Morbidelli_etal_2015Icar..258..418M},
    we set
    \begin{equation}
      v_{\mathrm{rel}} =
      \begin{dcases*}
	v_{\mathrm{head}} \, , &
	Bondi regime $\left(M_{\mathrm{em}} < M_{\mathrm{t}}\right)$ \\
	v_{\mathrm{shear}} \, , &
	Hill regime $\left(M_{\mathrm{em}} \geq M_{\mathrm{t}}\right)$ \\
      \end{dcases*}
      \label{eq:v_rel_model}
    \end{equation}
    where $v_{\mathrm{shear}}$ is the relative velocity due to Keplerian
    shear at the orbital separation $R_{\mathrm{eff}}$, and
    \begin{equation}
      \Delta M_{\mathrm{expec}} =
      \begin{dcases*}
	2R_{\mathrm{eff}}v_{\mathrm{rel}}\bar{\Sigma}_{\mathrm{p}}\times\Delta t \, , &
	$\left(\bar{H}_\mathrm{p} < R_{\mathrm{eff}}\right)$ \\
	\pi R_{\mathrm{eff}}^{2}v_{\mathrm{rel}}\frac{\bar{\Sigma}_{\mathrm{p}}}{\sqrt{2\pi}\bar{H}_{\mathrm{p}}}\times\Delta t \, , &
	$\left(\bar{H}_\mathrm{p} \geq R_{\mathrm{eff}}\right)$ \\
      \end{dcases*}
      \label{eq:dm_expec}
    \end{equation}
    where the overbar indicates the mean value taken
    over the respective cells and $\Delta t$ is the time step.
    Because $v_{\mathrm{rel}}$ is calculated self-consistently,
    the pebble accretion rate is approximately corrected for eccentric orbits
    ($v_{\mathrm{rel}}$ increases with the eccentricity,
    $M_{\mathrm{t}}$ increases as well and the embryo can
    experience a transition to the Bondi accretion regime
    which is less effective).
  \item The total available mass $\Delta M_{\mathrm{avail}}$.
    Assuming that pebbles have non-zero
    scale height $H_{\mathrm{p}}$ and that their vertical $z$-distribution
    is Gaussian (like for the gas; cf. Eq.~\ref{eq:density_distribution}),
    we calculate $\Delta M_{\mathrm{avail}}$ by numerically
    integrating the pebble fluid mass inside the overlap between the
    vertically spread disk of pebbles and
    the accretion sphere of radius $R_{\mathrm{eff}}$,
    located around the embryo which can generally be shifted in $z$ direction.
    The purpose of $\Delta M_{\mathrm{avail}}$ is mainly to account
    for 3D effects, e.g. inclined orbits can lead the accreting bodies
    away from their feeding zones.

    The pebble disk scale height
    is \citep{Youdin_Lithwick_2007Icar..192..588Y}
    \begin{equation}
      H_{\mathrm{p}} \simeq H\sqrt{\frac{\alpha_{\mathrm{p}}}{\tau}} \, ,
      \label{eq:pebble_scale_height}
    \end{equation}
    where $\alpha_{\mathrm{p}}=1\times10^{-4}$ parametrizes the turbulent stirring
    of the solids in the protoplanetary disk.
\end{itemize}
Finally, the mass transfered on the embryo in one time step
is 
\begin{equation}
   \Delta M_{\mathrm{em}} = \min(\Delta M_{\mathrm{expec}}, \Delta M_{\mathrm{avail}}) \, .
  \label{eq:pebble_accretion_rate}
\end{equation}
The pebble surface density in the cells below $R_{\mathrm{eff}}$
is reduced accordingly.
This instantaneous accretion rate $\Delta M_{\mathrm{em}}/\Delta t$
is also used to calculate the accretion heating $Q_{\mathrm{acc}}$
(Eq.~\ref{eq:accretion_heating}). The change in $\Sigma_{\mathrm{p}}$
due to accretion can propagate to radial distances interior to the embryo,
thus affecting the pebble mass flux.

\subsection{Numerical solution of the pebble fluid motion equation}

After the accretion step,
the hydrodynamic quantities describing the pebble
disk are evolved as follows. First, the Stokes number $\tau(r,\theta)$
is recalculated for each cell from the Eq.~(\ref{eq:stokes_epstein})
using the known dominant pebble size $R_{\mathrm{d}}$
and the quantities $\rho_{0}$ and $c_{\mathrm{s}}$
reflecting the state of the gaseous background.
Second, the velocity field $V_{r}$, $V_{\theta}$
is updated under the action of the source terms
standing on the right-hand side of the pebble fluid motion
Eq.~(\ref{eq:navierestokes_peb}).
Third, all the quantities are advected using the same transport
\textsc{fargo} algorithm as for the gas.

Regarding the source step, it is necessary to take into consideration
that pebbles are usually well coupled to the gas and
they have stopping times $t_{\mathrm{s}}$ much smaller than the typical
time step $\Delta t$ adopted for the explicit update of the gas dynamics.
Applying the same explicit integration for the pebble fluid
might require significant time step limitations.
In order to avoid this, we adopt a semi-implicit solution
as in \cite{Rosotti_etal_2016MNRAS.459.2790R} (see
Appendix~\ref{sec:semiimp_source_step} for a brief
overview of this method),
also including a particle diffusion term
related to turbulent mixing. This is
accounted for by adding the
following diffusive velocity
\citep{Clarke_Pringle_1988MNRAS.235..365C}
\begin{equation}
  \vec{V}_{\mathrm{D}} = -\frac{\nu}{\mathrm{Sc}}\frac{\Sigma}{\Sigma_{\mathrm{p}}}\nabla\frac{\Sigma_{\mathrm{p}}}{\Sigma} \, ,
  \label{eq:diffusive_veloc}
\end{equation}
to the pebble fluid velocity. The
Schmidt number $\mathrm{Sc}=1$ is considered,
representing the ratio of the gas diffusivity
  to the pebble diffusivity
  \citep[e.g.][]{Cuzzi_etal_1993Icar..106..102C,Youdin_Lithwick_2007Icar..192..588Y}.

\subsection{Boundary conditions}
\label{sec:bc}

The radial boundaries $r_{\mathrm{min}}$
and $r_{\mathrm{max}}$ are closed for all
hydrodynamic quantities. In addition,
we set wave-killing zones in the annuli adjacent to the inner
and outer boundary. These zones cover the intervals of
$r\in\left[r_{\mathrm{min}},1.2r_{\mathrm{min}}\right]$ and
$r\in\left[0.9r_{\mathrm{max}},r_{\mathrm{max}}\right]$. 
Inside these zones, the following equation
is solved each time the boundary condition is applied
\citep{Kley_Dirksen_2006A&A...447..369K,deValBorro_etal_2006MNRAS.370..529D}
\begin{equation}
  \frac{\mathrm{d}q}{\mathrm{d}t} = - \frac{q-q_{0}}{t_{\mathrm{damp}}}f\left(r\right) \, ,
  \label{eq:damping_boundary}
\end{equation}
where $q$ represents any hydrodynamic quantity and $q_{0}$ is its reference
value that is about to be reached by the damping. The characteristic
time scale is $t_{\mathrm{damp}}=0.1T_{\mathrm{orb}}$
\citep{Muller_Kley_2013A&A...560A..40M} with $T_{\mathrm{orb}}$
being the Keplerian orbital period at the corresponding (inner or outer) boundary.
By $f\left(r\right)$ we denote
a dimensionless ramp function which decreases from 1 at the boundary
to 0 at the end of the wave-killing zone \citep{deValBorro_etal_2006MNRAS.370..529D}.

The choice of $q_{0}$ for the gas disk
is the following:
The radial velocity $v_{r}$ is damped
to zero at the boundaries.
The remaining hydrodynamic quantities
characterizing the gas ($\Sigma$, $E$, $v_{\theta}$)
are damped towards the values they attain
at the end of the relaxation stage.
Owing to these conditions, any spiral wake that is invoked
by an embedded planet cannot reflect at the boundary.

The boundary conditions for pebbles are also imposed
within the wave-killing zones by damping 
$\Sigma_{\mathrm{p}}$, $V_{r}$ and $V_{\theta}$ towards the
initial steady-state solutions. Owing to these
conditions, the outer wave-killing zone behaves
like a pebble reservoir and the pebble disk does not
decay in time due to its inward drift. 

\subsection{Embryo-disk interaction}

In 2D simulations, a standard procedure
when simulating the planet-disk gravitational interactions
is to replace the real planetary potential
with a Plummer-type smoothed potential of a point mass
\citep{Morbidelli_etal_2008A&A...478..929M}
$\phi_{\mathrm{em}}= - GM_{\mathrm{em}}/\sqrt{s^2+z_{\mathrm{em}}^{2}+\epsilon^{2}}$,
where $s=\sqrt{(x-x_{\mathrm{em}})^2+(y-y_{\mathrm{em}})^2}$
is the midplane separation between a cell center
and an embryo with 3D coordinates $(x_{\mathrm{em}}, y_{\mathrm{em}}, z_{\mathrm{em}})$
and $\epsilon$ is the smoothing length, typically taken as
a fraction of the pressure scale height $H_{\mathrm{em}}$
at the embryo's orbit.
The reason for the smoothing is twofold. First, it is to
keep the otherwise diverging potential regular
for the gas parcels located close to the planet and second,
it is to mimic the interaction with columns of gas
instead of razor-thin midplane distribution.

However, we decided not to use the $\epsilon$-smoothed
potential in our case because of the following inconveniences.
As the embryo masses are typically $M_{\mathrm{em}}\approx 1\,M_{\mathrm{E}}$,
one can expect that the Hill sphere of the embryo
will be smaller than the vertical extent of the disk most of the time. This means
that the $\epsilon$-smoothing based on the thickness
would cause a significant \emph{underestimation} of the
embryo's gravitational influence already
outside the Hill sphere \citep{Kley_etal_2009A&A...506..971K}.
This could have at least two negative impacts on the reliability of our model:
The torques arising from the regions close to the planet
would be poorly reproduced and too many pebbles might
be able to cross the Hill sphere without being
accreted as they would drift in a shallower potential well.

To avoid these difficulties, we follow \cite{Klahr_Kley_2006A&A...445..747K}
and use the following deeper potential
    \begin{equation}
      \phi_{\mathrm{em}} =
      \begin{dcases*}
        -\frac{GM_{\mathrm{em}}}{d} \, , &
        $\left(d > r_{\mathrm{sm}}\right)$ \\
	-\frac{GM_{\mathrm{em}}}{d}\left[\left(\frac{d}{r_{\mathrm{sm}}}\right)^{4} - 2\left(\frac{d}{r_{\mathrm{sm}}}\right)^{3} + 2\frac{d}{r_{\mathrm{sm}}}\right] \, , &
	$\left(d \leq r_{\mathrm{sm}}\right)$ \\
      \end{dcases*}
      \label{eq:cubic_potential}
    \end{equation}
where $r_{\mathrm{sm}}=0.5R_{\mathrm{H}}$ is the
actually used (sufficiently small) smoothing length.
For the purpose of the embryo-disk interaction modelling,
we assume that the gas is stratified symmetrically above and beneath
the midplane, according to the distribution function
(\ref{eq:density_distribution}). Hereinafter, $d$ is 
the 3D separation between a point in the space
(located above or below a cell center) and the embryo.

Because the gas cells in our model our
  2D, we employ a method to vertically average the 3D
  potential given by Eq.~(\ref{eq:cubic_potential}) in the
  calculations. Adopting the approach outlined by
  \cite{Muller_etal_2012A&A...541A.123M} (see also their Appendix A),
the acceleration of 2D gas cells in the gravitational field
of the embryo can be obtained by calculating the specific
density of the force projected on the midplane
\begin{equation}
  F_{\mathrm{em}}(s) = - \int\rho\frac{\partial \phi_{\mathrm{em}}}{\partial s}\mathrm{d}z \, ,
  \label{eq:force_density}
\end{equation}
where $\phi_{\mathrm{em}}$ follows from Eq.~(\ref{eq:cubic_potential})
and $\rho(r,\theta,z)$ from
Eq.~(\ref{eq:density_distribution}).
As demonstrated in \cite{Muller_etal_2012A&A...541A.123M},
replacing the integral with a coarse sum over at least
10 vertical grid points per one side of the disk leads
to an accurate yet numerically feasible reproduction of the
realistic 3D interaction.

Eq.~(\ref{eq:density_distribution}) in principle
neglects the influence of embryos on the vertical
gas distribution in their vicinity. Although this effect
can (and should) be easily incorporated in fully isothermal models
\citep[as in][]{Muller_etal_2012A&A...541A.123M}, it is
not straightforward in our non-isothermal disk because
we only use an approximate treatment of the vertical radiation
transport, the model is convection-free, etc. Nevertheless,
we found by the means of numerical experiments that
even the simple $\rho(z)$ dependence
leads to results which agree with some of the advanced 3D simulations very well
(see Appendix \ref{sec:verification}). This justification is possible due to
the local nature of the pressure scale height $H$ in our model
and also owing to the mass range of embryos which we study --
they are not massive enough to perturb the disk scale height
significantly, nor do they form circumplanetary disks.
Absence of large gaseous structures gravitationally bound to the embryos
is also a motivation for including \emph{all} parts of the Hill sphere in
the disk-embryo torque computation.

In general, the orbits of embryos can become inclined or eccentric during mutual
close encounters, it is thus necessary to ensure the inclination damping
and the circularization of the orbit as it would operate in 3D disks.
Unfortunately, our 2D disk cannot support vertical waves and moreover,
Eq.~(\ref{eq:density_distribution}) always leads to a symmetric
density distribution with respect to the midplane which is certainly
not true if inclined perturbers are present.
An artificial vertical force is thus imposed
on the embryos in order to damp their orbital inclinations
in a fashion similar to realistic 3D disks
\citep{Tanaka_Ward_2004ApJ...602..388T}:
\begin{equation}
  F_{z} = \beta\frac{M_{\mathrm{em}}\Sigma\Omega_{\mathrm{K}}}{c_{\mathrm{s}}^{4}}\left(2A_{z}^{c}v_{z}^{\mathrm{em}}+A_{z}^{s}z_{\mathrm{em}}\Omega_{\mathrm{K}}\right) \, ,  
  \label{eq:force_damping}
\end{equation}
where $v_{z}^{\mathrm{em}}$ is the vertical component
of the planet's velocity, $A_{z}^{c}=-1.088$
and $A_{z}^{s}=-0.871$ are the coefficients given
by \cite{Tanaka_Ward_2004ApJ...602..388T}.
The parameter $\beta$ is problem-dependent
and has to be tuned so that the eccentricity
damping, provided naturally by the
potential (Eq.~\ref{eq:cubic_potential}), and the inclination
damping operate both on comparable time scales.

Finally, let us remark that the stellar potential is also modeled in terms
of the acceleration obtained by the vertical averaging procedure.
The evolution of pebbles in the gravitational field
follows the same recipe as for the gas (cf.
Eqs.~\ref{eq:cubic_potential} and \ref{eq:force_density})
but their scale height $H_{\mathrm{p}}$ is of course different
(Eq.~\ref{eq:pebble_scale_height}).

\begin{table}[!t]
\caption{A summary of the hydrodynamic model parameters
  introduced in Sec.~\ref{sec:model}.}
\label{tab:params}
\centering
\begin{tabular}{c c c}
parameter & notation & value/reference \\
\hline                                   
\hline                                   
gas surface density & $\Sigma$ & $750\left(\frac{r}{1\,\mathrm{AU}}\right)^{-0.5}\,\mathrm{g\,cm^{-2}}$ \\
kinematic viscosity & $\nu$ & $5\times10^{14}\,\mathrm{cm^{2}\,s^{-1}}$ \\
non-relaxed aspect ratio & $h$ & $H/r = 0.05$ \\
adiabatic index & $\gamma$ & $1.4$ \\
mean molecular weight & $\mu$ & $2.4\,\mathrm{g\,mol^{-1}}$ \\
mean Rosseland opacity & $\kappa$ & \cite{Bell_Lin_1994ApJ...427..987B} \\
vertical opacity drop & $c_{\kappa}$ & $0.6$ \\
stellar temperature & $T_{\star}$ & $4370\,\mathrm{K}$ \\
stellar radius & $R_{\star}$ & $1.5\,\mathrm{R_{\sun}}$ \\
disk albedo & $A$ & $0.5$ \\
radial grid resolution & $N_{r}$ & $1024$ \\
azimuthal grid resolution & $N_{\theta}$ & $1536$ \\
inner radial boundary & $r_{\mathrm{min}}$ & $2.8\,\mathrm{AU}$ \\
outer radial boundary & $r_{\mathrm{max}}$ & $14\,\mathrm{AU}$ \\
pebble radial mass flux & $\dot{M}_{\mathrm{F}}$ & $2\times10^{-4}\,\mathrm{M_{E}\,yr^{-1}}$ \\
pebble turbulent stirring & $\alpha_{\mathrm{p}}$ & $1\times10^{-4}$ \\
Schmidt number & $\mathrm{Sc}$ & $1.0$ \\
coagulation efficiency & $\epsilon_{\mathrm{p}}$ & $0.5$ \\
pebble bulk density & $\rho_{\mathrm{b}}$ & $1\,\mathrm{g\,cm^{-3}}$ \\
\hline
\end{tabular}
\end{table}

\subsection{Embryo-embryo interaction}
\label{sec:embryo_embryo_interaction}

  The mutual gravitational interaction
  among the massive bodies is solved using the
  \textsc{ias15} integrator 
\citep{Rein_Spiegel_2015MNRAS.446.1424R}
from the \textsc{rebound} package \citep{Rein_Liu_2012A&A...537A.128R}
which we interfaced with \textsc{fargo}. The integration
follows a 15-th order non-symplectic Runge--Kutta scheme
improved with the Gauss--Radau quadrature
\citep[see also][]{Everhart_1985dcto.proc..185E}.
There are several fundamental reasons for choosing this integrator
over more common symplectic integrators:
\begin{itemize}
  \item The time step $\Delta t$ in \textsc{fargo} is
    controlled by the hydrodynamic
    Courant--Friedrichs--Lewy (CFL) condition and the original code
    adopts the same time step to ensure that
    the planets and gas evolve synchronously.
    Some symplectic integration schemes can produce numerical 
    errors if the time step is not fixed.
  \item The N-body integrator must be capable of dealing
    with close encounters which are expected to occur
    in our simulations.
    \textsc{ias15} is convenient for this purpose because of its
    high-order accuracy and adaptive time-step subdivision.
  \item Although \textsc{ias15} is not symplectic in nature, it is reported
    to preserve the energy error within the double floating-point
    machine precision \citep{Rein_Spiegel_2015MNRAS.446.1424R}.
    Moreover, the energy error behaves like a random walk
    which we think is the best option for rather short
    time spans (compared to long-term integrations
    in celestial mechanics)
    that our simulations cover.
\end{itemize}
Additionally, the \textsc{rebound} package contains several routines
to detect and resolve collisions. In our runs,
we use the direct collision search and the embryos
are allowed to merge whenever they collide.
Merging is treated in the most simple way in which
the mass and momentum are conserved but the released
energy and possible mass loss are neglected.
The embryo
  radii, which are used to detect collisions,
  are inferred from the embryo masses, assuming
  the spherical shape and the uniform material density
  $3\,\mathrm{g\,cm^{-3}}$.

\subsection{Code performance}
\label{sec:code_performance}

The performance of our new RHD code of course depends on
the given machine architecture and the simulations usually
require parallel computation in order to be efficient.
Following the original \textsc{fargo} code,
our version supports distributed memory parallelism
utilizing MPI-based domain decomposition, shared memory
parallelism utilizing OpenMP,
or a combination of both. The simulations in this paper
were performed on clusters of Intel Xeon E5-2650 CPUs
(v2 and v4; with comparable core performance $\simeq33$
according to the SPECfp2006 benchmark) using MPI exclusively.
To provide a typical computation time
required for our simulations, here we present values
measured for a test simulation with the full two-fluid disk,
four embedded embryos and all implemented radiative processes.
The simulation spanned $50\,\mathrm{kyr}$ of evolution and required 
$\simeq5.4\,\mathrm{d}$ on 32 cores and $\simeq3\,\mathrm{d}$
on 96 cores.

\begin{figure}[!t]
\centering
\includegraphics[width=8.8cm]{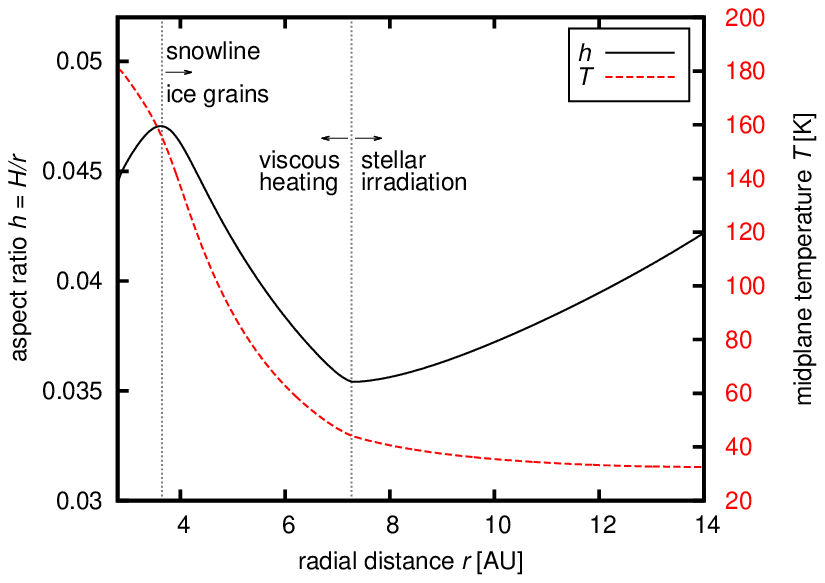}
\includegraphics[width=8.8cm]{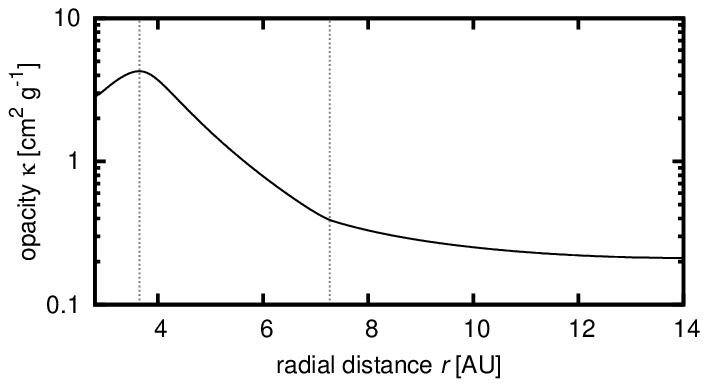}
\caption{\emph{Top}: Radial profile of the aspect ratio $h=H/r$ (black curve,
left vertical axis) and midplane temperature $T$ (red dashed curve, right vertical axis)
in our disk model. \emph{Bottom}: Radial profile of
the opacity $\kappa$.
The plots show the state reached after a relaxation, with all the
heating and cooling terms in balance. This is considered an equilibrium state,
prior to the follow-up simulations with embedded embryos. Vertical dotted lines
indicate important changes in the disk structure,
namely the snowline close to $r\simeq4\,\mathrm{AU}$ and
the transition to the flared stellar-irradiated outer
region near $r\simeq7\,\mathrm{AU}$.}
\label{fig:aspect_ratio}
\end{figure}

\section{Protoplanetary system simulations}
\label{sec:results}

\subsection{Equilibrium disk structure}
\label{sec:structure}

In this section, we discuss global characteristics
of the protoplanetary disk in thermal equilibrium,
before we actually start simulations with embedded
embryos. All the important hydrodynamic model parameters
were introduced one by one throughout Sec.~\ref{sec:model}
and we summarize all of them in Tab.~\ref{tab:params}
for the reader's convenience.

Fig.~\ref{fig:aspect_ratio} (top panel) shows the aspect ratio $h(r)=H(r)/r$
and the temperature radial profile $T(r)$ of the modelled disk.
We notice that $h$ first increases
with the radius, reaches a maximum at $r\simeq4\,\mathrm{AU}$,
drops again when moving to $r>4\,\mathrm{AU}$ and 
has another turn-over point at $r\simeq7\,\mathrm{AU}$.
The temperature $T$ on the other hand steadily decreases
outwards as a sequence of power-law functions with slopes
that change at radii corresponding to the inflection points in $h$.

We can follow the reasoning of \cite{Bitsch_etal_2013A&A...549A.124B}
to explain the changes in $h$ as well as
in $T$. Looking at the opacity profile $\kappa(r)$
(bottom of Fig.~\ref{fig:aspect_ratio}), we notice that it has
a maximum at $r\simeq4\,\mathrm{AU}$. This is related 
to the temperature rise up to $T\approx170\,\mathrm{K}$ at which
ice grains sublimate \citep{Bell_Lin_1994ApJ...427..987B},
a snowline is formed and 
silicate grains become the main source of the opacity.
The opacity maximum at $r\simeq4\,\mathrm{AU}$ prolongs
the radiative cooling time scale,
viscous friction deposits more heat in the midplane and 
creates a thermal pressure gradient which
puffs up the disk. Therefore the maximum of $h$ corresponds to the maximum
of $\kappa$.

The transition of $h$ at $r\simeq7\,\mathrm{AU}$ cannot be explained
in the same way because $\kappa$ is steadily decreasing in this region
(there is no change of the opacity regime), albeit with a shallower slope.
The transition is rather caused by the change of the dominant heating source.
Unlike at $r<7\,\mathrm{AU}$, where the viscous shear
is the main source of heating,
the stellar irradiation becomes more efficient
and prevails at $r>7\,\mathrm{AU}$. This is possible because
both $\Sigma$ and $\kappa$ are decreasing in the
outer disk and so is the vertical optical depth $\tau_{\mathrm{opt}}$.
Therefore starlight can penetrate deeper into the disk,
counteract the radiative cooling and slow down
the temperature decrease in the outer disk
which becomes flared.

\subsection{Dominant pebble properties}
\label{sec:pebble_properties}

\begin{figure}
\centering
\includegraphics[width=8.8cm]{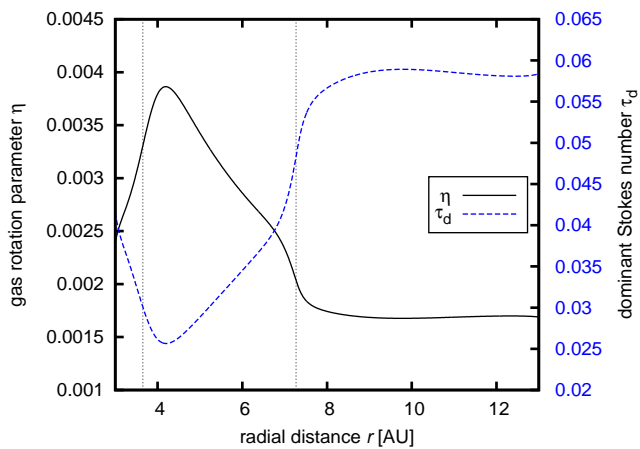}
\caption{Radial profile of the $\eta$ parameter
  (black curve, left vertical axis)
  which expresses the difference between the sub-Keplerian
  gas velocity and the Keplerian velocity,
  $v_{\theta}=(1-\eta)v_{\mathrm{K}}$.
  Initial radial profile of the dominant Stokes number
  $\tau_{\mathrm{d}}$ (blue dashed curve, right vertical axis)
  which characterizes aerodynamic properties of pebbles
  prevalent in the size-frequency
  distribution of solid particles.}
\label{fig:eta_stokes}
\end{figure}

The described transitions in the gas disk are
of a great importance for the remaining components of
the system -- both pebbles and embryos.
Let us turn our attention to pebbles first.
Fig.~\ref{fig:eta_stokes} shows the radial profile of
the gas rotation parameter $\eta$ (Eq.~\ref{eq:eta}).
The profile implies that the rotation curve
of the gas changes at the 
$4$ and $7\,\mathrm{AU}$ transitions.
For example, there is a rotation
slowdown in the inner part of the disk
due to stronger pressure support and viscous friction.

The rotation velocity of the gas is directly
related to the headwind felt by drifting pebbles.
Because the radial pebble mass flux through the disk is
assumed to be at a steady state, 
the radial distribution of the dominant Stokes number
$\tau_{\mathrm{d}}$ (Eq.~\ref{eq:tau_dominant_2})
must adapt to the $\eta$ profile in order to maintain the flux,
as shown by the blue dashed curve in Fig.~\ref{fig:eta_stokes}.
We recall that in our model, the initial $\tau_{\mathrm{d}}(r)$ profile sets
the dominant pebble sizes $R_{\mathrm{p}}(r)$ throughout the system for 
the rest of the simulation.
Going from large $r$ inwards, $R_{\mathrm{p}}$ first grows from
$7.5$ to $9\,\mathrm{cm}$, when crossing $r\simeq7\,\mathrm{AU}$ the
sizes begin to decrease down to $5\,\mathrm{cm}$ and finally
they increase at $r<4\,\mathrm{AU}$ up to $8\,\mathrm{cm}$.

However, the described variations of particle sizes
  and Stokes numbers are rather small, within a factor $\sim 2$
  in the region of interest. This is expected because
  the rotation curve transitions are smooth
  and the initial state of the pebble
  disk (Sec.~\ref{sec:init_pebbles})
  is based on the \cite{Lambrechts_Johansen_2014A&A...572A.107L}
  model which predicts the properties of the drifting
  pebbles to depend weakly on $\eta$ in smooth disks.

\subsection{Migration map}
\label{sec:migration_map}

\begin{figure}
\centering
\includegraphics[width=8.8cm]{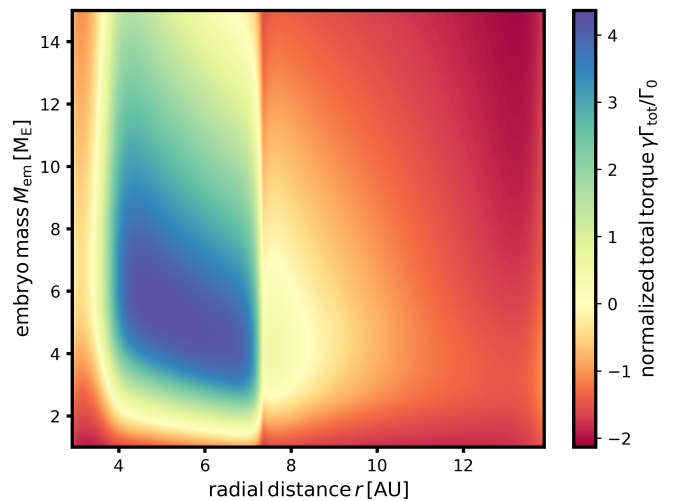}
\caption{Migration map based on the equilibrium state of the
protoplanetary disk. The color code shows the normalized 
value of the total torque $\gamma\Gamma_{\mathrm{tot}}/\Gamma_{0}$
acting on an embryo with the mass $M_{\mathrm{em}}$ (vertical axis)
placed on a circular orbit at the radial distance $r$ (horizontal axis)
in the disk. Calculated according to \cite{Paardekooper_etal_2011MNRAS.410..293P},
using the constant kinematic viscosity
$\nu=5\times10^{14}\,\mathrm{cm^{2}\,s^{-1}}$ and
the potential smoothing parameter $\epsilon=0.4H_{\mathrm{em}}$.}
\label{fig:torque_map}
\end{figure}

Let us also discuss the influence of the gas disk structure
on the orbital evolution of embedded planetary embryos.
In particular, we can estimate the expected direction
and rate of the Type-I migration of an embryo,
depending on its mass and location in the disk.
As in e.g. \cite{Kretke_Lin_2012ApJ...755...74K} or
\cite{Bitsch_etal_2013A&A...549A.124B},
we apply the analytical formulae from
\cite{Paardekooper_etal_2011MNRAS.410..293P}
on the azimuthally averaged profiles of the
equilibrium disk and compute the torque
acting on embryos. We do not list individual
steps of the torque calculation here, as there
are many, but note that the model of
\cite{Paardekooper_etal_2011MNRAS.410..293P}
is 2D and gives a prediction for low-mass planets
on fixed circular orbits, while accounting
for both Lindblad and corotation torques
in the non-linear regime, saturated and unsaturated
limits. The heating torque is not considered
in their model. Moreover, they used the
  $\epsilon$-smoothed Plummer-type potential
  for planet-disk interactions thus
  their torque formulae are parametric in
  the smoothing length $\epsilon$.

The resulting migration map, calculated for 
rather small $\epsilon=0.4H_{\mathrm{em}}$,
is shown in Fig.~\ref{fig:torque_map}.
The total torque $\Gamma_{\mathrm{tot}}$ felt by
embryos of various masses $M_{\mathrm{em}}$ is normalized
as $\gamma\Gamma_{\mathrm{tot}}/\Gamma_{0}$, where
\begin{equation}
  \Gamma_{0} = \left(\frac{q}{h_{\mathrm{em}}}\right)^{2}\Sigma_{\mathrm{em}} r_{\mathrm{em}}^{4}\Omega_{\mathrm{em}}^{2} \, ,
  \label{eq:gamma0}
\end{equation}
$q=M_{\mathrm{em}}/M_{\star}$ is the embryo-to-protostar
mass ratio and all of the remaining quantities
are calculated at the respective orbital radius $r_{\mathrm{em}}$.
It is important to emphasize that Fig.~\ref{fig:torque_map}
is only an auxiliary diagram
which does \emph{not} exactly represent the torque
felt by embryos in our simulations
\citep[see Appendix \ref{sec:verification} for a comparison
of torques with][]{Paardekooper_etal_2011MNRAS.410..293P}.
Despite of that, it is a useful tool for
getting a general picture of the expected migration rates
in different regions of the disk before actually
performing self-consistent simulations.

We notice there are two borderlines between
positive and negative torques in the disk.
The first is located at the snowline ($r\simeq4\,\mathrm{AU}$)
and the second is located at (roughly) $r\simeq7\,\mathrm{AU}$, i.e.
the transition between the viscously heated and stellar-irradiated
region. The outer one represents a zero-torque radius
where an accumulation (convergent migration) of embryos is expected to occur
because positive torques $\Gamma_{\mathrm{tot}}$
drive the embryos outwards while negative
torques inwards.

In the positive torque region, the negative Lindblad torque
is suppressed by the corotation torque.
The corotation torque generally arises as the gas parcels performing
U-turns exchange angular momentum with the embryo
and it is known to be determined by the vortensity
distribution which can be modified by advection along
the streamlines, or additional vorticity
can be produced by the temperature and entropy gradients
\citep{Baruteau_Masset_2008ApJ...672.1054B,Paardekooper_Papaloizou_2008A&A...485..877P}.
The latter is responsible for the strong positive torque
between the snowline and the stellar-irradiated region
because a suitable (negative) entropy gradient
is present due to the aspect ratio decrease.

The positive torque region should exist only for masses
$1.5\,\mathrm{M_{E}} \lesssim M_{\mathrm{em}} \lesssim 15\,\mathrm{M_{E}}$
for which the thermodynamic
conditions in the surrounding disk can sustain the corotation torque.
The corotation torque can be prevented from saturation when
the viscous and heat diffusion time scales
are shorter than the whole libration time scale (which decreases with increasing
embryo mass) but longer than the single U-turn
time scale \citep[e.g.][]{Pierens_2015MNRAS.454.2003P}.

\subsection{Case~I -- migration of non-accreting embryos in the gas disk only} 
\label{sec:caseI}

\begin{figure}[!ht]
\centering
\includegraphics[width=8.8cm]{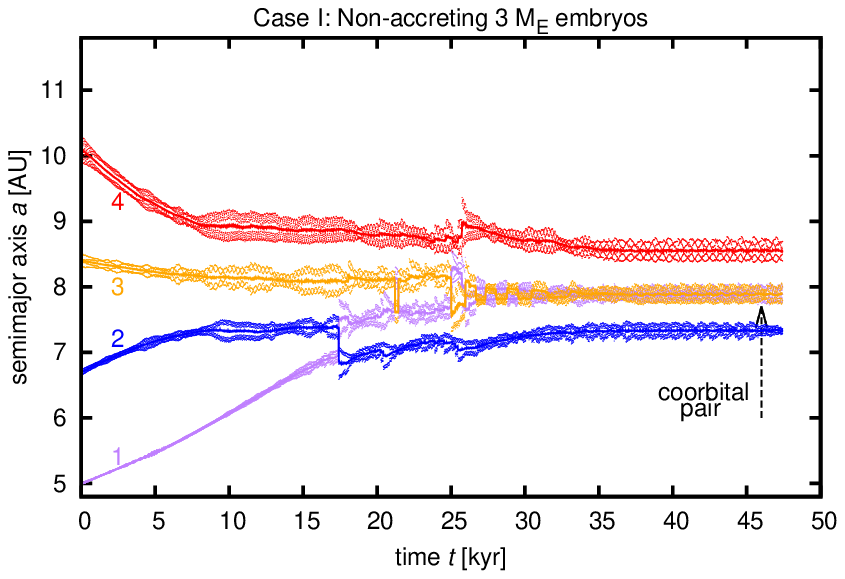}
\includegraphics[width=8.8cm]{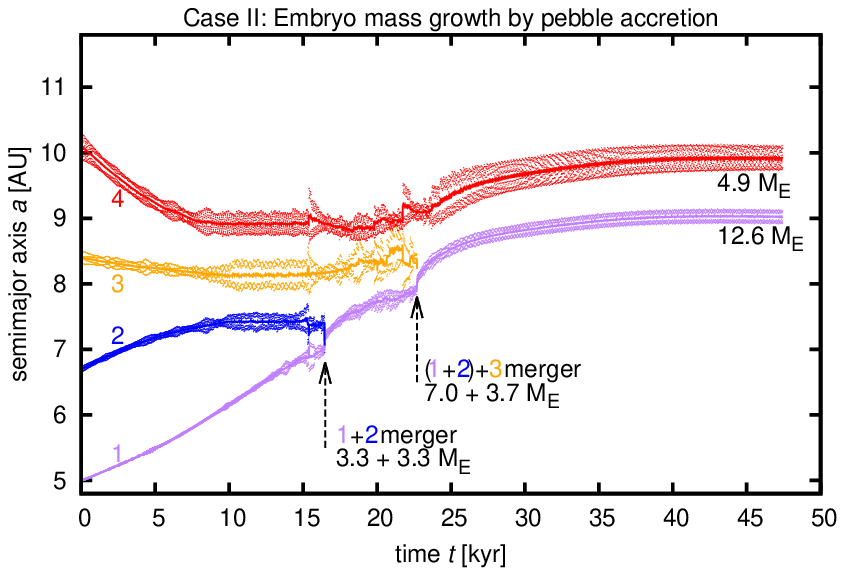}
\includegraphics[width=8.8cm]{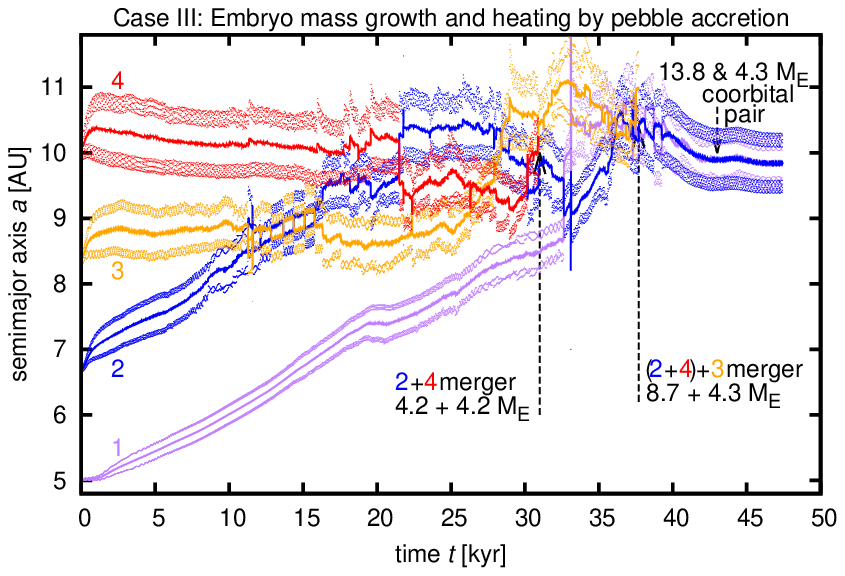}
\caption{Temporal evolution of semimajor axes $a(t)$, periastron
  distances $q_{\mathrm{p}}$ and apoastron distances $Q_{\mathrm{a}}$ of four
  embryos with the initial mass $3\,\mathrm{M_{E}}$ in three
  distinct simulation cases: Case~I neglecting the pebble disk (\emph{top}),
  Case~II including the pebble disk but only allowing for the mass growth of embryos
  by pebble accretion (\emph{middle}) and finally Case~III, considering also the effect
  of accretion heating (\emph{bottom}). Embryos are numbered from $1$ to $4$.
  Additional arrows and labels indicate mergers
  or coorbital pairs detected in the simulations, with corresponding embryo masses
  which can grow by pebble accretion (Cases~II and III) or merging.
  Striking differences are observed in Case~III as the migration
  rates are modified by the heating torque, orbits become
  moderately eccentric shortly after the simulation start and 
  the evolution is more violent compared to Cases~I and II.}
\label{fig:at_cases}
\end{figure}

Hereinafter we present and compare three different simulation
cases which start from the equilibrium disk and 
are numerically evolved for time spans covering
$t_{\mathrm{span}}\approx50\,\mathrm{kyr}$.
In all these simulations, we placed four embryos
with equal mass $M_{\mathrm{em}}=3\,\mathrm{M_{E}}$
on initially circular orbits with semimajor axes
equal to $a_{1}=5\,\mathrm{AU}$, $a_{2}=6.7\,\mathrm{AU}$,
$a_{3}=8.4\,\mathrm{AU}$ and $a_{4}=10.1\,\mathrm{AU}$;
the embryos being numbered inside out. The initial
inclinations were randomly chosen as small non-zero
values ($\lesssim 0.1\degr$).
The mass of the embryos is always introduced into
the system gradually in order to avoid shocks.
The same holds for the cases in which the embryos
act like the heat sources -- the released heat is
gradually amplified from zero towards the
self-consistently calculated value during
several initial orbits.

The simulation cases differ in the following manner.
In Case~I, we completely neglect the pebble disk,
thus the embryos interact only with the gaseous disk and
among themselves. Their masses remain fixed and
they do not release any heat into their vicinity.
In Case~II, the pebble disk is included and
the embryos are allowed to accrete from it, but
the corresponding accretion heating is still switched \emph{off}.
Therefore the heating torque cannot operate.
Finally, Case~III is the same as Case~II
except the accretion heating is switched \emph{on}.
Case~I represents a relatively standard scenario
\citep[comparable e.g. with][]{Pierens_2015MNRAS.454.2003P}
in which one can study interactions of multiple
embryos with the non-isothermal radiative disk.
We already made some predictions of the embryo
migration rates for this case in
Sec.~\ref{sec:migration_map}.

Fig.~\ref{fig:at_cases} (top panel) shows the temporal evolution
of the osculating semimajor axis $a$, periastron distance
$q_{\mathrm{p}}=a(1-e)$ and apoastron distance
$Q_{\mathrm{a}}=a(1+e)$ of embryos.
At the beginning, embryos~1 and 2
(purple and blue curves, respectively) migrate outwards
while embryos~3 and 4 (orange and red curves)
migrate inwards, in accordance with
the preliminary migration map (Fig.~\ref{fig:torque_map}).
After $\simeq8\,\mathrm{kyr}$ of convergent
migration towards the zero-torque radius,
the outermost three embryos get locked in mutual mean-motion
resonances which start to excite their orbital eccentricites.
The innermost embryo catches up with the resonant chain
at $\simeq17\,\mathrm{kyr}$ and shortly after its eccentricity
excitation it undergoes a close encounter with
the second embryo during which they switch positions in the disk.
As embryo~1 is scattered outwards, it interacts with
embryo~3 in a series of close encounters which, due to
damping effects of the surrounding disk, end up in a 
formation of a coorbital pair (1:1 commensurability).
The system remains stable for the rest of the simulation.

\subsection{Case~II -- introducing pebble disk and embryo growth by pebble accretion}
\label{sec:caseII}

\begin{figure}[!t]
\centering
\includegraphics[width=8.8cm]{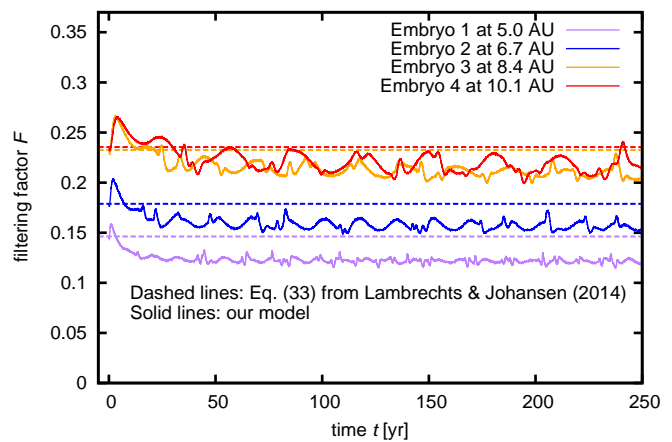}
\caption{Filtering factor $F$ measured for the embryos
at the beginning of Case II (solid curves); also applicable in Case III.
As a comparison (dashed lines), we plot the filtering factors calculated
at $t=0$ according to formula (33)
from \cite{Lambrechts_Johansen_2014A&A...572A.107L}.
The analytical prediction is in a good agreement with results
of our model.}
\label{fig:filtering}
\end{figure}

In Case~II, the pebble disk is considered and the embryos grow
by pebble accretion.
The pebble accretion rate onto individual embryos, which sets
their mass growth and eventually the amount of heat released
to their surroundings (Sec.~\ref{sec:caseIII}),
is shown in Fig.~\ref{fig:filtering}
in terms of the filtering factor $F$, defined as
\begin{equation}
  F \equiv \frac{\dot{M}_{\mathrm{em}}}{\dot{M}_{\mathrm{F}}} \, .
  \label{eq:filtering}
\end{equation}
We plot its temporal dependence with respect
to a fixed value of the radial pebble mass flux,
$\dot{M}_{\mathrm{F}}=2\times10^{-4}\,\mathrm{M_{E}}$.
We compare the filtering factor measured at the beginning
of Case~II with the
analytical formula from \cite{Lambrechts_Johansen_2014A&A...572A.107L}
which we applied on the equilibrium disk model.
At $t=0$, $F$ is in an excellent agreement with the analytical
prediction and at later times, the differences are
not larger than $3\%$.
Temporal oscillations of $F$ are due to the nature
of the accretion algorithm implementation. The expected
embryo mass change $\Delta M_{\mathrm{expec}}$
(Eq.~\ref{eq:dm_expec}) depends on
the instantaneous $\bar{\Sigma}_{\mathrm{p}}$ within
the accretion radius. The amount of removed pebbles per $\Delta t$
is not precisely balanced by the inflow of new pebbles
so the removal and inflow adapt to each other. If for example
 density waves are propagating near an accreting embryo,
  they can temporarily increase concentration of pebbles ($\bar{\Sigma}_{\mathrm{p}}$)
  and we observe an increase of $F$.
Such variations cannot be reproduced by the \cite{Lambrechts_Johansen_2014A&A...572A.107L}
model because it is not hydrodynamic.
We verified that the filtering factors measured
in Case~II are in agreement with those obtained later in
Case~III.
Finally, notice that the outermost embryo is the fastest grower
which is because $F\sim1/\eta$
\citep{Lambrechts_Johansen_2014A&A...572A.107L} and $\eta$
is smaller in the outer part of the disk (Fig.~\ref{fig:eta_stokes}).
However, the differences in $F$ between individual embryos
are rather marginal and the mass growth by pebble
accretion initially proceeds in the oligarchic fashion, as
expected \citep{Morbidelli_Nesvorny_2012A&A...546A..18M}.

The orbital evolution of embryos in Case~II
is shown in middle panel of Fig.~\ref{fig:at_cases}.
At first, the embryos evolve similarly to Case~I,
but the interaction among embryos~1 and 2
results in a merger at $t\simeq16.5\,\mathrm{kyr}$.
The resulting mass of the merger is $6.6\,\mathrm{M_{E}}$.
As the system adapts to the loss of one of its members
and to the suddenly increased mass of the merger,
embryo~3 is pushed slightly outwards and encounters embryo~4.
One of these events scatters embryo~3
inwards where it eventually collides with the previous merger.
The collision takes place at $t\simeq22.7\,\mathrm{kyr}$
and merges masses
$3.7\,\mathrm{M_{E}}$ (embryo~3) and $7\,\mathrm{M_{E}}$ (previous merger).
The remaining embryos are stabilized at somewhat distant
orbits in comparison with Case~I. The embryo masses
at the end of the simulation are $12.6\,\mathrm{M_{E}}$ (the inner
one) and $4.9\,\mathrm{M_{E}}$ (the outer one).
The outer embryo~4 gained $1.9\,\mathrm{M_{E}}$ 
by pebble accretion during the simulation time span.

Let us emphasize that as the mergers naturally
occur in the system of pebble-accreting embryos,
they immediately break the oligarchic growth of the
embryos by pebble accretion (instead of multiple
similar-sized embryos, a dominant massive core
is formed within the system).
In the light of this statement, models which
estimate the final planetary masses by tracking a
single pebble-accreting protoplanet
\citep[e.g][]{Bitsch_etal_2015A&A...582A.112B}
probably underestimate how
massive can the planets actually become,
at least near the zero-torque radii.

Because of possible strong sensitivity to the initial
conditions, it is debatable what is the significance of the differences
which we identified between Cases~I and II.
To partially answer this question,
we ran two more simulations for each case. In the first additional
set we increased the initial inclinations to about $\simeq1\degr$,
in the second additional set we started from a more closely-packed
system of embryos with orbital separations equal to
$4.5$ mutual Hill radius $R_{\mathrm{mH}}=0.5(a+a')[(q+q')/3]^{1/3}$.
In these additional simulations, Case~I always resulted in one merger
before the system became stabilized, whereas in Case~II, we always
detected two mergers. The larger number of mergers in Case~II
occurs because the resonant chains are destabilized more
often. The destabilization is provided by the mass growth
which changes the strength of the resonant forcing
and the streamline topology near the embryos, thus
modifying the acting torques.
At the same time, more massive embryos have a larger
encounter cross section. Yet our simulation statistics
it too poor to estimate corresponding probabilities
or merging in Cases~I and II.

\subsection{Case~III -- introducing heating by pebble accretion}
\label{sec:caseIII}

We now discuss Case~III, presented
in bottom panel of Fig.~\ref{fig:at_cases}.
The system evolves differently since the
simulation start compared to the previous cases.
First of all, the dispersion of both $q_{\mathrm{p}}$ and $Q_{\mathrm{a}}$
with respect to $a$ is much larger
in the presence of accretion heating.
In other words, the orbits of embryos are more eccentric.
We find $e\simeq0.02$ for the innermost embryo~1
and $e\simeq0.04$ for the outermost embryo~4 after $5\,\mathrm{kyr}$
of evolution, while the corresponding values in Case~II
were $e\simeq0.004$ and $e\simeq0.01$, respectively.
Moreover, the increased eccentricity is \emph{not} produced
by the resonant forcing, it is observable already before
the embryos form a closely-packed configuration.
Looking at the simulation beginning, we see a brief period
during which both the semimajor axis and orbital eccentricity
swiftly increase, especially for the three outer embryos.
It seems that this period of evolution must represent a 
transitional state of the system during which the hydrodynamic
background adjusts to the presence of the new heat source
and the orbits react accordingly.
The ability of the gas disk to circularize the orbits
is clearly reduced in this case
which is a new and unexpected phenomenon,
explored in detail in Sec.~\ref{sec:hot_trail}.

\paragraph{Modified disk torques.}
Another surprising feature is that the inner
embryos~1 and 2 are able to maintain outward
migration despite having moderate eccentricity.
We recall that the eccentricity growth
leads to shrinking of the horseshoe
region, and the corotation torque $\Gamma_{\mathrm{c}}$
in its unsaturated non-linear
limit depends on the half-width
of the horseshoe region $x_{\mathrm{hs}}$
\citep{Paardekooper_Papaloizou_2009MNRAS.394.2297P}
as $\Gamma_{\mathrm{c}} \sim x_{\mathrm{hs}}^{4}$
\citep{Fendyke_Nelson_2014MNRAS.437...96F}.
The positive contribution of $\Gamma_{\mathrm{c}}$
in the region of outward migration is thus expected
to vanish with increasing eccentricity
\citep{Bitsch_Kley_2010A&A...523A..30B}.
Yet, we observe that the migration of the inner embryos~1 and 2
is still directed outwards with a rate similar to Cases~I and II
and the torques even allow the embryos to penetrate into the outer disk.
As for the outer embryos~3 and 4,
their migration first proceeds inwards (except for a short initial phase)
but with significantly reduced migration rate.

It is worth noting that the zero-torque radius is somewhat
ignored by embryos in Case~III. As a result, we do not
see the embryos to become closely-packed around $\simeq7.5\,\mathrm{AU}$
like in the previous cases. Instead, embryo~2 swiftly penetrates
into the outer disk and interacts with embryo~3, and shortly after
that with embryo~4. Meanwhile, embryo~1 reaches the
expected location of the zero-torque radius and stays there
for a while, being stopped by interactions with embryo~3. But ultimately,
it continues outwards, migrating along with embryo~3 almost as a pair.

  Examining the excited orbital eccentricities
  properly, we notice that $e\simeq h$. Therefore one
  can expect significant modifications of the Lindblad
  torque \citep{Papaloizou_Larwood_2000MNRAS.315..823P,Cresswell_Nelson_2006A&A...450..833C}
  as the eccentric embryos exhibit radial excursions in
  the disk and variations of the orbital velocity,
  thus periodically exciting density waves
  propagating inwards and outwards during the orbit.
  In such a case, the Lindblad torque, which is usually
  negative, can become reduced, or even reversed.
  Regarding the heating torque, its contribution is positive.
  But we emphasize that because of the increased eccentricity
  and due to narrowing of the horseshoe region,
  we can expect the heating torque to operate
  in a mode which was \emph{not} described by
  \cite{Benitez-Llambay_etal_2015Natur.520...63B}
  who studied the heating torque for planets on fixed
  circular orbits. Here we summarize that the migration
  rate in Case~III is driven by the modified Lindblad and heating
  torques acting on eccentric orbits.
  Detailed investigation of the torques
  accompanying the accretion heating is provided in Sec.~\ref{sec:hot_trail_torque}.

\paragraph{Merging and resonant chain instabilities.}
Once the embryos become closely packed, they interact violently
because their eccentric orbits drive one another
into frequent close encounters. At $t\simeq12\,\mathrm{kyr}$,
embryos~2 and 3 get temporarily locked
in a coorbital resonance which is disrupted by
convergent migration towards the outer embryo~4.
The three outer embryos then strongly interact and
swap positions several times before there
is a first merger
of two $4.2\,\mathrm{M_{E}}$ embryos (blue and red) at $\simeq31\,\mathrm{kyr}$.
Three-body interactions of the remaining embryos
produce another merger at $\simeq37.7\,\mathrm{kyr}$
when $8.7\,\mathrm{M_{E}}$ embryo (blue)
and $4.3\,\mathrm{M_{E}}$ embryo (orange) collide.
The system is stabilized by formation of a coorbital pair,
having the final masses $13.8\,\mathrm{M_{E}}$
and $4.3\,\mathrm{M_{E}}$.

  Although the system evolves into a 1:1 orbital resonance
  at the end, it is not capable of establishing a
  global resonant chain during its evolution, apart from temporal
  resonant captures. This is different
  with respect to Cases~I and II where the system
  becomes resonant once the embryos become closely packed
  and stays that way except for occasional instabilities during
  encounters, orbital swapping and embryo merging.
  The decreased probability of resonant capture is again
  caused by excited eccentricities, as discussed
  e.g. by \cite{Batygin_2015MNRAS.451.2589B}.

  Regarding the possibility of mergers,
  their number is the same as in Case~II but they
  occur later during the evolution. This is slightly surprising
  because we already argued that close encounters are more frequent
  so a natural question arises -- why do not mergers
  appear sooner? To provide a basic statistical check
  as in Cases~I and II, we performed two additional simulations,
  the first with initially smaller orbital separations ($4.5\,R_{\mathrm{mH}}$)
  and the second with slightly larger inclinations ($\simeq1\degr$).
  The first simulation produced only one merger, the second
  produced none. At the same time, 
  we confirmed the
  strong eccentricity increase unrelated to mutual close encounters
  which became frequent as a consequence of the eccentricity growth.

  The reduced merging efficiency compared to Case~II is probably
  another consequence of larger eccentricities which lead
  to larger relative velocities during encounters and
  subsequently, merging is more difficult.
  Regarding the second additional simulation with zero mergers,
  we find that orbital inclinations are not reduced enough before the 
  close encounters start to occur. Due to larger encounter velocities,
  vertical stirring is observed, maintaining the inclinations above zero.
  Such an inclined orbital configuration is not suitable for merging.

  We remark that the influence of the accretion heating on the system's
  evolution and stability may be even more evident if higher
  number of embryos is considered, which is what
  we intend to study in the future (as proposed in Sec.~\ref{sec:discussion}).

In both Cases~II and III, we see that mergers
produce embryos massive enough to potentially
become giant planet cores. However, this subsequent evolution
is not covered in our simulations
as the gravitational attraction and subsequent collapse
of a massive gaseous envelope is a delicate
and not well understood process which is beyond the scope
of this paper
\citep[see e.g.][]{Ayliffe_Bate_2009MNRAS.393...49A,Machida_etal_2010MNRAS.405.1227M}.

\begin{figure}[!ht]
\centering
\includegraphics[width=8.8cm]{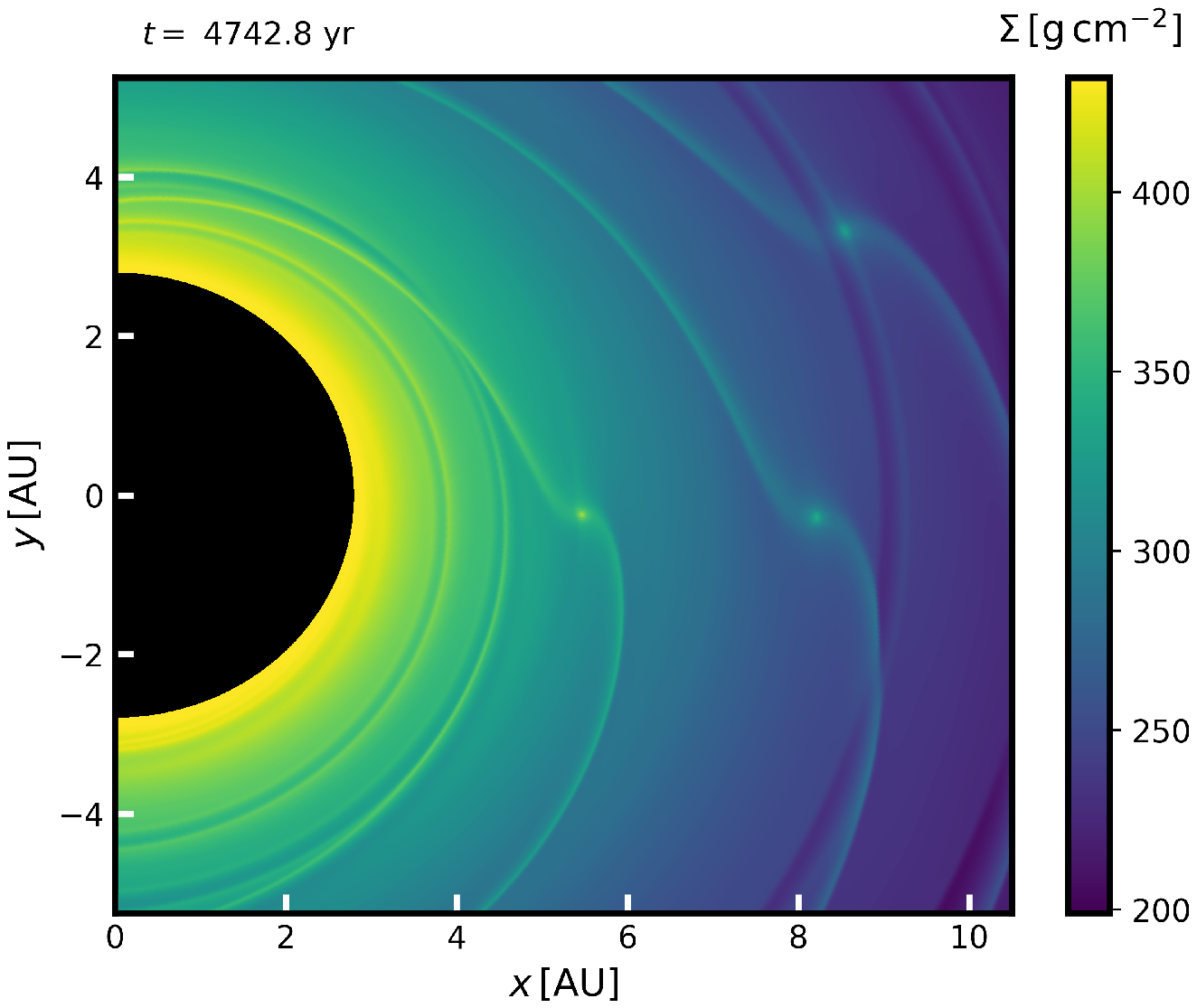}
\includegraphics[width=8.8cm]{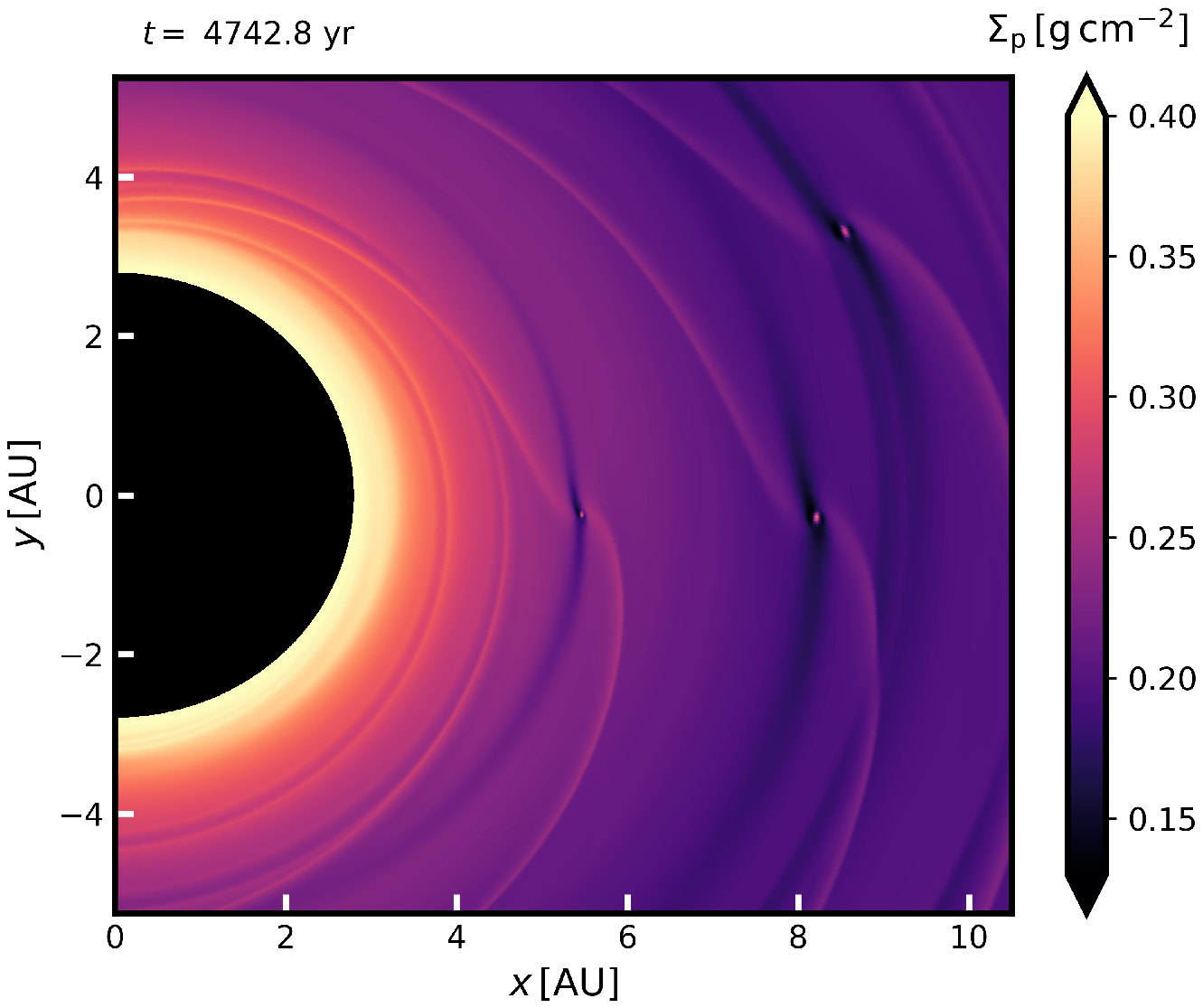}
\caption{A closeup of the gas surface density $\Sigma$ (\emph{top})
and pebble surface density $\Sigma_{\mathrm{p}}$ (\emph{bottom})
after $\simeq5\,\mathrm{kyr}$ of evolution in the simulation
with pebble accretion but \emph{without} accretion heating, i.e. Case~II.
The gaps in the pebble disk are opened by accreting planetary embryos.
A fourth embryo is also present in the
system but it is located outside the range.}
\label{fig:surfdens_noheating}
\end{figure}

\begin{figure}[!ht]
\centering
\includegraphics[width=8.8cm]{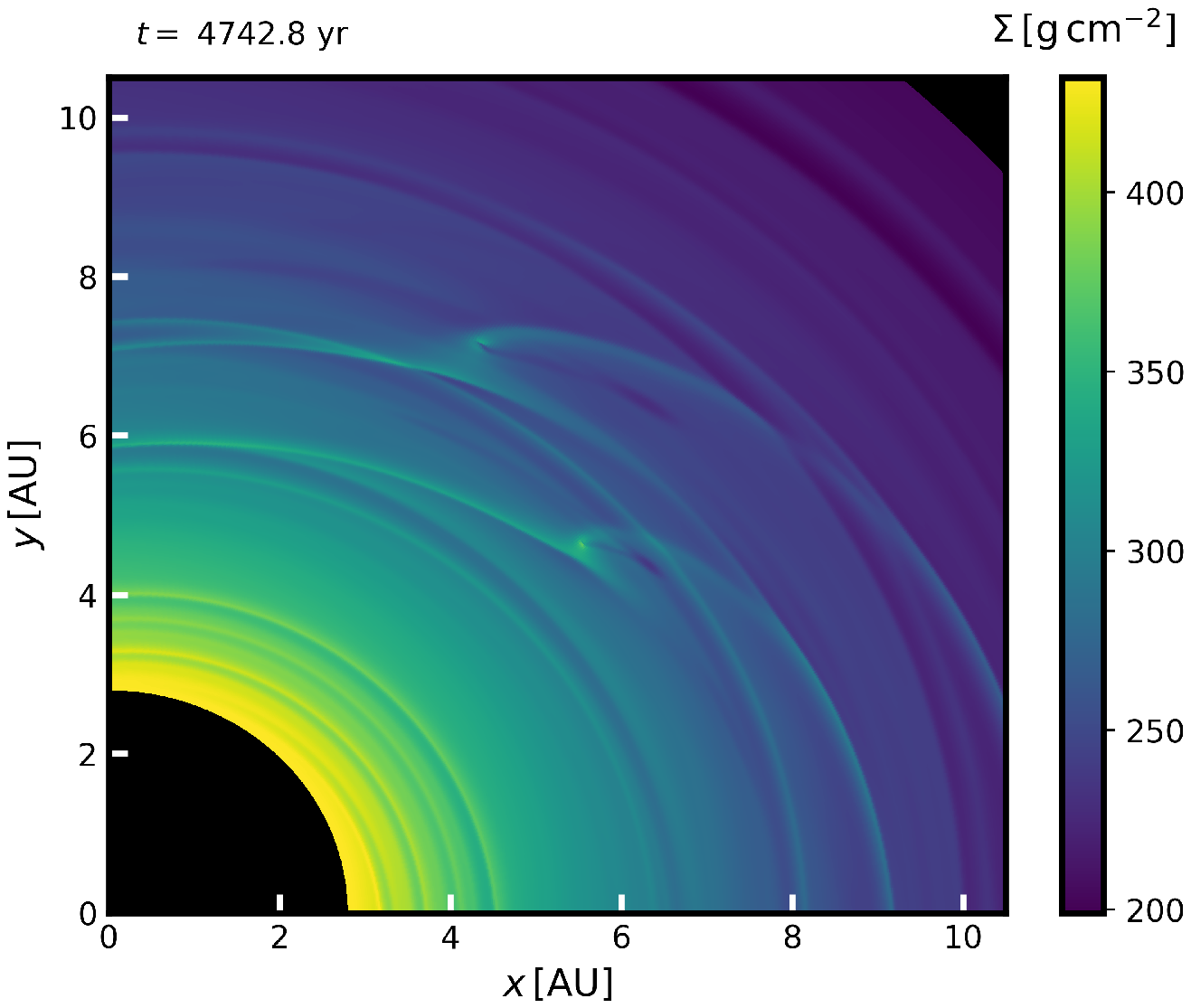}
\includegraphics[width=8.8cm]{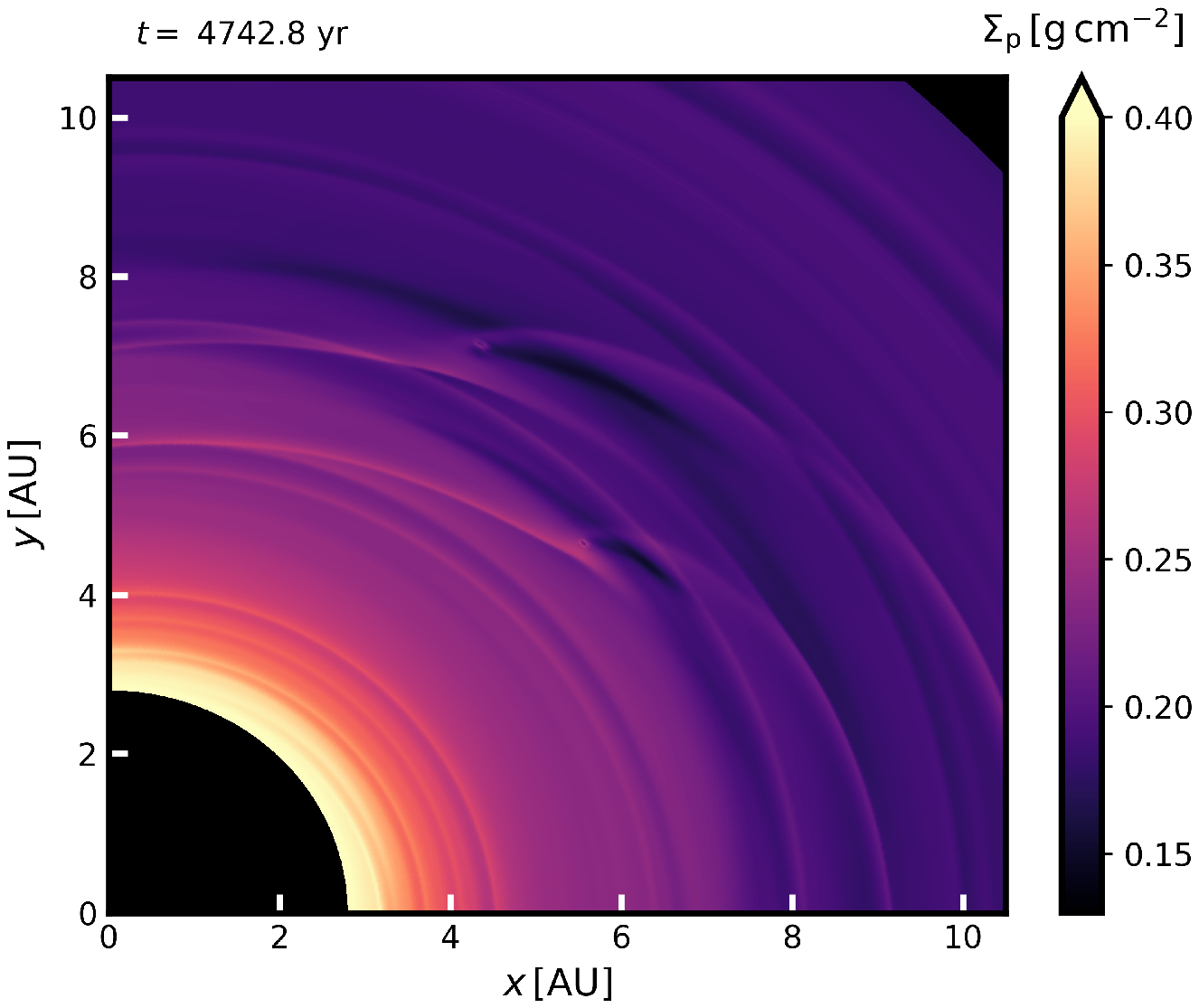}
\caption{Same as Fig.~\ref{fig:surfdens_noheating} but for the
simulation \emph{with} accretion heating (Case~III).
Two embryos are located at $x=5.55,y=4.65\,\mathrm{AU}$
and $x=4.35,y=7.17\,\mathrm{AU}$; two other embryos are
located outside the range.
The $\Sigma$ distribution shows
there are trails of underdense gas stretching outwards from the embryos,
trailing their orbital motion. The shape of cavities in the pebble component
is affected by the eccentric orbits of embryos.
Unlike in Fig.~\ref{fig:surfdens_noheating},
the concentration peak at the embryos' location is somewhat blurred
in both gas and pebbles.}
\label{fig:surfdens_heating}
\end{figure}

\paragraph{Gas and pebble surface density.}

To start the investigation of the unexpected
eccentricity growth related to accretion heating,
we first compare snapshots of the gas and pebble
surface density in Cases~II and~III.
Fig.~\ref{fig:surfdens_noheating}
shows $\Sigma$ and $\Sigma_{\mathrm{p}}$ in Case~II,
after $4.7\,\mathrm{kyr}$ of evolution.
The gas disk exhibits typical features --
embryos launch spiral arms and produce minor density variations
in their horseshoe regions.
The pebble disk is affected by the ongoing pebble accretion.
Accreting embryos carve partial gaps in the pebble component along
their orbits. The gap has two parts, one of them is trailing
and the other one is leading the orbital 
motion of an embryo (which is oriented counterclockwise in
all plots). The formation of these two
parts can be explained simply by the trajectories of
pebbles with respect to the embryo \citep{Morbidelli_Nesvorny_2012A&A...546A..18M}
-- those drifting from outside meet the embryo
head-on, and those which have drifted across
the embryo's orbit catch up with it from behind.
After a portion of the pebble flux is filtered out by
the embryo, there is a paucity of pebbles
behind it, slightly outside the embryo's orbit,
and another cavity is formed in the direction of orbital motion,
slightly inside the embryo's orbit.

Fig.~\ref{fig:surfdens_heating} shows 
$\Sigma$ and $\Sigma_{\mathrm{p}}$ in Case~III,
again in simulation time $4.7\,\mathrm{kyr}$.
We see that the shape of spiral arms is somewhat modified
which is to be expected as the embryos already orbit
with considerable eccentricities
\citep{Cresswell_etal_2007A&A...473..329C,Bitsch_Kley_2010A&A...523A..30B}.
The gaps in the pebble disk are slightly skewed 
and widened because the eccentric embryos
perform radial excursions while carving the gaps.
But looking at $\Sigma$, there is a strange feature --
underdense structures trailing the embryos, starting
at their locations and stretching slightly
to $r>r_{\mathrm{em}}$.
Explanation of these underdensities,
as well as investigation of the eccentricity growth,
is given in the following section.

\section{The hot trail effect -- the orbital eccentricity excitation due to accretion heating}
\label{sec:hot_trail}

In order to understand the process
leading to the eccentricity excitation
and also to the formation of underdense structures in the
gas distribution adjacent to the embryos,
we must first check whether we can recover
these phenomena in simulations with a single
embryo. This should verify whether the disk$\leftrightarrow$embryo
interaction \emph{alone} is sufficient
to raise the eccentricity, without
help of any additional perturbers.

Starting again with the equilibrium fiducial disk,
we placed a single $3\,\mathrm{M_{E}}$ embryo 
on an orbit with semimajor axis $a=6.5\,\mathrm{AU}$.
The orbit was initially circular in one case,
and $e_{0}=0.05$ was assigned to the embryo in another case.
Both the circular and the eccentric orbits
were evolved for several hundreds of years: (i) in
the gas disk only with fixed embryo mass, and (ii)
with pebble accretion and respective heating considered.
The embryo was allowed to fully interact with the disk,
i.e. the orbit was not held fixed.

\begin{figure}[!t]
\centering
\includegraphics[width=8.8cm]{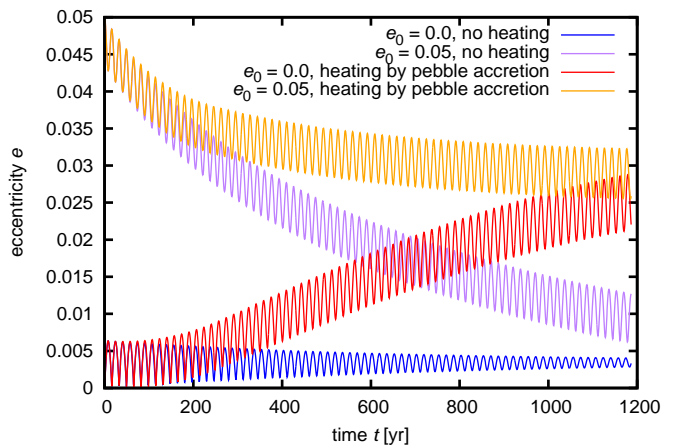}
\caption{Temporal evolution of the osculating eccentricity
  $e(t)$ for a single $3\,\mathrm{M_{E}}$ embryo in four distinct
  simulation setups. In the first two setups we neglect pebble accretion
  and the initial eccentricity is $e_{0}=0$ (blue curve) and $e_{0}=0.05$
  (purple curve). In the other two setups we consider pebble accretion
  and heating, the initial eccentricity being again $e_{0}=0$ (red curve)
  and $e_{0}=0.05$ (orange curve). Accretion heating reduces the eccentricity
  damping efficiency for the eccentric orbit and excites the eccentricity
  of the circular orbit.
}
\label{fig:et_hottrail}
\end{figure}

Let us first examine the eccentricity evolution
in these four simulation setups, as shown in Fig.~\ref{fig:et_hottrail}.
In simulations with fixed embryo mass,
the initially circular orbit oscillates around small eccentricity
values and the initially eccentric orbit is being damped
and almost circularized ($e=0.003$). On the other hand, $e$ in
simulations with accretion heating converges
to moderate non-zero value ($e=0.03$), even for the initially
circular orbit.
Therefore the eccentricity excitation and reduced eccentricity
damping which we identified in Sec.~\ref{sec:caseIII} are indeed reproduced.

The simulation with $e_{0}=0$ and heating
by pebble accretion is the most interesting one
because it proves that the embryo can gain and sustain
eccentricity solely due to forces arising
from the disk. We will thus discuss
this simulation in detail for the remainder
of this section.
Looking at the red curve in Fig.~\ref{fig:et_hottrail},
it is obvious that there are several
distinct stages during which the eccentricity
excitation rate changes.
We pick three characteristic
times $t\simeq180$, $360$ and $1130\,\mathrm{yr}$
at which we investigate the disk-embryo
interaction during one orbital period.
We will refer to these three evolutionary stages
as the \emph{onset}, \emph{growth} and \emph{saturation} phase for brevity.

In order to identify contributions from the disk 
responsible for $\mathrm{d}e/\mathrm{d}t$
variations, we employ the Gauss perturbation equation
for the eccentricity
\begin{equation}
  \frac{\mathrm{d}e}{\mathrm{d}t} = \frac{\sqrt{1-e^{2}}}{na}[\mathcal{R}\sin{f}+\mathcal{T}(\cos{f}+\cos{E})] \, ,
  \label{eq:gauss_dedt}
\end{equation}
where $n$ denotes the embryo's mean motion,
$\mathcal{R}$ and $\mathcal{T}$ are the radial
and tangential components of the perturbing
acceleration arising from the disk,
$f$ is the true anomaly and $E$ is the eccentric anomaly,
for which one can write
$\cos{E} = (e+\cos{f})/(1+e\cos{f})$.
Assuming that the variation of orbital elements
during one orbital period is negligible, we can 
limit ourselves to an analysis of the Gauss
factors inside the square brackets in Eq.~\ref{eq:gauss_dedt}.
We shall denote $G_{r} \equiv \mathcal{R}\sin{f}$
and $G_{\theta} \equiv \mathcal{T}(\cos{f}+\cos{E})$.

\subsection{Radial perturbation}
Fig.~\ref{fig:gauss_at} (top panel) shows the
values of $G_{r}$ acting on the embryo as it
travels along its orbit during the onset, growth
and saturation phases. Because $\mathcal{R}$ itself
is always negative and almost identical in all
the individual phases, $G_{r}$ also does not change
significantly. It is a $f$-periodic function and 
we find it to be typically an order of magnitude
stronger than $G_{\theta}$. Thus from
the dynamical point of view, it is responsible
for fast variations of the orbital eccentricity
which occur on the orbital time scale.
The varying $e(t)$ function corresponding to the onset
phase is overplotted in Fig.~\ref{fig:gauss_at} (dashed curve).
As the embryo moves from the periastron towards the apoastron, 
$G_{r}<0$ implies $\mathrm{d}e/\mathrm{d}t<0$ which
decreases $e$, and vice versa.
Because of $G_{r}$ symmetry, the respective changes
of the eccentricity average out and do \emph{not} lead
to secular variations.

The existence of non-zero
  radial acceleration $\mathcal{R}$ is due to the gas surface
  density profile of the surrounding disk which is in general
  an outward-decreasing power-law function. Consequently,
  within an arbitrary radius around the embryo, one can expect
  overabundance of gas inwards from the orbit, while the mass
  of the gas outwards is smaller.

\subsection{Azimuthal perturbation}
\begin{figure}[!h]
\centering
\includegraphics[width=8.8cm]{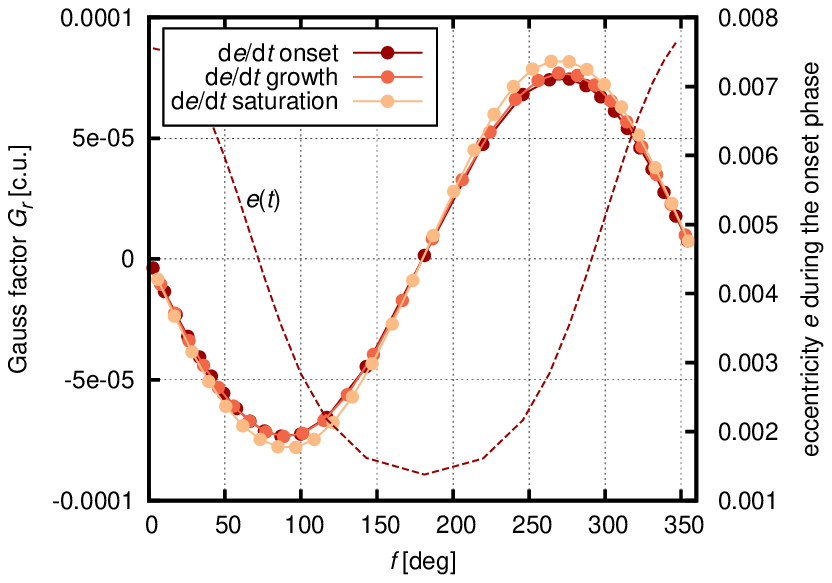}
\includegraphics[width=8.8cm]{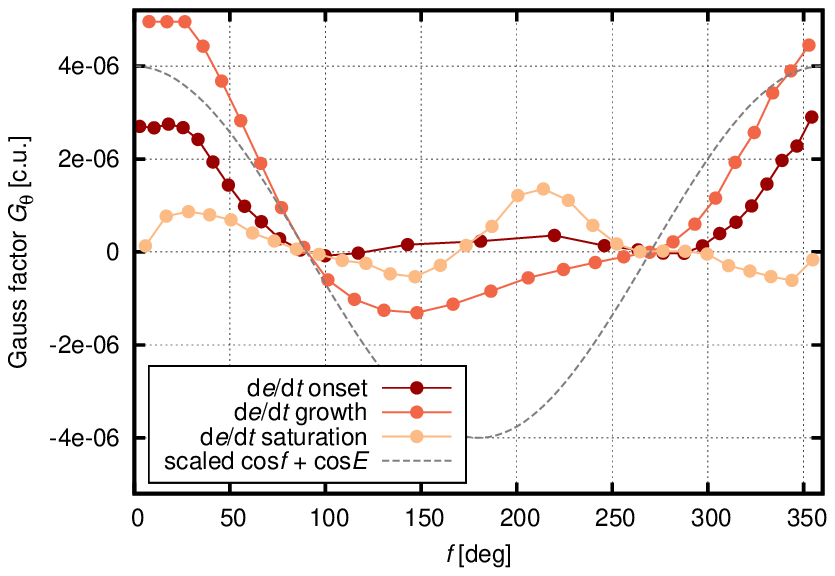}
\includegraphics[width=8.8cm]{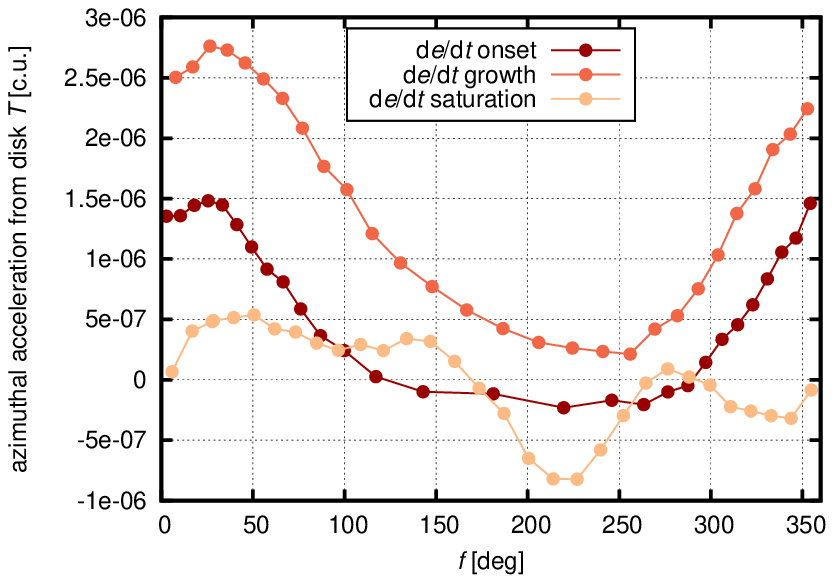}
\caption{Measures of the gravitational acceleration from the disk
  acting on the embryo, evolving from initially circular orbit in the presence
  of pebble accretion and the heating torque (i.e. red curve in Fig.~\ref{fig:et_hottrail}).
  The values are recorded during one orbital period (represented by the true
  anomaly $f$), at around $t\simeq180$, $360$ and $1130\,\mathrm{yr}$ of the simulation,
  i.e. during the onset, growth and saturation phase of the eccentricity
  excitation. \emph{Top}: Evolution
  of the Gauss factor $G_{r} \equiv \mathcal{R}\sin{f}$ (left
  vertical axis) and the osculating eccentricity $e$, which was recorded during the
  onset phase (right vertical axis). \emph{Middle}: Evolution
  of the Gauss factor $G_{\theta} \equiv \mathcal{T}(\cos{f}+\cos{E})$.
  The function $(\cos{f}+\cos{E})$ for $e=0.005$
  scaled to the axis range is also given for reference (gray dashed curve).
  \emph{Bottom}: The azimuthal acceleration $\mathcal{T}$ from the disk.
}
\label{fig:gauss_at}
\end{figure}

As argued above, $G_{r}$ is related to the 
orbital frequency in the $e$-oscillations and 
cannot cause the runaway growth of the eccentricity.
Consequently, $G_{\theta}$ must be responsible
for the secular changes and we plot it in middle
panel of Fig.~\ref{fig:gauss_at}.
In order to guide the eye, we overplot 
the $(\cos{f}+\cos{E})$ function for $e=0.005$, scaled down
to the figure range. It represents a dependence which
$G_{\theta}$ would follow if $\mathcal{T}$ was a constant
positive acceleration.
Examining the $G_{\theta}$ profile measured
in our simulation, we notice there are some \emph{asymmetries}
during the orbital period which can \emph{accumulate} in time
and cause $e$ to grow.

During the onset phase, $G_{\theta}$ is maximum when
the embryo is at periastron and shortly afterwards.
Then it decreases to zero as $f\rightarrow90\degr$,
stays at low positive values through the apoastron passage
and at $f\simeq290\degr$ it finally starts to increase
back to the maximum value. $G_{\theta}$ averaged over one orbital period
is positive which implies $\mathrm{d}e/\mathrm{d}t>0$,
in agreement with the onset of the eccentricity excitation
in Fig.~\ref{fig:et_hottrail}.

The azimuthal acceleration $\mathcal{T}$
related to $G_{\theta}$ is plotted in bottom panel of Fig.~\ref{fig:gauss_at}.
We see that the embryo feels strong positive acceleration
in the direction of its orbital motion around the periastron,
with the peak slightly shifted to $f\simeq30\degr$.
From $f\simeq110\degr$ to $f\simeq290\degr$,
$\mathcal{T}$ has a flat profile and it is negative.
In terms of the expected gas distribution, there must be an
accumulation of mass \emph{in front} of the embryo around the periastron.
For the rest of the orbit, this accumulation
should become weaker and from $f\simeq110\degr$ to $f\simeq290\degr$,
an excess of gas \emph{behind} the embryo's orbital motion is expected.

In the growth phase, the azimuthal acceleration $\mathcal{T}$ remains
positive for the \emph{entire} orbit, having a similar orbital evolution
as in the onset phase, with an enhanced peak near the periastron,
followed by decrease and plateau
towards the apoastron.
Consequently, $G_{\theta}$ has an increased amplitude but it also becomes negative from $f=90\degr$
to $270\degr$. Despite that, the averaged $G_{\theta}$
is again positive and so is $\mathrm{d}e/\mathrm{d}t$.
The shape of $\mathcal{T}(f)$ tells us that
we can expect the gas distribution around the embryo to
be denser ahead of the embryo for the \emph{entire} orbit.

During the saturation phase, the azimuthal acceleration $\mathcal{T}$
has a somewhat \emph{complex} dependence on $f$.
Its overall amplitude is smaller compared
to the previous phases by an order of magnitude. The acceleration $\mathcal{T}$
remains positive from periastron to apoastron and it is negative
through the remaining half of the orbit, apart from a short interval
at around $f\simeq275\degr$. Looking at the respective $G_{\theta}$
dependence, its shape is quite similar to a $\pi$-periodic function in $f$,
oscillating around zero,
having two maxima between the periastron and $f=90\degr$ and between
the apoastron and $f=270\degr$ and vice versa.

\subsection{Hydrodynamic explanation of the eccentricity excitation}
\begin{figure*}[!hpt]
  \centering
  \begin{tabular}{ccc}
    \begin{sideways} \hspace{2.0cm}periastron \end{sideways} &
    \includegraphics[width=6.5cm]{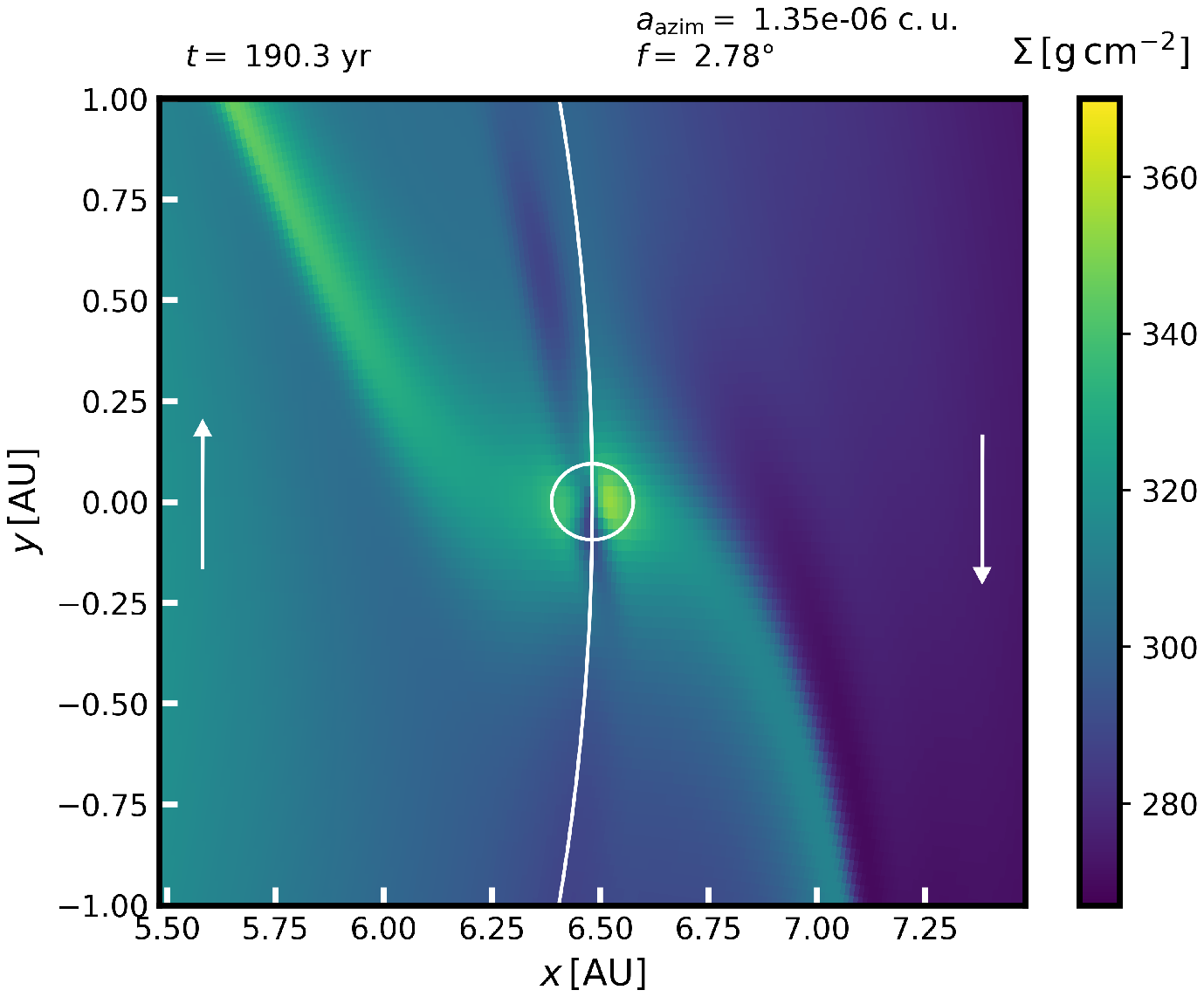} &
    \includegraphics[width=6.5cm]{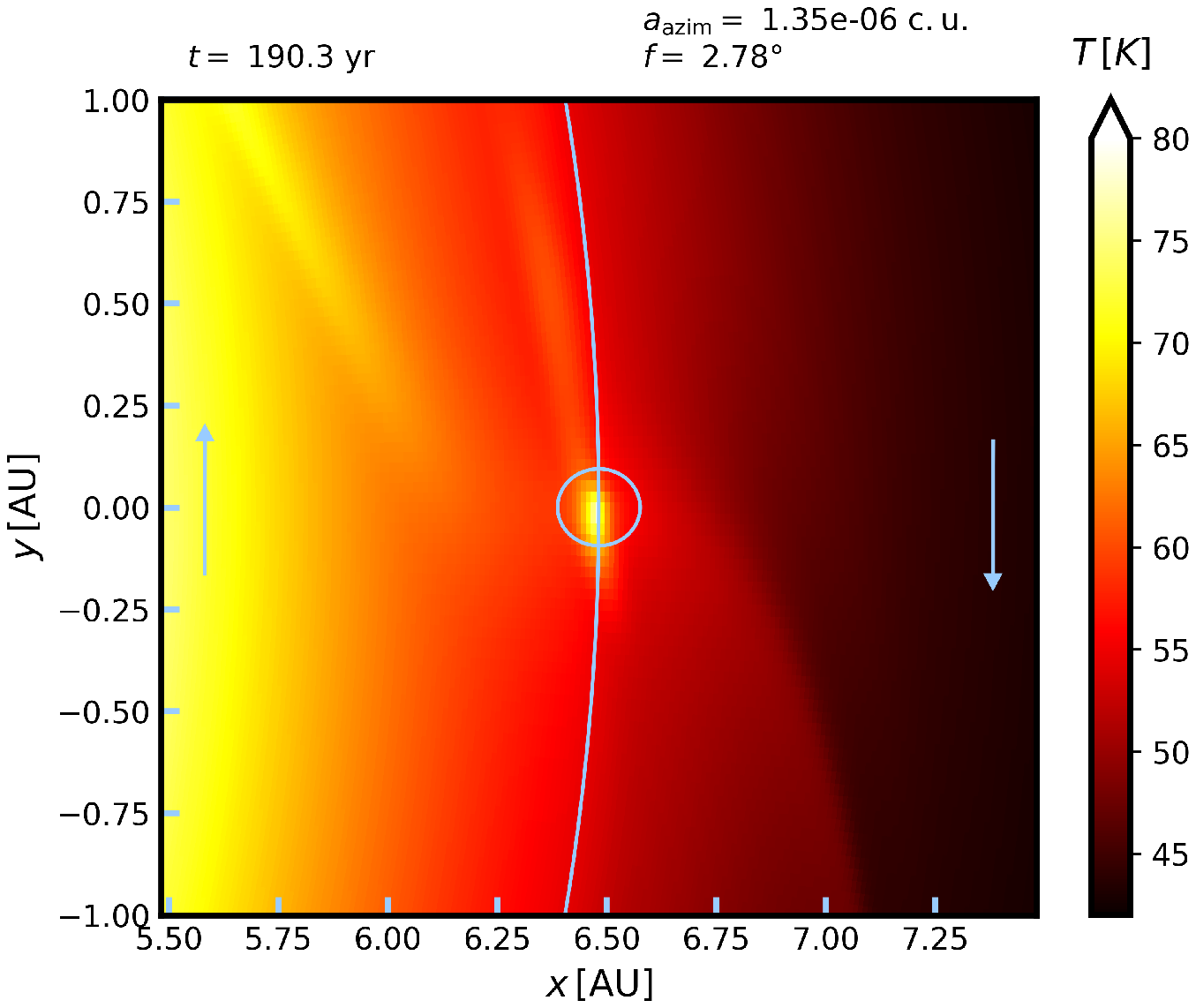} \\
    &
    \includegraphics[width=6.5cm]{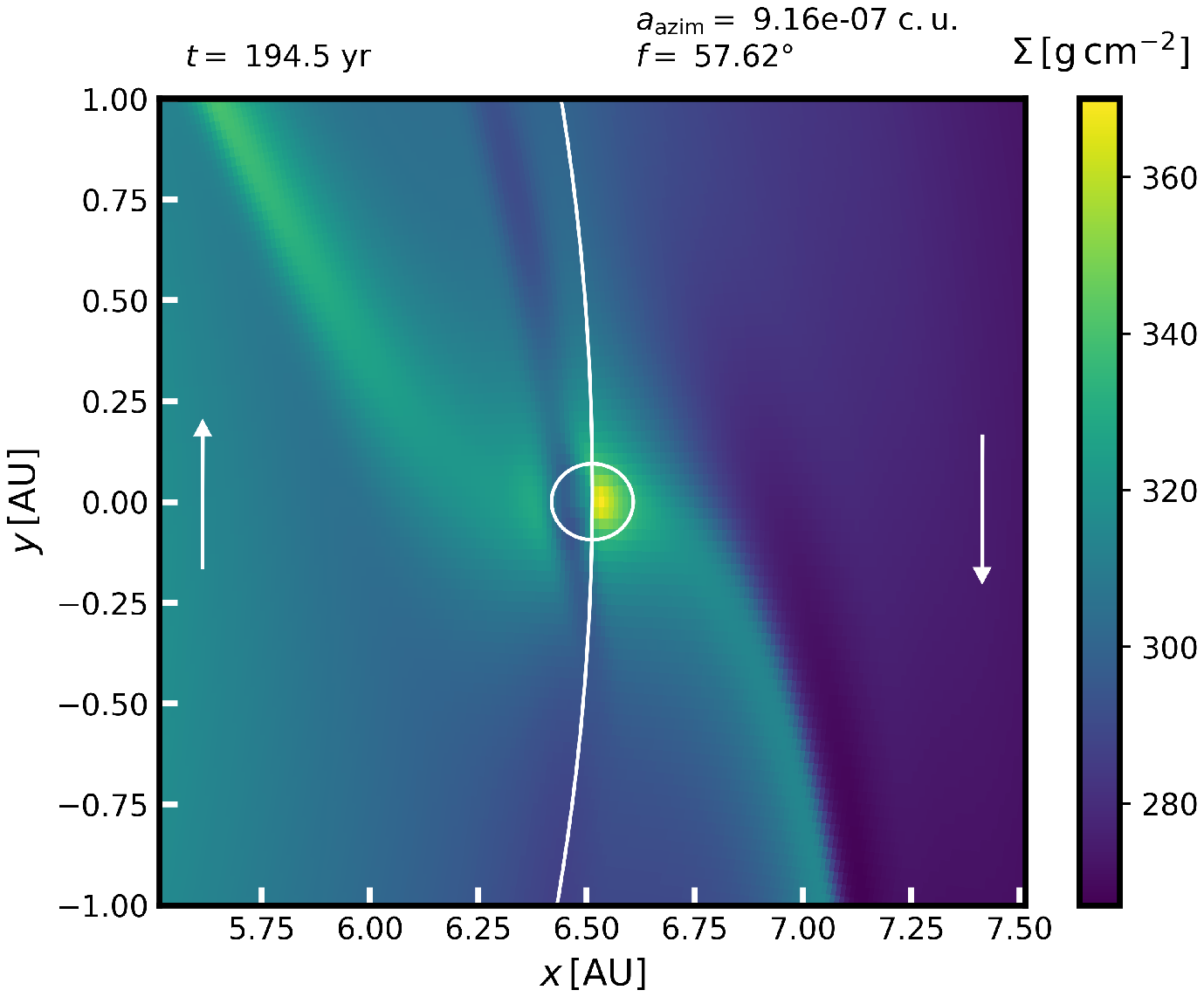} &
    \includegraphics[width=6.5cm]{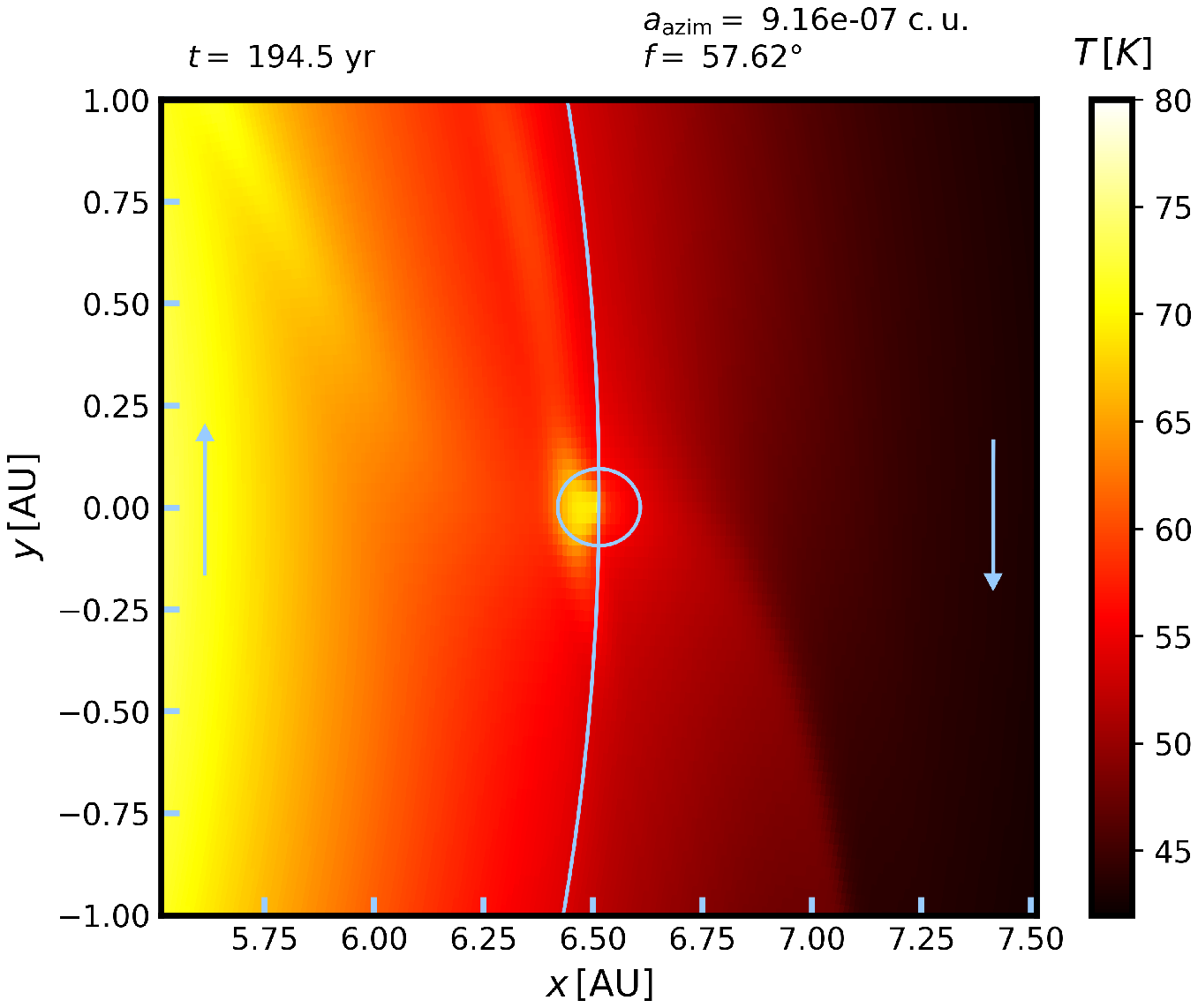} \\
    \begin{sideways} \hspace{2.0cm}apoastron \end{sideways} &
    \includegraphics[width=6.5cm]{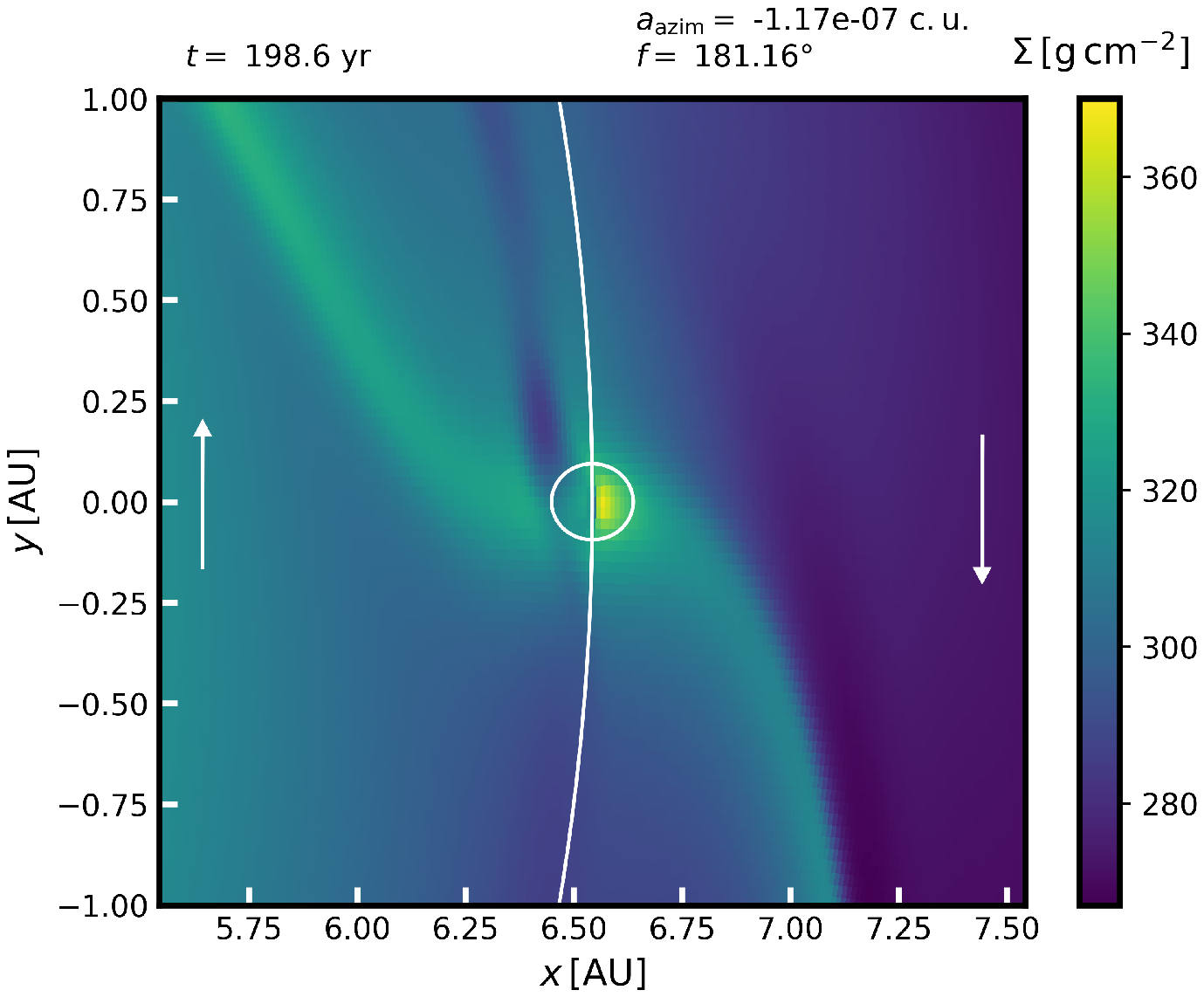} &
    \includegraphics[width=6.5cm]{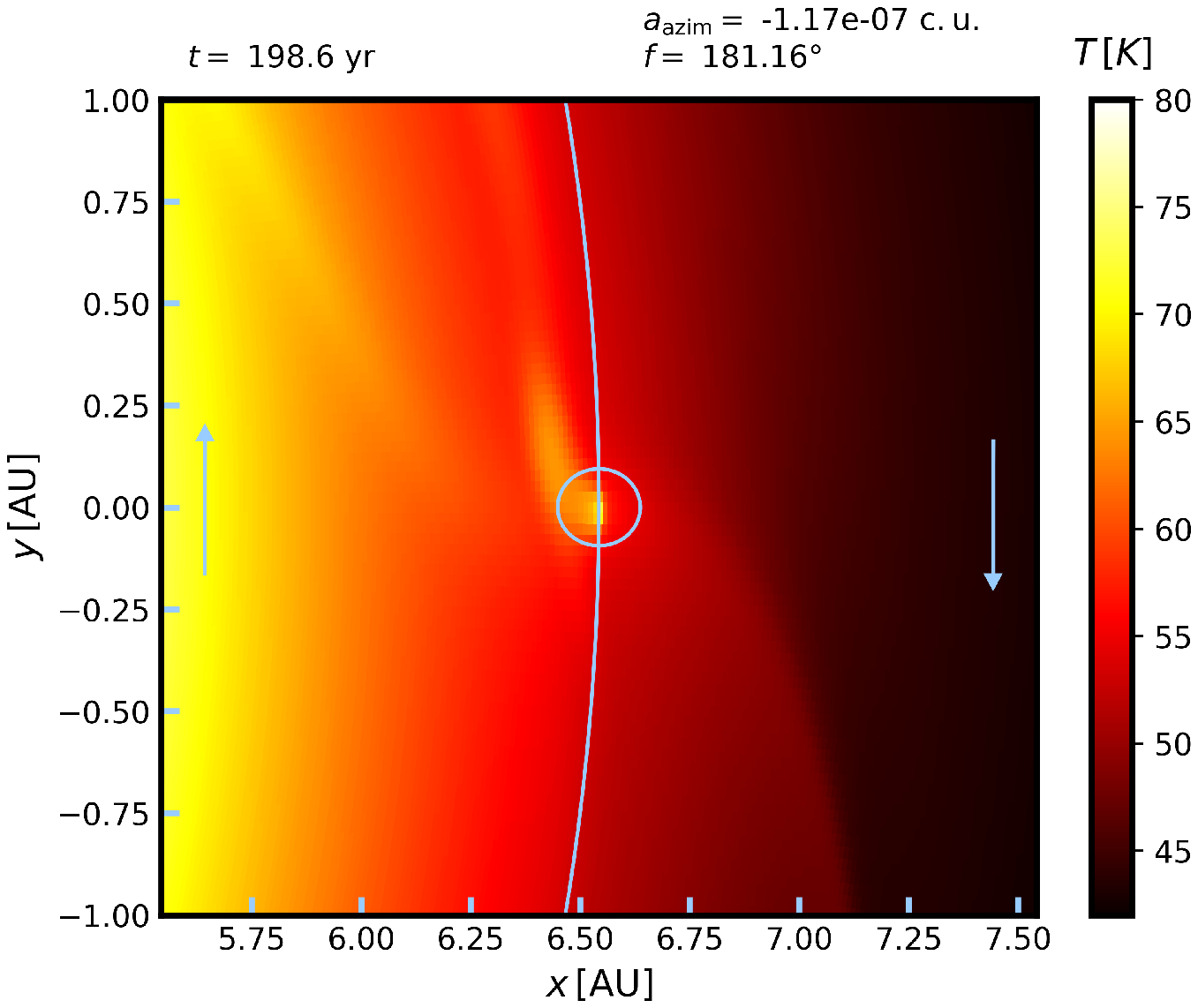} \\
    &
    \includegraphics[width=6.5cm]{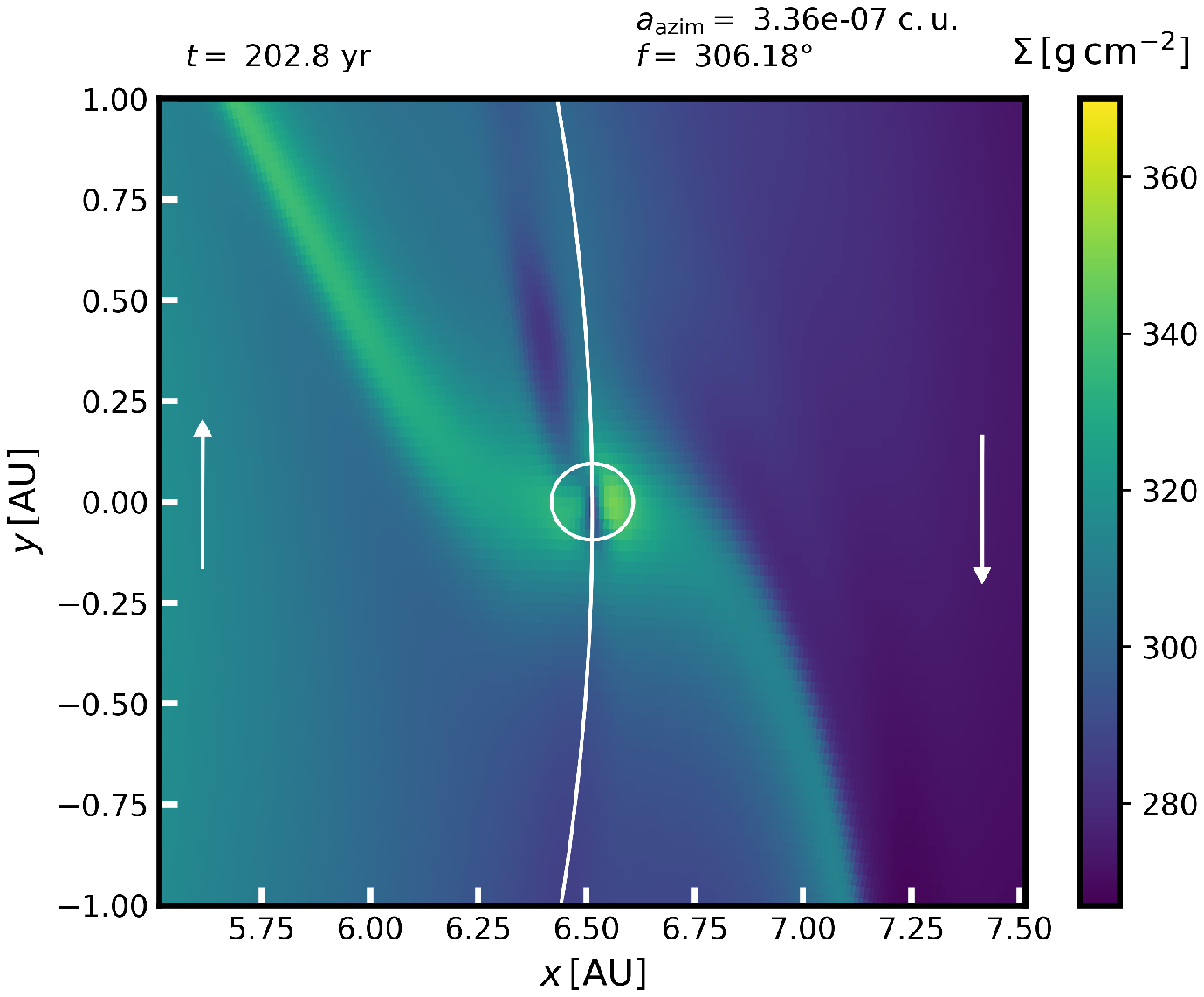} &
    \includegraphics[width=6.5cm]{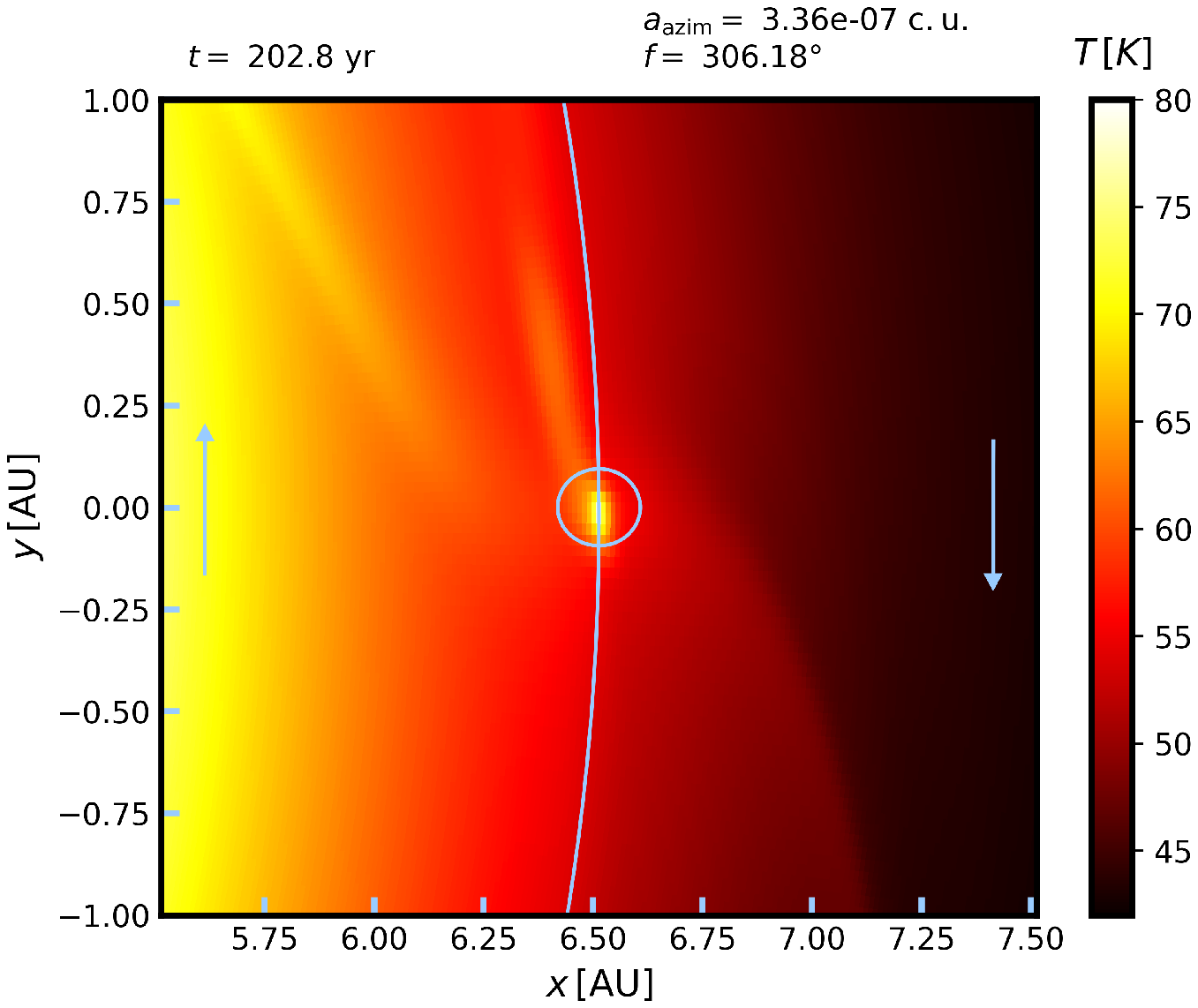} \\
  \end{tabular}
  \caption{Evolution of the gas surface density $\Sigma$ (left column) and
  temperature $T$ (right column) during one orbital period, recorded 
  within the onset phase of the eccentricity growth. Individual snapshots
  are labeled with the respective simulation time $t$, embryo's true anomaly $f$
  and azimuthal acceleration imposed by the disk, labeled here $a_{\mathrm{azim}}$.
  The figures are transformed to the corotating frame centered on the embryo.
  The Hill sphere and embryo's osculating orbit are plotted and we also indicate
  general directions of the gas flow with respect to the embryo by arrows.
  The orbital direction of the embryo is directed counterclockwise
  and the protostar is located at $(x=0,y=0)$.
  The \emph{top} row depicts the situation in the periastron, while the \emph{third} row
  corresponds to the apoastron. The \emph{second} row is recorded approximately
  halfway from periastron to apoastron, and vice versa for the \emph{bottom} row.
  }
  \label{fig:hot_trail_onset}
\end{figure*}

\begin{figure*}[!hpt]
  \centering
  \begin{tabular}{ccc}
    \begin{sideways} \hspace{2.0cm}periastron \end{sideways} &
    \includegraphics[width=6.5cm]{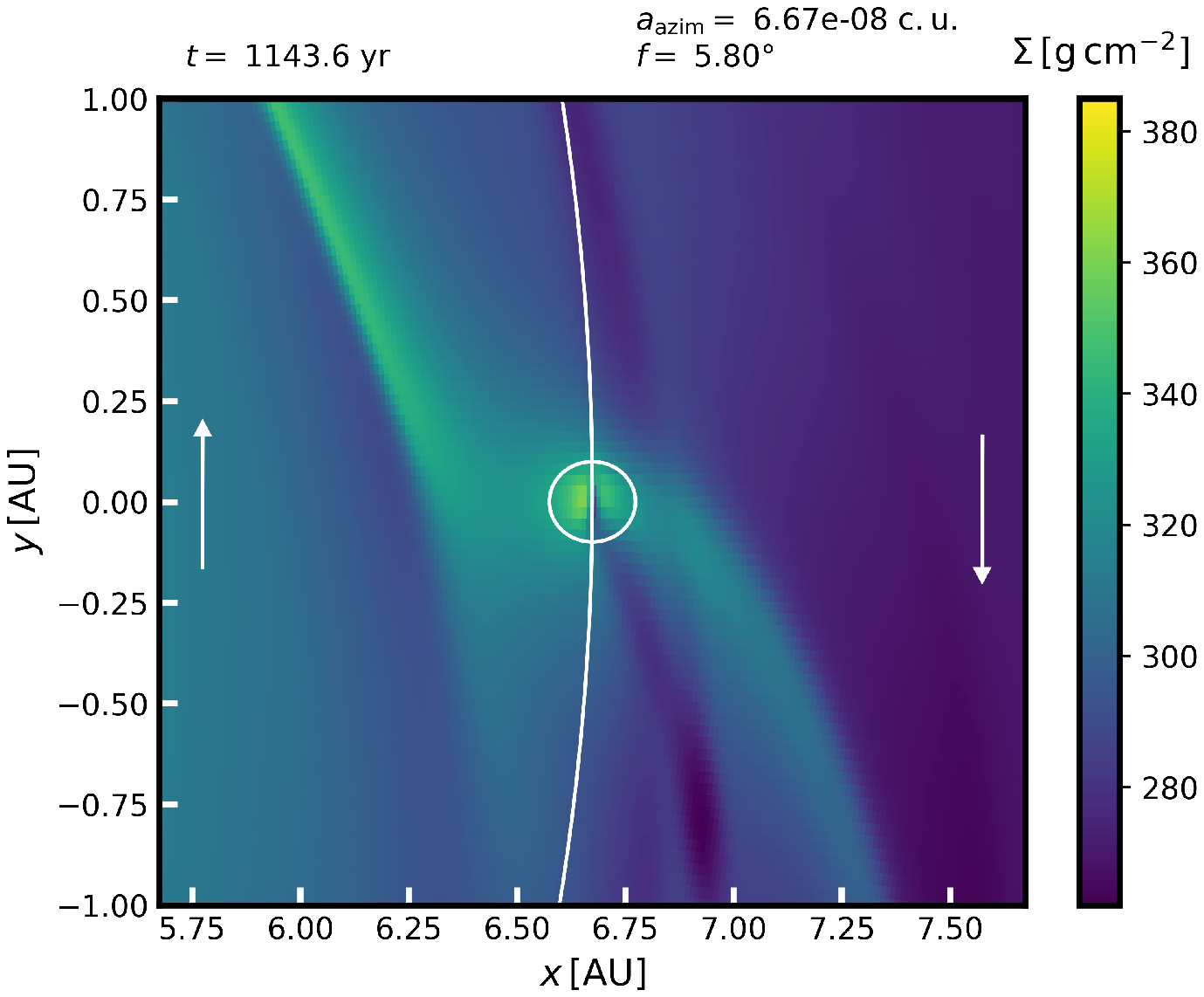} &
    \includegraphics[width=6.5cm]{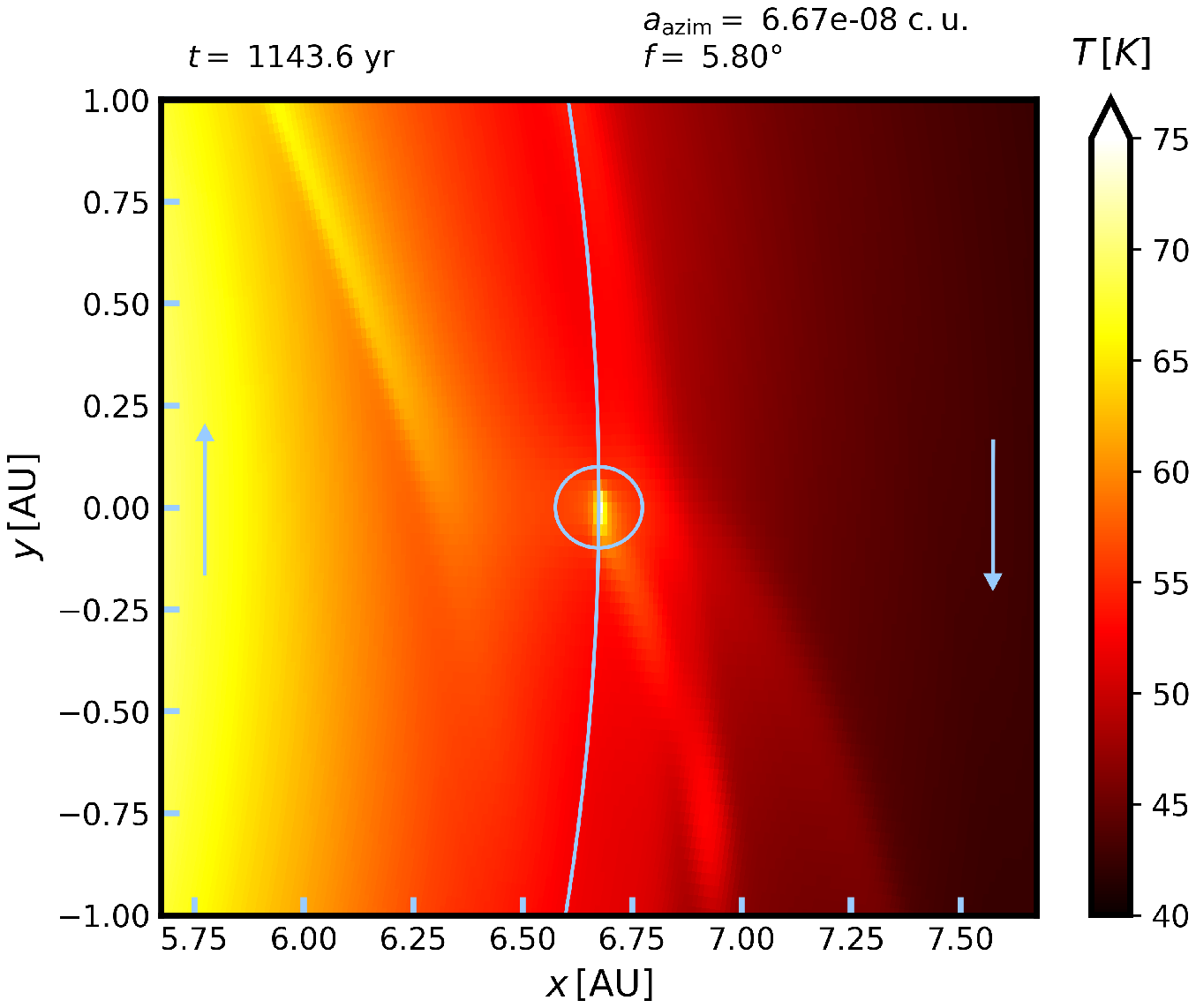} \\
    &
    \includegraphics[width=6.5cm]{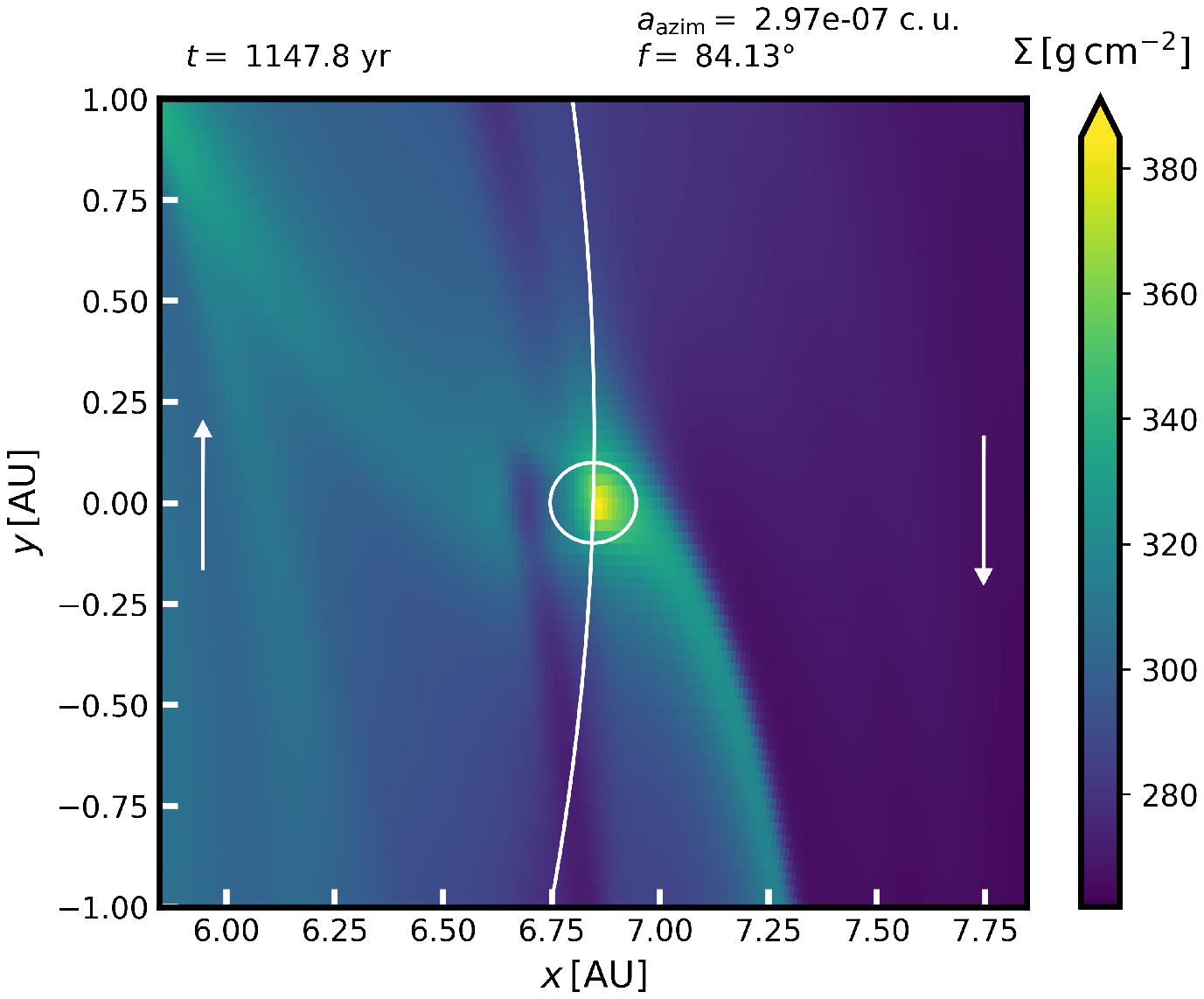} &
    \includegraphics[width=6.5cm]{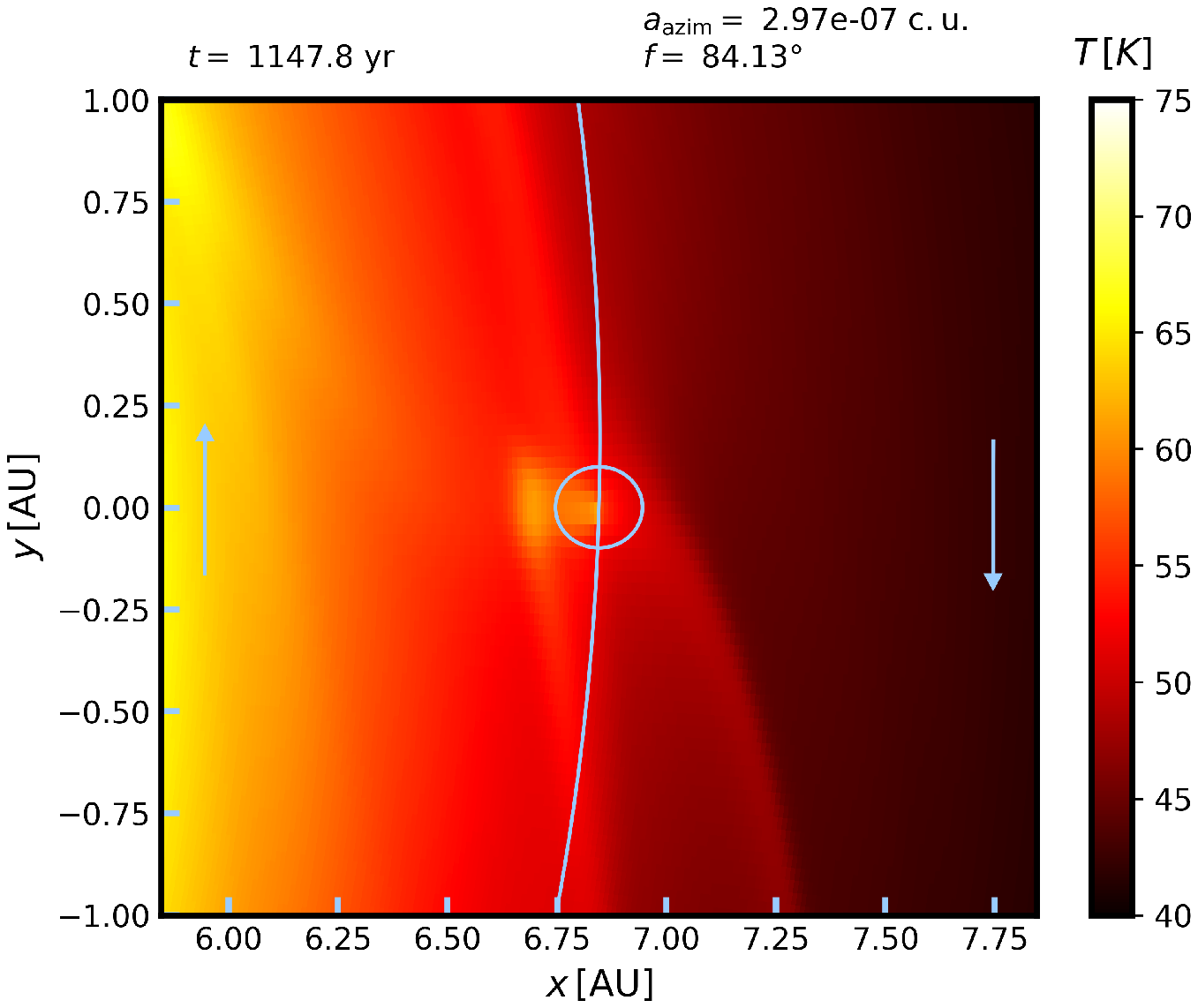} \\
    \begin{sideways} \hspace{2.0cm}apoastron \end{sideways} &
    \includegraphics[width=6.5cm]{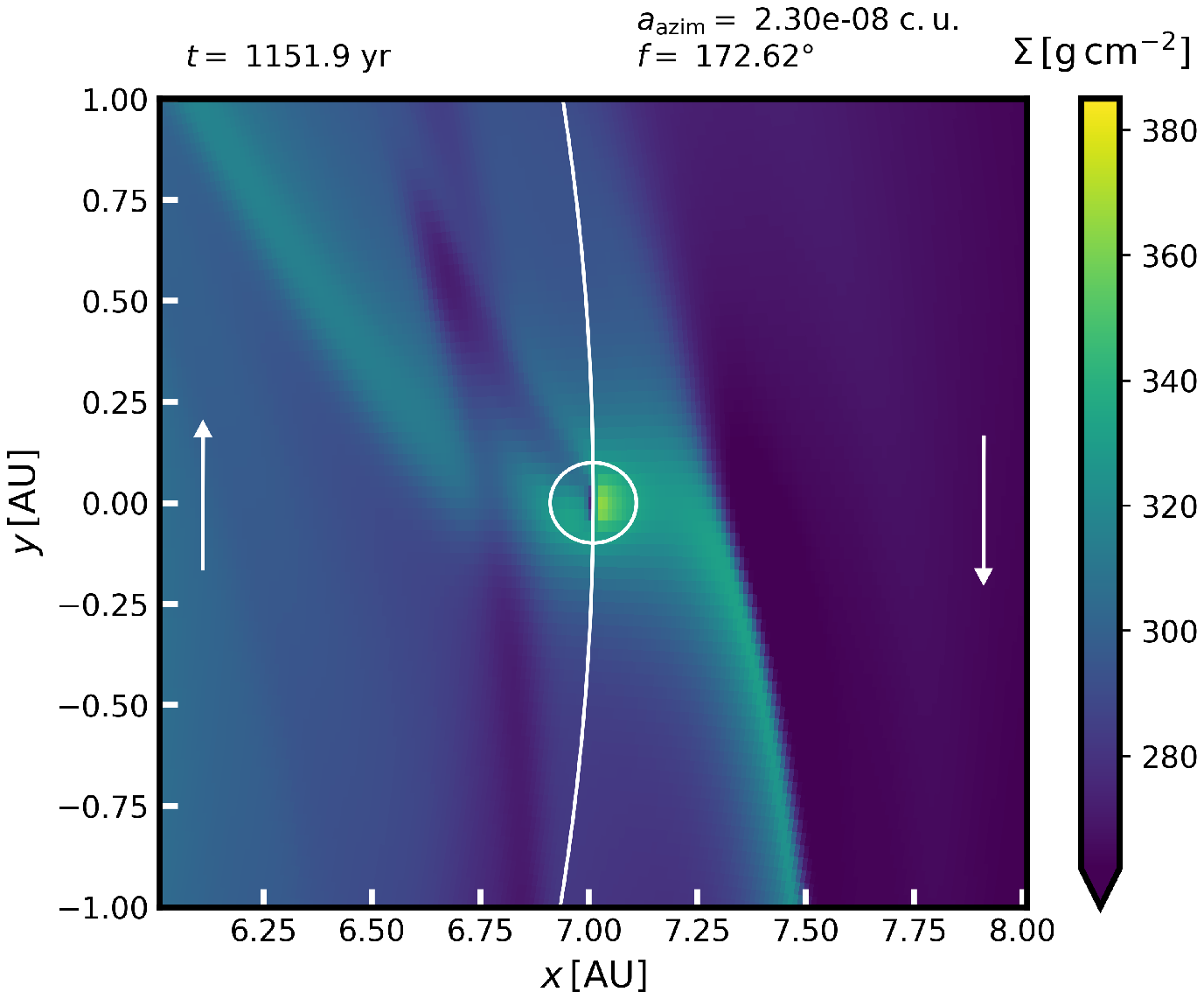} &
    \includegraphics[width=6.5cm]{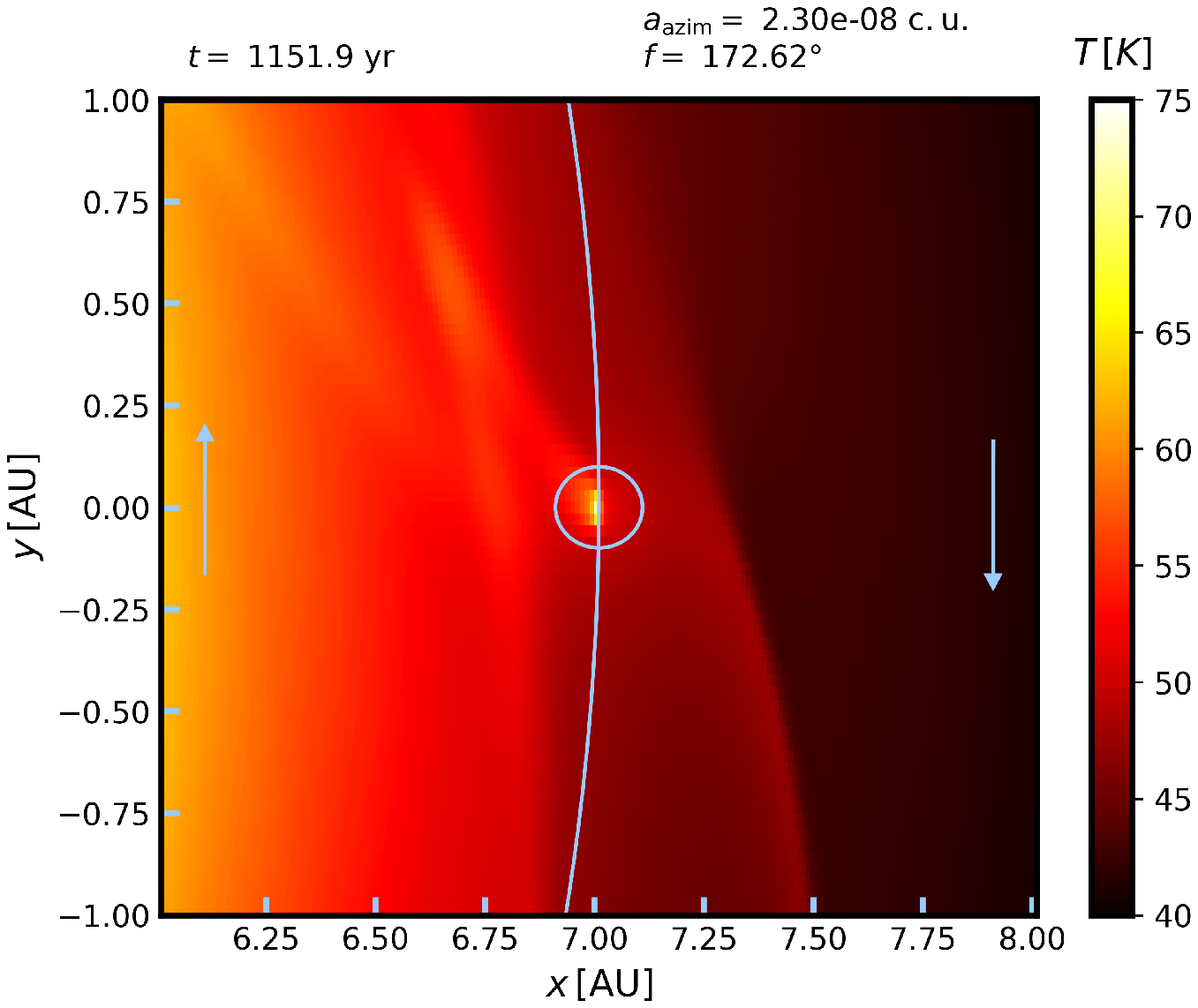} \\
    &
    \includegraphics[width=6.5cm]{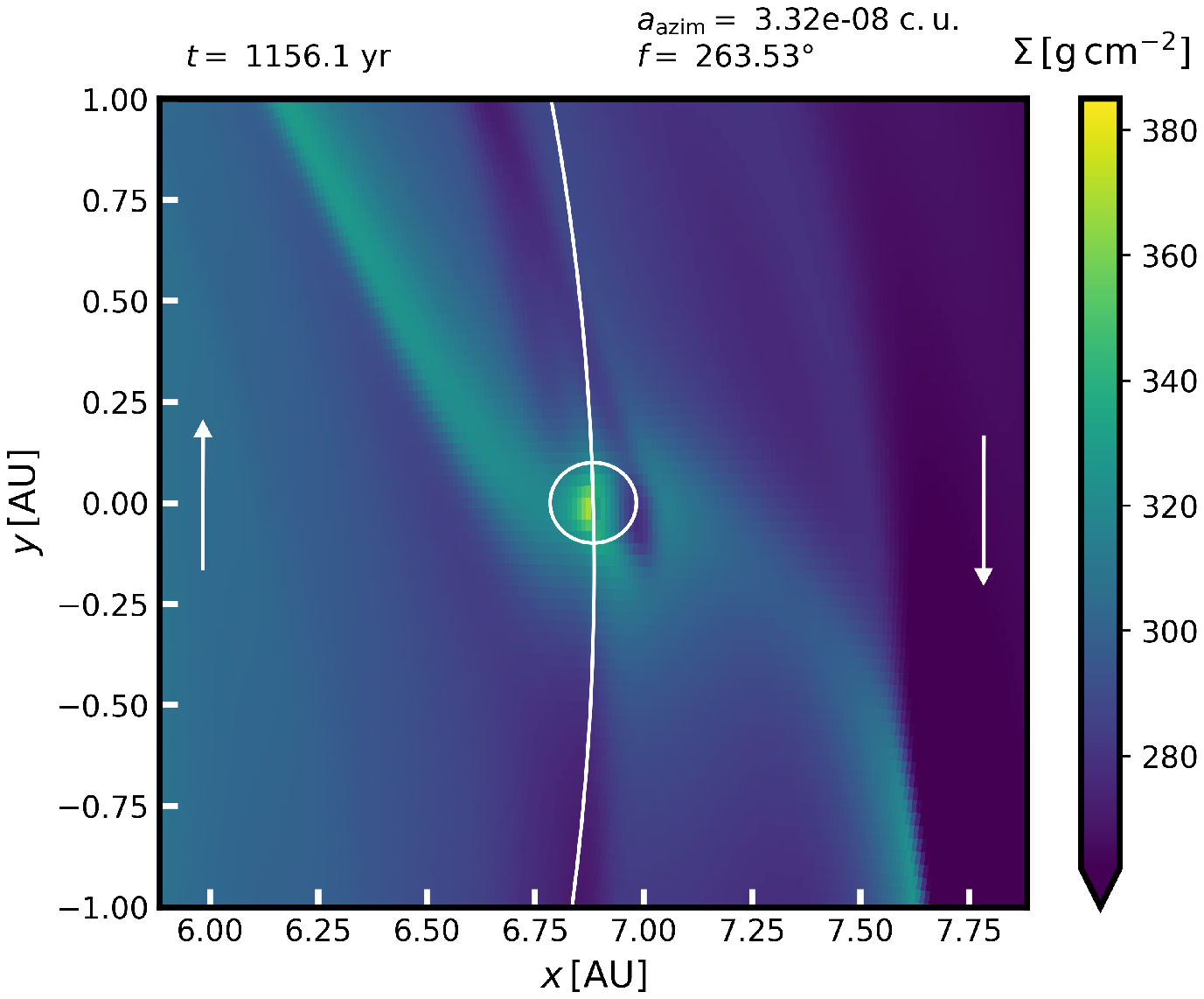} &
    \includegraphics[width=6.5cm]{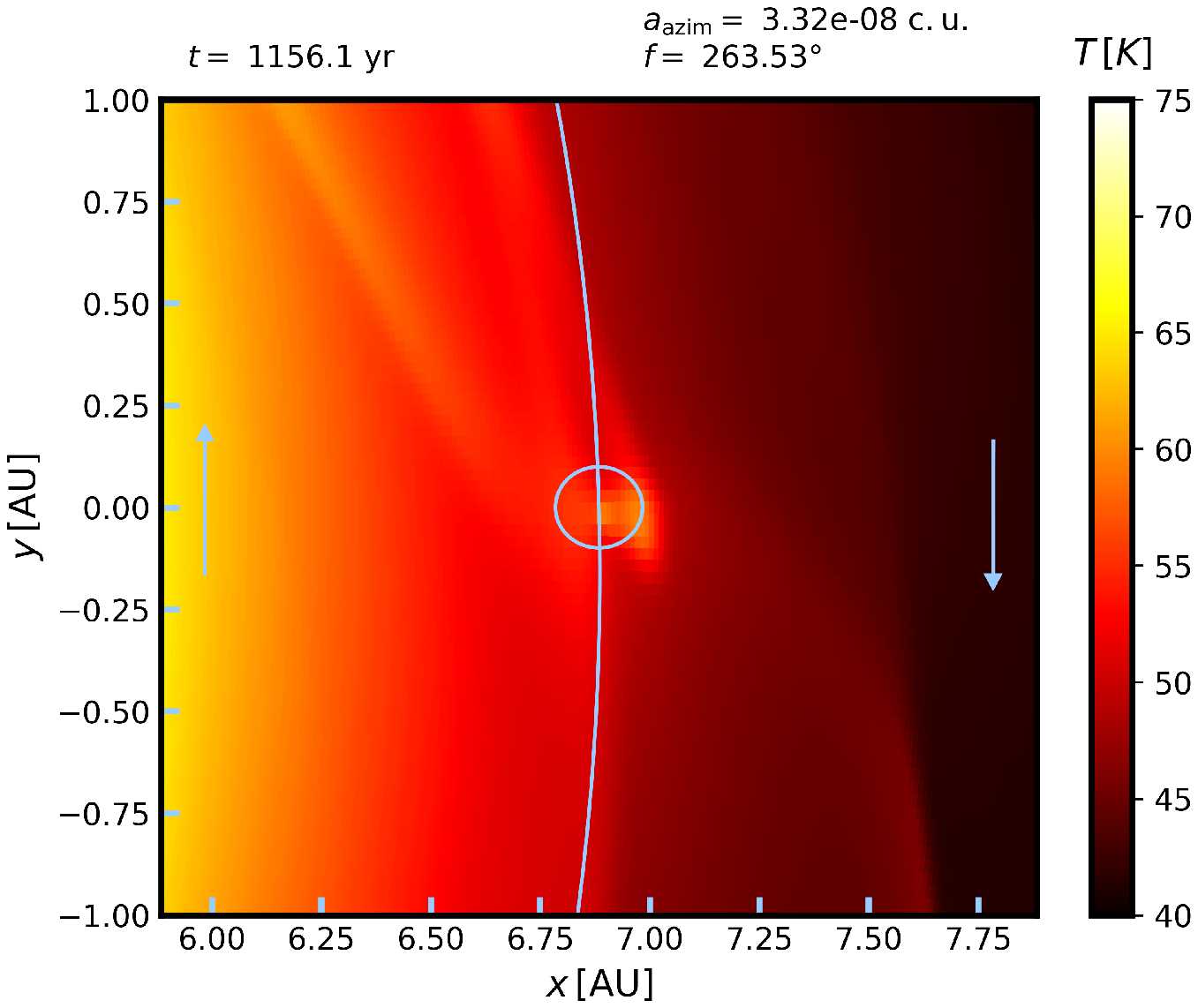} \\
  \end{tabular}
  \caption{Same as Fig.~\ref{fig:hot_trail_onset} but the hydrodynamic
  quantities are recorded within the saturation phase of the eccentricity
  excitation.}
  \label{fig:hot_trail_satur}
\end{figure*}

In the following, we explain the eccentricity
excitation from the hydrodynamic point of view.
For this purpose, we present series of figures
capturing the gas density $\Sigma$ and temperature $T$ 
distribution in the embryo's vicinity, corresponding
to the onset phase (Fig.~\ref{fig:hot_trail_onset})
and the saturation phase (Fig.~\ref{fig:hot_trail_satur}).

Let us first recall the advection-diffusion
problem which causes the standard mode of the heating
torque on \emph{fixed circular} orbits according
to \cite{Benitez-Llambay_etal_2015Natur.520...63B}.
The embryo heats the gas near its position and the gas becomes
overheated and therefore underdense\footnote{We remind the reader that
our model can only produce an underdensity in terms of the surface
density $\Sigma$.}, in order to maintain the
pressure balance with the surroundings. The heated gas is being
advected by the nearby flows and in the meantime, its internal energy
changes by the radiative diffusion.
For a circular orbit of the embryo,
the gas from the outer part of the disk approaches the embryo head-on, it
is being heated and forms an underdense lobe behind the embryo.
The gas from the inner disk which is moving faster than
the embryo approaches from behind, forming an underdense
lobe in front of the embryo.
Because the gas velocity is sub-Keplerian, the corotation
between the embryo and the gas is shifted slightly inwards,
thus there is a prevalence of gas which approaches as the headwind
and the underdense lobe behind the embryo is dominant.

For an embryo which is allowed to move freely in the disk, 
we already saw that the orbit is never perfectly circular.
It periodically gains a small eccentricity ($\sim10^{-3}$)
due to the $G_{r}$ forcing (Fig.~\ref{fig:gauss_at}).
Thus the embryo makes small radial excursions in the disk (see the
changing range of the $x$-axis in Fig.~\ref{fig:hot_trail_onset})
as it performs a small epicyclic motion. The heat source located at
the embryo's position trails this epicyclic motion.
In the temperature map, the epicyclic motion manifests itself as
a `hot trail', attached to the temperature maximum, which wobbles
around between the individual snapshots.
We thus call this new phenomenon {\bf the hot trail effect}.

Looking at the $\Sigma$ profiles in Fig.~\ref{fig:hot_trail_onset},
we see that in the periastron, there are again two underdense lobes,
similar to the circular case. The deep lobe attached behind the embryo 
represents the dominant paucity of material.
The less pronounced and more stretched lobe
in front of the embryo is rather a leftover
of the dominant lobe which was displaced by the epicyclic motion
during the previous orbit. This is proved by the
sequence of $\Sigma$ profiles -- as the embryo travels
towards the apoastron, its radial distance increases thus the
dominant lobe is left at $r<r_{\mathrm{em}}$
and subsequently moves ahead of the embryo due to the transport
by the interior flows which move faster than the embryo.
In the meantime, the less dominant leftover lobe is being lost
by the Keplerian shear and diffusive effects.

Near the apoastron, the embryo has the lowest orbital velocity.
  If we, for example, consider the eccentricity $e=0.003$ (typical value
  due to the $G_{r}$ forcing),
  the orbital velocity in the apoastron with
  respect to the Keplerian velocity is $v_{\mathrm{apo}}=(1-0.003)v_{\mathrm{K}}$.
  At the corresponding orbital distance $r\simeq6.5\,\mathrm{AU}$,
  the gas orbital velocity is $v_{\theta} = (1-0.0026)v_{\mathrm{K}}$
  (cf.~Fig.~\ref{fig:eta_stokes}). The headwind
  therefore significantly vanishes and no additional lobe can be formed
  behind the embryo. The embryo is left with the lobe formed
  at the periastron which has already been transported by the flows
  interior to the orbit.

The position of the dominant underdense lobe is the key factor which
determines the resulting
azimuthal acceleration $\mathcal{T}$ acting on the embryo. In the periastron,
there is a paucity of mass \emph{behind} the planet, so the acceleration
is in the orbital direction. In the apoastron, the lobe is located
ahead of the embryo, but it is also radially displaced ($r<r_{\mathrm{em}}$)
with respect to the embryo. As a consequence, $\mathcal{T}$ is negative
but its magnitude is much smaller compared to that at periastron,
where the underdense lobe is adjacent to the embryo.
This asymmetry between the periastron and apoastron
causes the eccentricity excitation.

During the growth phase (not shown in figures), the situation
  is similar to the onset phase. But as $e$ continuously grows,
  the lobe at the periastron becomes prolonged because the relative velocity
  of the embryo with respect to the gas increases. As a consequence,
  the azimuthal acceleration $\mathcal{T}$ measured in the periastron
  of the growth phase is larger compared to the onset phase.
  The relative velocities become large enough for the embryo 
  to start feeling tailwind near the apoastron, which delivers
  heat to the lobe positioned ahead of the embryo at that time.
  But because the gas is sub-Keplerian, the relative velocity
  is always larger in the periastron than in the apoastron thus
  the positive eccentricity pumping during the periastron passage
  still prevails and the runaway eccentricity growth continues.

The eccentricity cannot grow indefinitely, however,
but its excitation saturates at a certain level.
The hydrodynamic state at the saturation phase
is given in Fig.~\ref{fig:hot_trail_satur}
where we see that the hot trail spans a larger 
portion of the embryo's surroundings because the radial
excursion (the epicycle) of the embryo has already
increased significantly. As a consequence, the underdense structures
are more distant from the embryo and the Hill sphere
can refill with gas which is yet-to-be heated
and which blurs asymmetries in the embryo's vicinity,
responsible for the eccentricity excitation.
At the same time, the underdense structures
are strongly affected by the Keplerian shear because
their radial extension is considerable.

At the saturation phase, the eccentricity growth stops
right before exceeding the local value of the aspect
ratio $h\simeq0.036$. For $e\gtrsim h$, the relative motions
could lead to the reversal of normally negative Lindblad torque
\citep{Papaloizou_Larwood_2000MNRAS.315..823P}.
\cite{Cresswell_Nelson_2006A&A...450..833C}
found that the Lindblad torque transition for $e\gtrsim h$ 
is accompanied by very efficient eccentricity damping leading
to a strong energy loss which can outweigh the angular momentum gain.
This efficient damping is finally able to prevent the hot trail
from exciting the eccentricity even more.
But for lower $e$, the hot trail
effect dominates -- otherwise the eccentricity would not grow
in the first place.

\begin{figure}
\centering
\includegraphics[width=8.8cm]{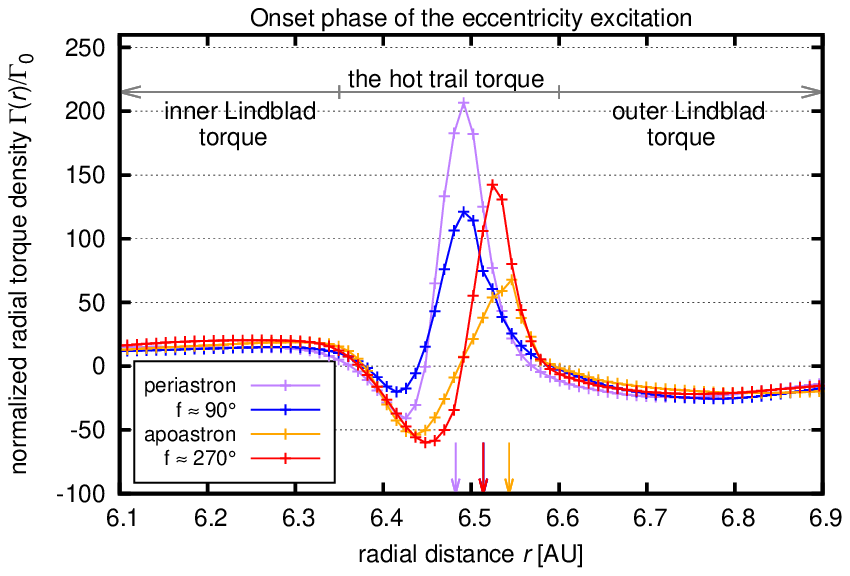}
\includegraphics[width=8.8cm]{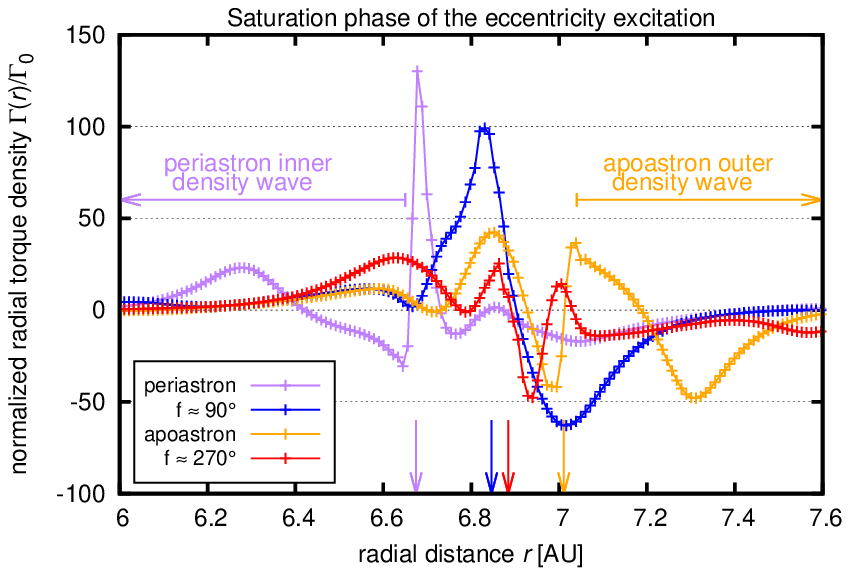}
\caption{
  The radial torque density $\Gamma(r)$ acting
  on the embryo during the onset (top) and saturation (bottom) phases,  
  normalized to $\Gamma_{0}$. The individual curves
  represent measurements in the periastron (purple), apoastron
  (orange) and in-between. The vertical arrows indicate the instantaneous
  radial distance of the embryo corresponding to the individual
  curves. The horizontal arrows and labels approximately distinguish
  some of the important torque contributions discussed in the text.
  To avoid misinterpretation, we remark that the hot trail torque
  is acting in the bottom panel as well but it spans different
  radial extent for each curve and thus cannot be marked unambiguously.
}
\label{fig:hot_trail_torque}
\end{figure}

\subsection{Torque distribution}
\label{sec:hot_trail_torque}
The periodic changes of $\Sigma$ and of the related
$\mathcal{T}$ are also reflected
in the variations of the torque $\Gamma_{\mathrm{tot}}$
felt by the embryo during its orbit.
Fig.~\ref{fig:hot_trail_torque} shows the normalized radial torque distribution
$\Gamma(r)/\Gamma_{0}$ which relates to the total torque $\Gamma_{\mathrm{tot}}$ as
\begin{equation}
  \Gamma_{\mathrm{tot}}=\int\limits_{r_{\mathrm{min}}}^{r_{\mathrm{max}}}\Gamma(r)\mathrm{d}r \, .
  \label{eq:radial_torque}
\end{equation}
Fig.~\ref{fig:hot_trail_torque} generally demonstrates
which parts of the disk are responsible for positive and 
negative torques and how does the magnitude of these
torques change with radial separation from the embryo.

During the onset phase (Fig.~\ref{fig:hot_trail_torque}, top panel),
the shape of $\Gamma(r)/\Gamma_{0}$ is
similar to the calculations of \cite{Benitez-Llambay_etal_2015Natur.520...63B}
(cf. their Fig.~1). In the periastron, it exhibits a negative
peak at $r<r_{\mathrm{em}}$ which is smaller than
a positive peak at $r>r_{\mathrm{em}}$.
As the embryo travels along its orbit,
the difference compared to \cite{Benitez-Llambay_etal_2015Natur.520...63B}
is in the position of the profile with respect to the embryo
(which is labeled by arrows)
and in the asymmetry between the positive and negative peak.
The asymmetry is pronounced in the periastron and disappears
in the apoastron, in accordance with our previous findings.

During the saturation phase (Fig.~\ref{fig:hot_trail_torque}, bottom panel),
$\Gamma(r)/\Gamma_{0}$ becomes wavy and complex.
It corresponds to the hot trail strongly distorted by the Keplerian
shear, which is produced by a large epicycle.
Compared to the onset phase, the torque contribution arising from
the density waves is modified. Let us focus on the situation in periastron
  first. Looking at Fig.~\ref{fig:hot_trail_satur}, we notice that
  the gas surface density exhibits a pronounced inner density wave.
  The underdense structure related to the hot trail effect
  is located at $r>r_{\mathrm{em}}$ thus the dominant contribution
  to $\Gamma(r)$ at $r<r_{\mathrm{em}}$ must be related 
  to the inner density wave.

  The contribution from the inner density wave is labeled in
  Fig.~\ref{fig:hot_trail_torque} (bottom panel).
  Although the inner Lindblad torque is purely positive
  for circular orbits, we can see that it has both positive
  and negative contributions for the eccentric orbit during
  the saturation phase. In the apoastron, the situation is similar
  (but the outer density wave is more pronounced). This implies
  that the orbit is indeed close to the state of the Lindblad
  torque reversal and proves our aforementioned argument
  about what phenomenon finally stops the eccentricity growth.

\section{Future improvements and observational signatures}
\label{sec:discussion}

\paragraph{Additional free parameters.}
Regarding the discussion in this paper, we essentially restricted
ourselves to switching pebble accretion and the accretion heating on and off,
in order to understand the basic physics of the hot trail effect
and to simplify the discussion. It is clear, however, that our
problem has a number of additional \emph{free} parameters. In particular:
the number of embryos (up to say $10^1$),
initial embryo masses (of the order of $10^0\,\mathrm{M_{E}}$),
initial spacing of embryos (multiples of $R_{\mathrm{mH}}$),
embryo positions in the disk and with respect to the zero-torque radius,
the radial pebble flux $\dot M_{\rm F}$,
gas surface density~$\Sigma_0$ and its slope,
viscosity~$\nu$ (or~$\alpha$), turbulent stirring of solids $\alpha_{\rm p}$,
or stellar luminosity $L_\star$, etc.
Even if we have only 2~values per parameter, the resulting number
of models is so high, we are unable to compute a full matrix.
Nevertheless, it is certainly possible to compute differences (derivatives)
with respect to the fiducial model and that is the actual work
postponed to the next paper.

\paragraph{Possible model improvements.}
We can outline a number of opportunities for the hydrodynamic model extensions,
e.g. full 3D treatment, implementation of gas accretion, deposition of pebbles in
various layers of protoatmospheres, gas self-gravity, stochastic forcing
by turbulent flows \citep{Pierens_etal_2013A&A...558A.105P},
independently evolved dust component as the main opacity constituent, etc. 

Moreover, as we demonstrated that the hot trail effect reduces
  the ability of the surrounding disk to damp the orbital
  eccentricity, it is also possible that the inclination damping
  is somehow modified if a full 3D disk is considered. In our 2D model,
  the inclination damping is provided by Eq.~(\ref{eq:force_damping})
  which is not self-consistent but based on a model
  which neglects the accretion heating \citep{Tanaka_Ward_2004ApJ...602..388T}.
  We also plan to refine this part of the model in the future.

\paragraph{Observational signatures.}
From the observational point of view,
the imprints of various migration histories and 
orbital excitations should be recognizable
in the observed exoplanetary systems, but they
can be successfully understood only when the effects described
in this paper are taken into account in future works
dealing with this issue.

There can possibly be observational signatures
of e.g. mergers or multiple embryos on
closely-packed orbits in the datasets of the campaigns
involved in the direct protoplanetary disk
imaging (e.g. by ALMA).
We have already started to investigate this possibility
and plan to publish the study in a separate paper.

In our case, most if not all observational circumstances should be
determined by 3D radiative transfer in the dust continuum.
The optical thickness for the typical \cite{Bell_Lin_1994ApJ...427..987B} opacity
$\kappa \simeq 10^{0}\,\mathrm{cm^{2}}\,\mathrm{g^{-1}}$ and the surface density
$\Sigma \simeq 10^{2}\,\mathrm{g}\,\mathrm{cm^{-2}}$ is
$\tau_{\mathrm{opt}} \simeq \kappa\Sigma \simeq 10^2 \gg 1$.
We thus definitely need a good enough description of the disk
atmosphere, far from the midplane.

In order to become observable,
it seems that protoplanets must open considerably large gaps
in the gas disk \citep{Rosotti_etal_2016MNRAS.459.2790R}.
Partially opened gaps are probably not observable because
these are still optically thick; the density contrast has to be at least $10^2$.
The close encounters between embryos in our simulations
lead to an asymmetry, but only present for a short time interval. 
As argued in \cite{Rosotti_etal_2016MNRAS.459.2790R},
the threshold mass for detection is about $12\,\mathrm{M_{E}}$ in sub-mm.
Moreover, for VLT/SPHERE or Gemini/GPI instruments, the protostar should be
more massive ($M_{\star} \simeq 2\,\mathrm{M}_{\sun}$) to become at least a Herbig Ae star,
because of current flux limitations.

\section{Conclusions}
\label{sec:conclusions}

In this paper, we studied the orbital
evolution of four $3\,\mathrm{M_{E}}$ embryos
embedded in a region of a protoplanetary
disk where the convergent migration is expected
to occur under the influence of the standard Type-I
torques.
In our simulations, however, we
considered that the embryos rapidly accrete
mass from the pebble disk (modeled
hydrodynamically). 
Three classes of simulations were performed:
Case~I as a reference scenario in which pebble
accretion is completely neglected, Case~II in
which pebble accretion leads to the mass growth
of embryos and Case~III in which embryos
also become heated by the deposition of pebbles.
We investigated the impact of the additional
processes on the migration and mutual interactions
of the embryos.
The simulations were performed using a new 
state-of-the-art and rather self-consistent
hydrodynamical model,
which we extensively described and verified.

We found that in both Cases~I and II, the system 
evolves through a sequence of resonant chains,
first of which is usually established around the zero-torque
radius.
As the embryos gain non-zero eccentricity (typically
ranging from $0.004$ to $0.01$) due to perturbations
from the mean-motion resonances, occasional close
encounters are possible, leading to mutual scattering
(sometimes accompanied by a swap of orbits) or embryo merging.

We reported that merging of embryos is more probable
in Case~II in which the mass growth by pebble accretion is accounted for.
The reason is that the resonant chain is destabilized
more often as the masses of embryos responsible 
for the resonant forcing (e.g. of eccentricities) evolve. 
Additional forcing is provided as the streamline
topology around the embryos changes with the evolving
masses, thus imposing a slightly different disk torque.

In Case~III, the positive heating torque
changes the expected migration rates. As a result,
the embryos somewhat ignore the zero-torque radius
and are driven into mutual interactions preferentially
in the outer part of the disk, rather than being
packed in a resonant chain around the zone of convergence.

Close encounters occur frequently in Case~III
and cover a longer period of the evolution.
We realized that the encounters are facilitated by an eccentricity increase 
($e\simeq h$, typically ranging from $0.02$ to $0.04$)
\emph{prior to} resonant perturbations
by means of a new `hot trail' effect.
The effect is due to variable
gravitational acceleration arising from
the gas in the vicinity of each embryo, which is 
periodically modified by formation and advection
of an overheated and thus underdense lobe
trailing the epicyclic motion of the embryo.
The effect was independently reported
by \cite{Eklund_Masset_2017arXiv170401931E}
\citep[see also][]{Masset_VelascoRomero_2017MNRAS.465.3175M}
while our research was ongoing \citep{Chrenko_Broz_2016DPS....4831803C}.
The hot trail effect reduces the ability of the surrounding
disk to damp the eccentricities and circularize the orbits.
Despite that more encounters pose more opportunities
for merging, we actually found that merging is 
less frequent compared to Case~II
probably because of larger encounter velocities on the eccentric orbits.

The eccentricity excitation by the hot trail effect
stalls when $e\simeq h$ because
the Lindblad torque acting on eccentric
orbits is modified and can actually operate in a mode close to its reversal
(from negative to positive)
\citep{Papaloizou_Larwood_2000MNRAS.315..823P,Cresswell_etal_2007A&A...473..329C,Bitsch_Kley_2010A&A...523A..30B}.
Because the transition to the reversed Lindblad torque
would require the embryo to cross the orbital
resonances at which it excites the density waves,
strong eccentricity damping occurs and the eccentricity
growth saturates. Nevertheless, the eccentricity does
not decrease and it is maintained by the hot trail effect.
Note that many $N$-body
models \citep[e.g.][and many others]{Sandor_etal_2011ApJ...728L...9S,Izidoro_etal_2015ApJ...800L..22I,Coleman_Nelson_2016MNRAS.457.2480C}
usually employ a strong eccentricity damping prescriptions
\citep[e.g.][]{Cresswell_Nelson_2006A&A...450..833C,Cresswell_Nelson_2008A&A...482..677C}
derived from hydrodynamic models which neglected the
accretion heating. We suggest that these analytic
damping rates should be carefully refined for future
applications because they could be inaccurate
in cases when the protoplanets undergo any
kind of strong accretion.

Orbital excitation
of embryos heated by pebble accretion prevents formation of
a global resonant chain, except for short transient
periods. An interesting overlap of this result can be found
with recent developments in the analytical theory.
For example, \cite{Batygin_2015MNRAS.451.2589B}
used the Hamiltonian formalism
to study the probability of the resonant capture for
migrating low mass planets and compared his predictions
with the occurrence of the first-order mean-motion
resonances in exoplanetary systems.
He found that the resonant capture probability
is greatly diminished (and thus the observed non-resonant systems
can be explained) if a pre-encounter orbital excitation
$e\gtrsim 0.02$ is considered. Our model
thus provides a natural way of exciting the eccentricity
enough to prevent resonant locking
and may have important implications for explaining
the structure of exoplanetary systems.

Mergers large enough to possibly become giant planet
cores with masses $\simeq13\,\mathrm{M_{E}}$ were found
in both Cases~II and III.
We emphasize that merging caused by fast
migration and accretion in convergence zones breaks
the otherwise oligarchic nature of the embryo growth
by pebble accretion.

We conclude that orbital instabilities,
eccentricity excitations and (possibly) mergers
naturally accompany evolution of pebble-accreting
embryos and may have an important impact
on shaping the final architecture of any planetary system.
This is a major result compared to previous
models which neglected self-consistent hydrodynamics, accretion or heating.
But in order to find general implications, a larger
statistical sample of simulations is required
because we expect a strong dependence on the initial
conditions (possibly on the initial number and masses
of embryos, their position within the disk,
accretion rate related to the pebble mass
flux and heating efficiency influenced by the opacity).

\begin{acknowledgements}

We thank Alessandro Morbidelli,
Steven N.~Shore and David Nesvorn{\'y} for helpful
discussions during this project.
We are very grateful to Bertram Bitsch
who kindly provided his code for calculating
the migration maps.
We also thank an anonymous referee for valuable
comments.
The work of OC and MB has been supported
by Charles University in Prague (project GA~UK no.~128216; project SVV-260441).
The work of MB was supported by the Grant Agency of
the Czech Republic (grant no.~13-01308S).
Access to computing and storage facilities owned by parties and projects
contributing to the National Grid Infrastructure MetaCentrum,
provided under the programme `Projects of Large Research, Development,
and Innovations Infrastructures' (CESNET LM2015042), is greatly appreciated.

\end{acknowledgements}

\bibliographystyle{aa}
\bibliography{references}

\begin{appendix}

\section{Numerical scheme of the energy equation solver}
\label{sec:implicit_scheme}

This appendix summarizes our approach to
modelling non-isothermal disks, which undergo
heating and cooling, within the framework
of the original 2D \textsc{fargo} code. 
Here we elaborate the numerical update
of the internal energy due to the considered
source terms (Sec.~\ref{sec:model}).
Following  the
formalism of \cite{Stone_Norman_1992ApJS...80..753S},
the advection term is treated separately in the transport step.

Starting with the energy equation (Eq.~\ref{eq:energy}),
our aim is to derive an implicit numerical
scheme. 
The reason for this is to avoid a possible
time step restriction which could arise
in the case of an explicit solution
due to the Courant-Friedrichs-Lewy
condition related to the radiative transport.
As we discussed in Sec.~\ref{sec:model},
we assume that the specific internal energy
is entirely thermal
thus we can write $E=\Sigma c_{V}T$,
where $c_{V}$ is the specific heat at constant
volume. Within the one-temperature approach,
the radiation field with the energy
density $4\sigma_{\mathrm{R}}T^{4}/c$
only contributes to the energy transport
via the radiative diffusion term.
In order to obtain the implicit scheme,
we rewrite Eq.~(\ref{eq:energy})
for the temperature only and we drop
the advection term which is treated separately
\begin{equation}
  \frac{ \partial \Sigma c_{V}T}{ \partial t} = - \Sigma\frac{R}{\mu}T\nabla\cdot\vec{v}
  + Q_{\mathrm{visc}} + Q_{\mathrm{irr}} + Q_{\mathrm{acc}} - Q_{\mathrm{vert}} + 2H\nabla\cdot D\nabla T \, ,
\label{eq:ener_01}
\end{equation}
where $D=16\lambda\sigma_{\mathrm{R}}T^{3}/\left(\rho_{0}\kappa\right)$
is the diffusion coefficient.

For simplicity, let us first discretize
the diffusion term and return to the
other source terms later on.
Because \textsc{fargo} is designed as a staggered-mesh code,
all scalar quantities are cell-centered
whereas components of vector quantities
are face-centered. 
In the following, the differential
operators are written in polar coordinates,
integers $i$ and $j$ represent the indices
of radial zones and azimuthal sectors,
respectively:
\begin{equation}
  \begin{split}
  & \frac{T_{i,j}^{n+1} - T_{i,j}^{n}}{\Delta t} =
  \left(\frac{2H}{\Sigma c_{V}}\right)_{i,j}\frac{1}{r_{i}^{\mathrm{c}}} \\
      & \times \Biggl[
        \frac{1}{\left(\Delta r\right)_{i}^{\mathrm{f}}}
             \left(
	     r_{i+1}^{\mathrm{f}}\bar{D}^{r}_{i+1,j}\frac{T_{i+1,j}-T_{i,j}}{\left(\Delta r\right)_{i+1}^{\mathrm{c}}} - 
	     r_{i}^{\mathrm{f}}\bar{D}^{r}_{i,j}\frac{T_{i,j}-T_{i-1,j}}{\left(\Delta r\right)_{i}^{\mathrm{c}}}
	     \right) \\
      & + \frac{1}{\Delta \theta}
             \left(
	     \bar{D}_{i,j+1}^{\theta}\frac{T_{i,j+1}-T_{i,j}}{r_{i}^{\mathrm{c}}\Delta \theta} - 
	     \bar{D}_{i,j}^{\theta}\frac{T_{i,j}-T_{i,j-1}}{r_{i}^{\mathrm{c}}\Delta \theta}
             \right)
        \Biggl] \, .
  \end{split}
  \label{eq:discret_diffpart}
\end{equation}
Here $r_{i}^{\mathrm{c}}$ denotes the radial coordinate
of a cell centre, $r_{i}^{\mathrm{f}}$
is the radius of an inner radial cell interface
and $\Delta \theta$ denotes the angular width of
sectors which is identical for all cells.
The additional quantities naturally occur because
of the staggered-grid formalism:
\begin{equation}
  \bar{D}_{i,j}^{r} = \frac{1}{2}\left(D_{i,j} + D_{i-1,j}\right) \, ,
  \label{eq:diff_coef_r}
\end{equation}
\begin{equation}
  \bar{D}_{i,j}^{\theta} = \frac{1}{2}\left(D_{i,j} + D_{i,j-1}\right) \, ,
  \label{eq:diff_coef_t}
\end{equation}
\begin{equation}
  \left(\Delta r\right)^{\mathrm{c}}_{i} = r^{\mathrm{c}}_{i} - r^{\mathrm{c}}_{i-1} \, ,
  \label{eq:deltar_c}
\end{equation}
\begin{equation}
  \left(\Delta r\right)^{\mathrm{f}}_{i} = r^{\mathrm{f}}_{i+1} - r^{\mathrm{f}}_{i} \, .
  \label{eq:deltar_f}
\end{equation}
Obviously, $\left(\Delta r\right)^{\mathrm{c}}_{i}= \left(\Delta r\right)^{\mathrm{f}}_{i}$
in the case of an equidistant radial spacing.

The implicit form
can now be obtained by putting
$T^{n+1}_{i,j}\equiv T_{i,j}$
and by placing all $T_{i,j}$-dependent
terms on one side of the left-hand side, while
moving the remaining terms to the right-hand side.
Because any non-linear terms in temperature
would make the problem difficult to invert,
we shall linearize the equation.
To do so, the diffusion coefficients are evaluated
using the hydrodynamic quantities from the beginning
of the sub-step.

Concerning the remaining source terms and their
linearity, $Q_{\mathrm{visc}}$, $Q_{\mathrm{acc}}$
and $Q_{\mathrm{irr}}$ terms
are temperature independent.
The compressional heating term is linear in temperature
thus it can be easily incorporated in the
left-hand side.
The vertical radiative cooling term $Q_{\mathrm{vert}}$
is proportional to $T^{4}$ but it can be linearized,
as e.g. in \cite{Commercon_etal_2011A&A...529A..35C}
or \cite{Bitsch_etal_2013A&A...549A.124B}.
If the temperature changes over $\Delta t$
are sufficiently small, we can rewrite Eq.~\ref{eq:qminus} as
\begin{equation}
  \begin{split}
    (Q_{\mathrm{vert}})_{i,j} & = \frac{2\sigma_{\mathrm{R}}}{(\tau_{\mathrm{eff}})_{i,j}}(T_{i,j}^{\mathrm{n}})^{4}\left(1+\frac{T-T^{\mathrm{n}}}{T^{\mathrm{n}}}\right)^{4}_{i,j} \\
    & \approx \frac{2\sigma_{\mathrm{R}}}{(\tau_{\mathrm{eff}})_{i,j}}\left[4(T^{\mathrm{n}})^{3}T - 3(T^{\mathrm{n}})^{4}\right]_{i,j} \equiv (Q'_{\mathrm{vert}}T - Q''_{\mathrm{vert}})_{i,j}  \, .
  \end{split}
  \label{eq:qminus_linearized}
\end{equation}

After some algebraic rearrangements,
we can formally write
\begin{equation}
  \begin{split}
  A_{i,j}T_{i,j} &+ B_{i,j}T_{i+1,j} + C_{i,j}T_{i-1,j} + D_{i,j}T_{i,j+1} + E_{i,j}T_{i,j-1} \\
  &= T_{i,j}^{\mathrm{n}} + \Delta t\left(\frac{Q_{\mathrm{visc}} + Q_{\mathrm{irr}}  + Q_{\mathrm{acc}} + Q''_{\mathrm{vert}}}{\Sigma c_{V}}\right)_{i,j} \, ,
  \end{split}
  \label{eq:implicit_scheme}
\end{equation}
which is a linear matrix equation.
To solve this linear problem, we 
use the successive over-relaxation (SOR)
method with odd-even ordering. 
Our implementation is parallelized
by the domain splitting
which is complementary to the radial grid decomposition
of the original \textsc{fargo} code. The optimization
of the over-relaxation parameter
is done similarly to \cite{Kley_1989A&A...208...98K}.

\section{Steady-state motion equations of a pebble}
\label{sec:motion_pebble}  

Here we reproduce the derivation of the
Eqs.~(\ref{eq:vr_pebbles})
and (\ref{eq:vt_pebbles}) which are used
to initialize the velocity field
of the pebble disk. The approach is well known
and closely follows the derivation of
\cite{Adachi_etal_1976PThPh..56.1756A},
with one clarification.

Let us study a system consisting of a pebble
with negligible mass which orbits a massive
primary $M_{\star}$ and experiences the
aerodynamic friction acceleration $F_{\mathrm{D}}$
in the gaseous environment at the same time.
We further assume that the motion is confined
in one plane and no vertical perturbations
are present.

The dynamical equation for the pebble
takes form
\begin{equation}
  \frac{\mathrm{d}^{2}\vec{r}}{\mathrm{d}t^{2}} = -\frac{GM_{\star}}{r^{3}}\vec{r} + \vec{F_{\mathrm{D}}} \, .
  \label{eq:pebble_motion_eq}
\end{equation}
Transforming into polar coordinates,
one obtains
\begin{equation}
  \frac{\partial V_{r}}{\partial t} + V_{\mathrm{r}}\frac{\partial V_{r}}{\partial r} - \frac{V_{\theta}^{2}}{r} =
  -\frac{GM_{\star}}{r^{2}} - \frac{F_{\mathrm{D}}}{v_{\mathrm{rel}}}\left(V_{r} - v_{r}\right) \, ,
  \label{eq:vr_pebble_polar_full}
\end{equation}
\begin{equation}
  \frac{\partial V_{\theta}}{\partial t} + V_{\mathrm{r}}\frac{\partial V_{\theta}}{\partial r} - \frac{V_{r}V_{\theta}}{r} =
  - \frac{F_{\mathrm{D}}}{v_{\mathrm{rel}}}\left(V_{\theta} - v_{\theta}\right) \, ,
  \label{eq:vt_pebble_polar_full}
\end{equation}
where we utilize the fact that the friction
force is directed against the relative velocity
vector, having the magnitude
$v_{\mathrm{rel}}=\sqrt{\left(V_{r}-v_{r}\right)^{2}+\left(V_{\theta}-v_{\theta}\right)^{2}}$.
Unlike \cite{Adachi_etal_1976PThPh..56.1756A},
we retain the $v_{r}$ component of the flow and
allow for the radial transport in the gaseous
disk \cite[see also][]{Guillot_etal_2014A&A...572A..72G}.

Let us simplify the equations above by assuming
a steady-state situation, $\partial_{t}=0$. Furthermore,
we only allow the drag force to cause small
perturbations in pebble's azimuthal velocity,
compared to the local Keplerian rotation.
We thus decompose $V_{\theta}=v_{\mathrm{K}}+V'_{\theta}$,
using $\left|V'_{\theta}\right|\lesssim\delta\ll v_{\mathrm{K}}$.
Similarly, the radial velocity of the pebble itself
is considered to be highly sub-Keplerian
$\left|V_{r}\right|\lesssim\delta\ll v_{\mathrm{K}}$.
We assume that the spatial derivatives
of $V'_{\theta}$ and $V_{r}$ are also as small as $\delta$.

In the equation (\ref{eq:vr_pebble_polar_full}),
the first and the second term on the left-hand side
can be neglected in our approximation,
while the third term can be rearranged
using the $V_{\theta}$ decomposition. Consequently
\begin{equation}
  v_{\mathrm{K}}^{2} + 2v_{\mathrm{K}}V'_{\theta} + \mathcal{O}\left(\delta^{2}\right)= v_{\mathrm{K}}^{2} + \frac{F_{\mathrm{D}}}{v_{\mathrm{rel}}}r\left(V_{r}-v_{r}\right) \, ,
  \label{eq:intermed1}
\end{equation}
which is obviously equivalent to
\begin{equation}
  2\Omega_{\mathrm{K}}V'_{\theta} = \frac{F_{\mathrm{D}}}{v_{\mathrm{rel}}}\left(V_{r}-v_{r}\right) \, .
  \label{eq:intermed2}
\end{equation}

Concerning Eq.~(\ref{eq:vt_pebble_polar_full}),
the first term on the left-hand side
can be again discarded but the radial derivative
has to be performed, leading to
\begin{equation}
  V_{r}\frac{\partial v_{\mathrm{K}}}{\partial r} + \frac{V_{r}v_{\mathrm{K}}}{r} + \mathcal{O}\left(\delta^{2}\right) = -\frac{F_{\mathrm{D}}}{v_{\mathrm{rel}}}\left(v_{\mathrm{K}}+V'_{\theta}-v_{\theta}\right) \, .
  \label{eq:intermed3}
\end{equation}
A useful simplification of the right-hand side
can be made using the $\eta$ parameter,
describing sub-Keplerian rotation of the gas
as $v_{\theta}=\left(1-\eta\right)v_{\mathrm{K}}$,
yielding
\begin{equation}
  \frac{1}{2}\Omega_{\mathrm{K}}V_{r} = - \frac{F_{\mathrm{D}}}{v_{\mathrm{rel}}}\left(V'_{\theta}+\eta v_{\mathrm{K}}\right) \, .
  \label{eq:intermed4}
\end{equation}

Recalling the Stokes number
definition
$\tau = t_{\mathrm{s}}\Omega_{\mathrm{K}} = v_{\mathrm{rel}}\Omega_{\mathrm{K}}/F_{\mathrm{D}}$,
one can rewrite the set of Eqs.~(\ref{eq:intermed2}) and (\ref{eq:intermed4})
as
\begin{equation}
  V_{r} = -\frac{2}{\tau} \left(V'_{\theta}+\eta v_{\mathrm{K}}\right) \, ,
  \label{eq:intermed5}
\end{equation}
\begin{equation}
  V'_{\theta} =  \frac{1}{2\tau}\left(V_{r} - v_{r}\right) \, .
  \label{eq:intermed6}
\end{equation}
Final arithmetic rearrangements are required
to eliminate $V'_{\theta}$ from $V_{r}$
and then plug them both back into the $V_{\theta}$
decomposition. The resulting set of equations
directly describes steady-state velocities
of the drifting pebble \citep{Guillot_etal_2014A&A...572A..72G}
\begin{equation}
  V_{r} = - \frac{2\tau}{1+\tau^{2}}\left(\eta v_{\mathrm{K}} - \frac{1}{2\tau}v_{r}\right) \, ,
  \label{eq:final_vr}
\end{equation}
\begin{equation}
  V_{\theta} = v_{\mathrm{K}} - \frac{1}{1+\tau^{2}}\left(\eta v_{\mathrm{K}} - \frac{\tau}{2}v_{r}\right) \, .
  \label{eq:final_vt}
\end{equation}

\section{Semi-implicit source term update of the pebble fluid}
\label{sec:semiimp_source_step}
In order to perform the source step \citep{Stone_Norman_1992ApJS...80..753S}
for the fluid of pebbles and avoid severe
time step limitations due to small friction time scales,
we do not use the explicit integration scheme for pebbles 
and utilize the semi-implicit approach of \cite{Rosotti_etal_2016MNRAS.459.2790R}
instead.

Let us rewrite the fluid motion Eqs.~(\ref{eq:navierestokes})
and (\ref{eq:navierestokes_peb}) in a symbolic notation
and without advection, which is solved separately. We have
\begin{equation}
  \frac{ \partial \vec{v}}{\partial t} = \vec{a}_{\mathrm{g}} \, ,
  \label{eq:navstok_short}
\end{equation}
\begin{equation}
  \frac{ \partial \vec{V}}{\partial t} = \vec{a}_{\mathrm{p}} - \frac{\Omega_{\mathrm{K}}}{\tau}\left(\vec{V}-\vec{v}\right) \, ,
  \label{eq:navstok_short_peb}
\end{equation}
where $\vec{a}_{\mathrm{p}}$ is the non-drag
acceleration of the pebble fluid
and $\vec{a}_{\mathrm{g}}$ is now understood as
the total acceleration acting on the gas which is
calculated explicitly at time $t$.
Note that the drag back-reaction term is
contained in $\vec{a}_{\mathrm{g}}$ and it is also
evaluated explicitly. This
is justified if the solid-to-gas
ratio remains low (which is what we expect in our
simulations).
Under these assumptions, an analytical solution
for the pebble fluid velocity update can be found
\citep{Rosotti_etal_2016MNRAS.459.2790R}:
\begin{equation}
  \begin{split}
    \vec{V}^{n+1} &= \vec{V}^{n}\exp\left(-\Delta t\frac{\Omega_{\mathrm{K}}}{\tau}\right)
  + \vec{a}_{\mathrm{g}}\Delta t \\
  & + \left[\vec{v}^{n} + \left(\vec{a}_{\mathrm{p}}-\vec{a}_{\mathrm{g}}\right)\frac{\tau}{\Omega_{\mathrm{K}}}\right]\left[1-\exp\left(-\Delta t\frac{\Omega_{\mathrm{K}}}{\tau}\right)\right] \, .
  \end{split}
  \label{eq:semiimp_source_step}
\end{equation}
The solution conveniently provides a smooth transition between two limiting
cases: When $\Delta t \ll \tau/\Omega_{\mathrm{K}}$, the solution is equivalent
to the explicit integration. If on the other hand $\Delta t \gg \tau/\Omega_{\mathrm{K}}$,
the solution turns into a form known as the short friction time approximation
\citep[e.g.][]{Johansen_Klahr_2005ApJ...634.1353J}.

To ensure the numerical stability, a CFL condition, additional
to the one which controls the gas evolution,
must be imposed on the time step $\Delta t$. The
condition is given by
\begin{equation}
  \Delta t = C\frac{\Delta x_{r,\theta}}{\max\left(V,V-v\right)_{r,\theta}} \, ,
  \label{eq:CLF_peb}
\end{equation}
where $\Delta x$ is the cell size in the radial (index~$r$)
or azimuthal (index~$\theta$) direction
and $C=0.5$ is the Courant number.

\section{Verification of the code}
\label{sec:verification}

\paragraph{Embryo-disk interaction in radiative disks.}
Here we try to reproduce several recent advanced
simulations of the embryo-disk interactions
using our new hydrodynamic code.
These test runs are compared against the original results
in order to provide a verification of
our code and some benchmarks.
Note that most of the comparison models
are 3D whereas our code is essentially 2D.
The results of the verification runs therefore prove
that we are indeed able to capture many aspects
of 3D models if the physics is treated carefully.
In the following, the stellar irradiation is always
neglected as well as the pebble disk,
and the opacity drop factor $c_{\kappa}=0.6$
is introduced into the simulation parameters.
Comparison figures are always provided in the unit
systems corresponding to the original works.

\begin{figure}
\centering
\includegraphics[width=8.8cm]{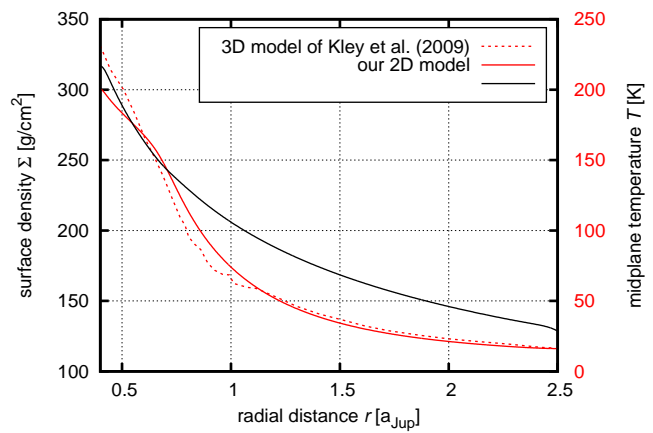}
\caption{The equilibrium gas surface density
$\Sigma(r)$ (black curve, left vertical axis) and temperature $T(r)$ profile
(red curve, right vertical axis) in a radiative disk according to
the setup from \cite{Kley_etal_2009A&A...506..971K}, as it was
reproduced by our code.
Temperature profile obtained by the original 3D model 
of \cite{Kley_etal_2009A&A...506..971K} is given by the red dashed
curve for comparison.
The obtained disk is indeed in good agreement
with the comparison simulation
and serves as the hydrodynamic background
for verification runs of the disk-embryo interaction.
}
\label{fig:kleysetup_profiles}
\end{figure}

First, we present a reproduction of an equilibrium
gas disk corresponding to the initial
setup of \cite{Kley_etal_2009A&A...506..971K}
who performed simulations using the 3D \textsc{nirvana}
code. The comparison of the radial temperature profile $T(r)$
is given in Fig.~\ref{fig:kleysetup_profiles}. The surface density profile
$\Sigma(r)$ is also displayed for reference (without a comparison
curve for clarity of the figure).
We see that $T(r)$ is in a good agreement
with the 3D model, apart from variations in the inner disk.
These are missing mostly because our 2D model does not
support vertical convection.

\begin{figure}
\centering
\includegraphics[width=8.8cm]{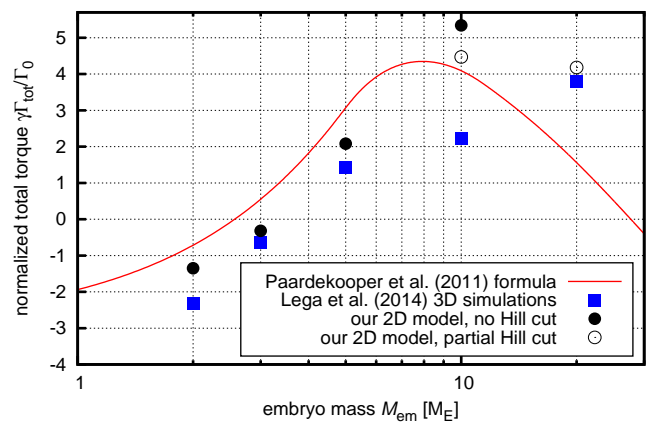}
\caption{A comparison of the normalized total torque
  $\gamma\Gamma_{\mathrm{tot}}/\Gamma_{0}$ acting
  on embryos of various masses $M_{\mathrm{em}}$, moving on fixed
  circular orbits in the disk shown in Fig.~\ref{fig:kleysetup_profiles}.
  The results achieved with our code are shown by black circles,
  or open circles if the Hill cut was applied.
  Values obtained by 3D calculations of \cite{Lega_etal_2014MNRAS.440..683L}
  are represented by blue squares. Formula from \cite{Paardekooper_etal_2011MNRAS.410..293P}
  applied to the equilibrium disk profile (with the potential
  smoothing parameter $\epsilon=0.4$) is given by the red curve.
  We consider the differences between our model
  and the comparison simulations to be acceptable.
}
\label{fig:torquecomparison_lega_paardek}
\end{figure}

\begin{figure}
\centering
\includegraphics[width=8.8cm]{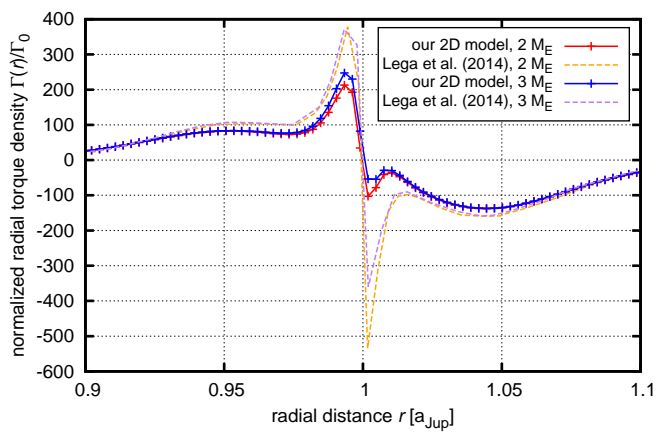}
\caption{The normalized radial torque density $\Gamma(r)/\Gamma_{0}$ acting
on $2\,\mathrm{M_{E}}$ and $3\,\mathrm{M_E}$ embryos as obtained by
our code (red and blue curve, respectively). Results
of the original 3D experiment from \cite{Lega_etal_2014MNRAS.440..683L}
are given for comparison (orange dashed curve for $2\,\mathrm{M_{E}}$
and purple dashed curve for $3\,\mathrm{M_E}$). As the cold finger
structure is not entirely reproduced by our code, the 
torque density peaks are less pronounced. However, the overall
torque (i.e. the integral of $\Gamma(r)$ over~$r$) is still
in very good agreement with the 3D model
(cf.~Fig.~\ref{fig:torquecomparison_lega_paardek}).
}
\label{fig:torquedens_nocut_lowmass}
\end{figure}

\begin{figure}
\centering
\includegraphics[width=8.8cm]{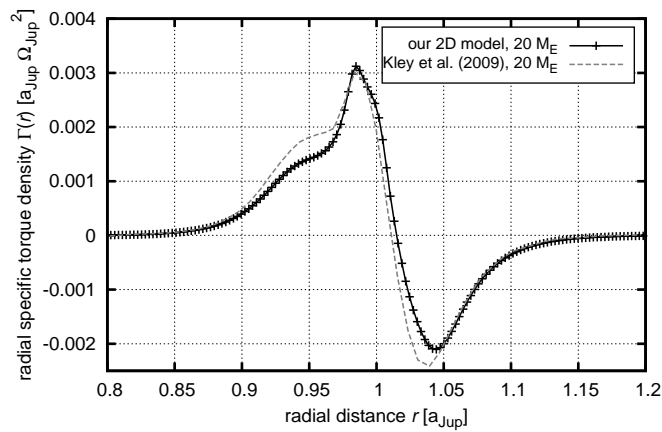}
\caption{The radial density of the specific torque $\Gamma(r)$ acting
  on a $20\,\mathrm{M_{E}}$ embryo as calculated by our code (black curve).
  The comparative profile from the original 3D experiment of
  \cite{Kley_etal_2009A&A...506..971K} is represented by the gray dashed line.
  Again, the agreement is very good.
}
\label{fig:torquedens_hillcut}
\end{figure}

We use exactly this equilibrium disk to compare the embryo-disk
interactions for various masses $M_{\mathrm{em}}$.
Since this work is focused on low-mass embryos,
we perform tests with $M_{\mathrm{em}}=2, 3, 5, 10$ and $20\,\mathrm{M_{E}}$.
This range of masses was studied by \cite{Lega_etal_2014MNRAS.440..683L}
who utilized the 3D \textsc{fargoca} code and conveniently, the same
equilibrium disk model was used
in their work. The embryo mass $M_{\mathrm{em}}=20\,\mathrm{M_{E}}$
was also studied by \cite{Kley_etal_2009A&A...506..971K}.
It is customary to exclude part of the gas
enclosed by the Hill sphere from the torque calculation (a so-called Hill cut)
if the planet is massive enough to form a distinct circumplanetary
disk. However, the determination of the threshold mass is not straightforward.
Thus we always perform the Hill cut for $M_{\mathrm{em}}=20\,\mathrm{M_{E}}$
and for $M_{\mathrm{em}}=10\,\mathrm{M_{E}}$ we perform
two simulations with and without the Hill cut.
For lower masses, no gas is excluded from calculations.

After placing the embryos on fixed
circular orbits with $a=a_{\mathrm{Jup}}=5.2\,\mathrm{AU}$, we evolved the system
for several tens of orbits until
the torque converged to a stationary value.
In Fig.~\ref{fig:torquecomparison_lega_paardek}, we compare
the measured normalized torques with results of \cite{Lega_etal_2014MNRAS.440..683L}
as well as with the torque--mass dependence given by the formulae
of \cite{Paardekooper_etal_2011MNRAS.410..293P}, applied to the
equilibrium disk. For low-mass embryos, the agreement
seems good enough. The torque in our model is generally between
the prediction of \cite{Paardekooper_etal_2011MNRAS.410..293P}
and the result of the 3D model from \cite{Lega_etal_2014MNRAS.440..683L}.
The torque on the $M_{\mathrm{em}}=10\,\mathrm{M_{E}}$ embryo
differs the most, nevertheless the result is improved when the Hill cut is applied.
For the medium-mass embryo $M_{\mathrm{em}}=20\,\mathrm{M_{E}}$, we
see that the value is in agreement with \cite{Lega_etal_2014MNRAS.440..683L}
which is a desirable result as 3D models generally
lead to the torque which is larger than the prediction
by \cite{Paardekooper_etal_2011MNRAS.410..293P}
by a factor of $3$ to $4$ \citep{Bitsch_Kley_2011A&A...536A..77B}
for the medium-mass embryos.

\cite{Lega_etal_2014MNRAS.440..683L} also
discovered the so-called cold finger structure near low-mass
embryos. These overdensity structures are responsible
for a modification of the radial torque density profile,
it is thus worth checking whether we can
find these modifications using our code as well.
In Fig.~\ref{fig:torquedens_nocut_lowmass}, we
plot the normalized radial torque density $\Gamma(r)/\Gamma_{0}$
(Eq.~\ref{eq:radial_torque}) for $2\,\mathrm{M_{E}}$
and $3\,\mathrm{M_{E}}$ embryos, compared to corresponding
results from \cite{Lega_etal_2014MNRAS.440..683L}.
It is obvious that the strong positive and negative
peaks are less pronounced in our case. As the cold finger is
responsible for the enhancement of these peaks, the effect
is not entirely recovered by our code.
We conclude that this is due to the local nature of the cold finger effect.
In our model, the gas flow around an embryo follows the velocity field
affected by the vertically-averaged potential and the resulting compressional
heating is not strong enough for the cold finger effect to fully develop.
Nevertheless, the overall torque magnitude obtained by our model
is still viable (Fig.~\ref{fig:torquecomparison_lega_paardek})
as the asymmetry of the positive versus negative contributions
is preserved to a satisfactory level.

Finally, we compare the torque for the upper end
of the tested embryo mass spectrum.
Fig.~\ref{fig:torquedens_hillcut} shows the radial
specific torque density (not normalized) for $M_{\mathrm{em}}=20\,\mathrm{M_{E}}$
compared to the result of \cite{Kley_etal_2009A&A...506..971K}.
The agreement is very good in this case, with slight
departures from the 3D model.

\paragraph{The heating torque.}

\begin{figure}
\centering
\includegraphics[width=8.8cm]{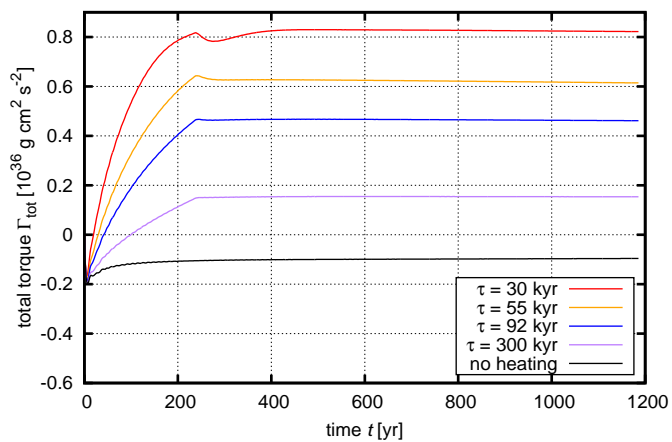}
\caption{
  The total torque $\Gamma_{\mathrm{tot}}$
  measurement in the experiment according to \cite{Benitez-Llambay_etal_2015Natur.520...63B},
  reproduced using our 2D code. 
  The $3\,\mathrm{M_{E}}$ embryo is either non-accreting (black curve)
  or growing with the doubling time $\tau$ (see the legend).
  The positive heating torque gets stronger
  with high accretion rates corresponding to short doubling times.
}
\label{fig:benitez}
\end{figure}

In order to assess how the heating torque is recovered
by our code, we repeated the numerical experiment
from \cite{Benitez-Llambay_etal_2015Natur.520...63B}.
Their setup is different from the verification
runs above, namely the surface density profile is different
and the opacity is assumed constant, $\kappa=1\,\mathrm{cm^2\,g^{-1}}$.
Therefore, we prevented any vertical opacity drop ($c_{\kappa}=1$)
in our test.
The stellar irradiation and pebble disk are again excluded.
We use grid resolution $N_{r}=738$ and $N_{\theta}=1382$,
unlike \cite{Benitez-Llambay_etal_2015Natur.520...63B} who
used $512$ cells in radius and $1024$ cells in azimuth
but also included colatitude.

An embryo with $M_{\mathrm{em}}=3\,\mathrm{M_E}$ is embedded in the disk
at $a_{\mathrm{Jup}}$
after the relaxation phase and the static torque is measured.
The source of the mass growth and accretion heating
is simply parametrized using the embryo mass doubling time
$\tau = M_{\mathrm{em}}/\dot{M}_{\mathrm{em}}$.
We studied cases with fixed embryo mass and with
$\tau=30$, $55$, $92$ and $300\,\mathrm{kyr}$.
Shorter $\tau$ means higher accretion rate and should correspond
to stronger heating torque.

The results of our test are shown in Fig.~\ref{fig:benitez}
which can be directly compared with the original experiment
in \cite{Benitez-Llambay_etal_2015Natur.520...63B} (cf. their
Fig.~2). First, it is important to notice that the steady-state
torque on the embryo in the absence of heating is less negative in our case.
This essentially corresponds to Fig.~\ref{fig:torquecomparison_lega_paardek},
where we found that the torque acting on the low-mass embryos in our model is always
more positive than in 3D models. Another reason might
be related to the midplane resolution which
is slightly better in our test, thus we cover the embryo's horseshoe
region with more cells. According to \cite{Lega_etal_2014MNRAS.440..683L},
increasing the resolution of the horseshoe region makes
the torque more positive.

Because the torque in the absence of heating is less negative
compared to \cite{Benitez-Llambay_etal_2015Natur.520...63B},
it is then easier for even the low accretion rates and respective
luminosities to revert the migration because the heating torque
does not have to compete with strong negative counteracting torques.

Finally, the torque scaling with increasing accretion rate is
more efficient in our model than in the original 3D model.
We notice that the total difference between the torque with
$\tau=30\,\mathrm{kyr}$ and the torque without accretion is
$\Delta \Gamma \approx 0.9\times10^{36}\,\mathrm{g\,cm^{2}\,s^{-2}}$,
compared to $\Delta \Gamma \approx 0.6\times10^{36}\,\mathrm{g\,cm^{2}\,s^{-2}}$
found by the 3D modelling.
The slight discrepancy is again caused by the vertically averaged flow
field around the planet (as already discussed for the cold finger effect) and
also due to the simplified treatment of the radiative diffusion which in our case is
acting only in the midplane and it is replaced by an approximation of the 
radiation escape in the vertical direction.
Yet we consider the heating torque to be reproduced
accurately enough and we shall strive in future works
to achieve an improved agreement with the 3D model.

\end{appendix}

\end{document}